%% file: sss05.tex
\documentclass[aps,prd,preprint,tightenlines,superscriptaddress,showpacs,byrevtex]{revtex4}
\usepackage{graphicx}
\usepackage{dcolumn}
\usepackage{bm}
\usepackage{pstricks}
\usepackage{pst-node}

\input{defs.tex}

\input{defs-sss05.tex}

\begin{document}

\preprint{\vbox{
                 \hbox{BELLE-CONF-0569}
                 \hbox{LP2005-204}
}}

\input{title-sss05.tex}

\input{author-conf2005.tex}

\input{abstract-sss05.tex}

\maketitle

\input{main-sss05.tex}

\section*{Acknowledgments}
\input{lp05ack.tex}


\input{bib-sss05.tex}
\end{document}

%% file: defs.tex
\def\bz{{B^0}}
\def\bzb{{\overline{B}{}^0}}
\def\bp{{B^+}}
\def\bm{{B^-}}
\def\kl{K_L^0}
\def\dE{{\Delta E}}
\def\mb{{M_{\rm bc}}}
\def\Dt{\Delta t}
\def\Dz{\Delta z}
\def\fol{f_{\rm ol}}
\def\fsig{f_{\rm sig}}
\newcommand{\sinbb}{{\sin2\phi_1}}
\def\sinbbeff{{\sin2\phi_1^{\rm eff}}}

\newcommand{\fCP}{f_{CP}}

\newcommand{\ftag}{f_{\rm tag}}

\newcommand{\cala}{{\cal A}_f}
\newcommand{\cals}{{\cal S}_f}

\newcommand{\dm}{\Delta m_d}
\newcommand{\dmd}{\dm}
\def\taubz{{\tau_\bz}}
\def\taubp{{\tau_\bp}}
\def\ks{{K_S^0}}
\newcommand{\btosqq}{b \to s\overline{q}q}
\newcommand{\btosss}{b \to s\overline{s}s}
\def\btoccs{b \to c\overline{c}s}
\newcommand*{\dwl}{\ensuremath{{\Delta w_l}}}
\newcommand*{\fq}{\ensuremath{q}}
\def\kz{{K^0}}
\def\kp{{K^+}}
\def\km{{K^-}}
\def\fzero{{f_0}}
\def\fx{{f_X(1300)}}
\def\pip{{\pi^+}}
\def\pim{{\pi^-}}
\def\piz{{\pi^0}}
\def\kstarz{{K^{*0}}}
\def\kstarp{{K^{*+}}}

\def\kl{{K_L^0}}
\def\bbar{{\overline{B}}}
\def\ufs{{\Upsilon(4S)}}
\def\nev{{N_{\rm ev}}}
\def\nsig{{N_{\rm sig}}}

\def\nsigmc{{N_{\rm sig}^{\rm MC}}}
\def\nbkg{{N_{\rm bkg}}}

\def\jpsi{{J/\psi}}

\def\rhoz{{\rho^0}}

\newcommand{\dslnu}{D^{*-}\ell^+\nu}
\newcommand{\bzdslnu}{\bz \to \dslnu}

\def\sperp{{S_{\perp}}}
\def\lsig{{\cal L}_{\rm sig}}
\def\lbkg{{\cal L}_{\rm bkg}}
\def\rsigbkg{{\cal R}_{\rm s/b}}
\def\calf{{\cal F}}
\def\rkpi{{\cal R}_{K/\pi}}

\def\mgg{M_{\gamma\gamma}}
\def\ppizcms{p_\piz^{\rm cms}}

\def\pbstar{p_B^{\rm cms}}

\def\eeff{\ensuremath{\epsilon_\textrm{eff}}}

\def\etap{{\eta'}}

\def\kspm{{K_S^{+-}}}
\def\kszz{{K_S^{00}}}

\def\rsig{{R_{\rm sig}(\Dt)}}

%% file: defs-sss05.tex
\def\nbb{386}
\def\nbblastsummer{275}
\def\nbbthisyear{111}
\def\nbbsvdone{152}
\def\nbbsvdtwo{234}
\def\lint{357}
\def\lintlastsummer{253}
\def\lintthisyear{104}
\def\lintsvdone{140}
\def\lintsvdtwo{217}

\def\efftot{{0.30\pm 0.01}}

\def\efftotdsone{{0.287 \pm 0.005}}

\def\sinbbWA{+0.726}
\def\sinbbERR{0.037}
\def\sinbbWAResult{\sinbbWA\pm\sinbbERR}

   \def\Pphiks{0.57}     
   \def\Nsigphiks{180\pm 16}

   \def\Pphikl{0.12}     
   \def\Nsigphikl{78\pm 13}
   \def\Pfzeroks{0.47}    
   \def\Nsigfzeroks{145\pm 16}
   \def\Petapks{0.61}    
   \def\Nsigetapks{830 \pm 35}

   \def\Petapkl{0.26}    
   \def\Nsigetapkl{187\pm 18}
   \def\Pomegaks{0.17}    
   \def\Nsigomegaks{68\pm13}
   \def\Pkspiz{0.25}     
   \def\Nsigkspiz{344\pm 30}

   \def\NsigkspizH{248\pm 20}

   \def\NsigkspizL{96\pm 23}
   \def\Pksksks{0.64}    
   \def\Nsigksksks{105 \pm 12}

   \def\Pkspmkspmkspm{0.70}    
   \def\Nsigkspmkspmkspm{88\pm 10}

   \def\Pkspmkspmkszz{0.43}
   \def\Nsigkspmkspmkszz{16\pm 6}
   \def\Pkpkmks{0.55}    
   \def\Nsigkpkmks{536 \pm 29}
   \def\Pjpsiks{0.98} 
   \def\Nsigjpsiks{5264 \pm 73} 

   \def\Pjpsikl{0.60}    
   \def\Nsigjpsikl{4792 \pm 105}
\def\SphikzVal{+0.44}     \def\SphikzStat{0.27}     \def\SphikzSyst{0.05}
\def\sinbbphikzVal{\SphikzVal} 
\def\sinbbphikzStat{\SphikzStat} 
\def\sinbbphikzSyst{\SphikzSyst}
\def\AphikzVal{+0.14}     \def\AphikzStat{0.17}     \def\AphikzSyst{0.07}
\def\SphiksVal{+0.19}     \def\SphiksStat{0.32}     
 
\def\AphiksVal{+0.12}     \def\AphiksStat{0.20}     
    \def\mSphiklStat{0.59} 
\def\SphiklVal{-1.54}    \def\SphiklStat{\mSphiklStat}
 
\def\AphiklVal{+0.38}     \def\AphiklStat{0.36}  
\def\SfzeroksVal{-0.47} \def\SfzeroksStat{0.36} \def\SfzeroksSyst{0.08}
\def\sinbbfzeroksVal{+0.47} 
\def\sinbbfzeroksStat{\SfzeroksStat} 
\def\sinbbfzeroksSyst{\SfzeroksSyst}
\def\AfzeroksVal{-0.23} \def\AfzeroksStat{0.23} \def\AfzeroksSyst{0.13}
\def\SetapkzVal{+0.62}  \def\SetapkzStat{0.12}  \def\SetapkzSyst{0.04}
\def\sinbbetapkzVal{\SetapkzVal}  \def\sinbbetapkzStat{\SetapkzStat}  \def\sinbbetapkzSyst{\SetapkzSyst}
\def\AetapkzVal{-0.04}  \def\AetapkzStat{0.08}  \def\AetapkzSyst{0.06}
\def\SetapksVal{+0.60}  \def\SetapksStat{0.14} 
  
\def\AetapksVal{-0.04}  \def\AetapksStat{0.09} 
 \def\mSetapklStat{0.29}
\def\SetapklVal{-0.73} \def\SetapklStat{\mSetapklStat}
 
\def\AetapklVal{-0.02}  \def\AetapklStat{0.18} 
\def\SomegaksVal{+0.95} \def\SomegaksStat{0.53} \def\SomegaksSyst{^{+0.12}_{-0.15}}
\def\sinbbomegaksVal{\SomegaksVal} 
\def\sinbbomegaksStat{\SomegaksStat}
\def\sinbbomegaksSyst{\SomegaksSyst}
\def\AomegaksVal{+0.19} \def\AomegaksStat{0.39} \def\AomegaksSyst{0.13}
\def\SkspizVal{+0.22}   \def\SkspizStat{0.47}   \def\SkspizSyst{0.08}
\def\sinbbkspizVal{\SkspizVal}   
\def\sinbbkspizStat{\SkspizStat}   
\def\sinbbkspizSyst{\SkspizSyst}
\def\AkspizVal{+0.11}   \def\AkspizStat{0.18}   \def\AkspizSyst{0.08}
\def\SksksksVal{-0.58}  \def\SksksksStat{0.36}  \def\SksksksSyst{0.08}
\def\sinbbksksksVal{+0.58}  
\def\sinbbksksksStat{\SksksksStat}  
\def\sinbbksksksSyst{\SksksksSyst}
\def\AksksksVal{+0.50}  \def\AksksksStat{0.23}  \def\AksksksSyst{0.06}
\def\SkpkmksVal{-0.52}  \def\SkpkmksStat{0.16}  \def\SkpkmksSyst{0.03}
\def\sinbbkpkmksVal{+0.60}  
\def\sinbbkpkmksStat{0.18}  
\def\sinbbkpkmksSyst{0.04}
\def\sinbbkpkmksFevenErr{^{+0.19}_{-0.12}}
\def\AkpkmksVal{-0.06}  \def\AkpkmksStat{0.11}  \def\AkpkmksSyst{0.07}
\def\FcpkpkmksVal{0.93}
\def\FcpkpkmksStat{0.09}
\def\FcpkpkmksSyst{0.05}
\def\xifkpkmksVal{+0.86}
\def\xifkpkmksStat{0.18}
\def\xifkpkmksSyst{0.09}

\def\FcpkpkmksResultSS
  {\FcpkpkmksVal\pm\FcpkpkmksStat\mbox{(stat)}\pm\FcpkpkmksSyst\mbox{(syst)}}
\def\xifkpkmksResult{\xifkpkmksVal\pm\xifkpkmksStat\pm\xifkpkmksSyst}
\def\xifkpkmksResultSS
  {\xifkpkmksVal\pm\xifkpkmksStat\mbox{(stat)}\pm\xifkpkmksSyst\mbox{(syst)}}
\def\SjpsikzVal{+0.652}  \def\SjpsikzStat{0.039}  \def\SjpsikzSyst{0.020}

\def\AjpsikzVal{+0.010}  \def\AjpsikzStat{0.026}  \def\AjpsikzSyst{0.036}
\def\SjpsiksVal{+0.668}  \def\SjpsiksStat{0.047}

\def\AjpsiksVal{-0.021}  
\def\AjpsiksStat{0.034}  

\def\mSjpsiklStat{0.069} 
\def\SjpsiklVal{-0.619} 
\def\SjpsiklStat{\mSjpsiklStat}

\def\AjpsiklVal{+0.049}  
\def\AjpsiklStat{0.039}  
\def\SphikzResult{\SphikzVal\pm\SphikzStat\pm\SphikzSyst}

\def\sinbbphikzResult{\sinbbphikzVal\pm\sinbbphikzStat\pm\sinbbphikzSyst}

\def\AphikzResult{\AphikzVal\pm\AphikzStat\pm\AphikzSyst}

\def\SphiksResultStat{\SphiksVal\pm\SphiksStat}

\def\AphiksResultStat{\AphiksVal\pm\AphiksStat}

\def\SphiklResultStat{\SphiklVal\pm\SphiklStat}

\def\AphiklResultStat{\AphiklVal\pm\AphiklStat}
\def\SfzeroksResult{\SfzeroksVal\pm\SfzeroksStat\pm\SfzeroksSyst}
\def\sinbbfzeroksResult{\sinbbfzeroksVal\pm\sinbbfzeroksStat\pm\sinbbfzeroksSyst}

\def\AfzeroksResult{\AfzeroksVal\pm\AfzeroksStat\pm\AfzeroksSyst}

\def\SetapkzResult{\SetapkzVal\pm\SetapkzStat\pm\SetapkzSyst}
\def\sinbbetapkzResult{\sinbbetapkzVal\pm\sinbbetapkzStat\pm\sinbbetapkzSyst}

\def\AetapkzResult{\AetapkzVal\pm\AetapkzStat\pm\AetapkzSyst}

\def\SetapksResultStat{\SetapksVal\pm\SetapksStat}

\def\AetapksResultStat{\AetapksVal\pm\AetapksStat}

\def\SetapklResultStat{\SetapklVal\pm\SetapklStat}

\def\AetapklResultStat{\AetapklVal\pm\AetapklStat}
\def\SomegaksResult{\SomegaksVal\pm\SomegaksStat\SomegaksSyst}
\def\sinbbomegaksResult{\sinbbomegaksVal\pm\sinbbomegaksStat\sinbbomegaksSyst}

\def\AomegaksResult{\AomegaksVal\pm\AomegaksStat\pm\AomegaksSyst}

\def\SkspizResult{\SkspizVal\pm\SkspizStat\pm\SkspizSyst}
\def\sinbbkspizResult{\sinbbkspizVal\pm\sinbbkspizStat\pm\sinbbkspizSyst}

\def\AkspizResult{\AkspizVal\pm\AkspizStat\pm\AkspizSyst}

\def\SksksksResult{\SksksksVal\pm\SksksksStat\pm\SksksksSyst}
\def\sinbbksksksResult{\sinbbksksksVal\pm\sinbbksksksStat\pm\sinbbksksksSyst}

\def\AksksksResult{\AksksksVal\pm\AksksksStat\pm\AksksksSyst}

\def\SkpkmksResult{\SkpkmksVal\pm\SkpkmksStat\pm\SkpkmksSyst}
\def\sinbbkpkmksResult{\sinbbkpkmksVal\pm\sinbbkpkmksStat\pm\sinbbkpkmksSyst\sinbbkpkmksFevenErr}

\def\AkpkmksResult{\AkpkmksVal\pm\AkpkmksStat\pm\AkpkmksSyst}

\def\SjpsikzResult{\SjpsikzVal\pm\SjpsikzStat\pm\SjpsikzSyst}
\def\SjpsikzResultSS
  {\SjpsikzVal\pm\SjpsikzStat\mbox{(stat)}\pm\SjpsikzSyst\mbox{(syst)}}

\def\AjpsikzResult{\AjpsikzVal\pm\AjpsikzStat\pm\AjpsikzSyst}
\def\AjpsikzResultSS
  {\AjpsikzVal\pm\AjpsikzStat\mbox{(stat)}\pm\AjpsikzSyst\mbox{(syst)}}

\def\SjpsiksResultStat{\SjpsiksVal\pm\SjpsiksStat}

\def\AjpsiksResultStat{\AjpsiksVal\pm\AjpsiksStat}

\def\SjpsiklResultStat{\SjpsiklVal\pm\SjpsiklStat}

\def\AjpsiklResultStat{\AjpsiklVal\pm\AjpsiklStat}
\def\SphikzSystVtx{0.02}
\def\SphikzSystFbtg{0.01}
\def\SphikzSystResol{0.02}
\def\SphikzSystPhys{<0.01}
\def\SphikzSystFit{0.01}
\def\SphikzSystBG{0.01}
\def\SphikzSystBGdt{0.03}
\def\SphikzSystTagIntrfr{<0.01}
\def\AphikzSystVtx{0.04}
\def\AphikzSystFbtg{0.01}
\def\AphikzSystResol{0.01}
\def\AphikzSystPhys{<0.01}
\def\AphikzSystFit{0.01}
\def\AphikzSystBG{0.01}
\def\AphikzSystBGdt{0.02}
\def\AphikzSystTagIntrfr{0.05}
\def\SfzeroksSystVtx{0.02}
\def\SfzeroksSystFbtg{0.01}
\def\SfzeroksSystResol{0.02}
\def\SfzeroksSystPhys{0.01}
\def\SfzeroksSystFit{0.02}
\def\SfzeroksSystBG{0.02}
\def\SfzeroksSystBGdt{0.07}
\def\SfzeroksSystTagIntrfr{<0.01}
\def\AfzeroksSystVtx{0.05}
\def\AfzeroksSystFbtg{0.01}
\def\AfzeroksSystResol{0.02}
\def\AfzeroksSystPhys{<0.01}
\def\AfzeroksSystFit{0.01}
\def\AfzeroksSystBG{0.01}
\def\AfzeroksSystBGdt{0.09}
\def\AfzeroksSystTagIntrfr{0.07}
\def\SetapkzSystVtx{0.02}
\def\SetapkzSystFbtg{<0.01}
\def\SetapkzSystResol{0.02}
\def\SetapkzSystPhys{<0.01}
\def\SetapkzSystFit{0.02}
\def\SetapkzSystBG{0.02}
\def\SetapkzSystBGdt{<0.01}
\def\SetapkzSystTagIntrfr{<0.01}
\def\AetapkzSystVtx{0.04}
\def\AetapkzSystFbtg{0.01}
\def\AetapkzSystResol{<0.01}
\def\AetapkzSystPhys{<0.01}
\def\AetapkzSystFit{0.01}
\def\AetapkzSystBG{0.01}
\def\AetapkzSystBGdt{<0.01}
\def\AetapkzSystTagIntrfr{0.04}
\def\SomegaksSystVtx{0.02}
\def\SomegaksSystFbtg{0.04}
\def\SomegaksSystResol{0.05}
\def\SomegaksSystPhys{0.01}
\def\SomegaksSystFit{^{+0.01}_{-0.09}}
\def\SomegaksSystBG{0.09}
\def\SomegaksSystBGdt{0.01}
\def\SomegaksSystTagIntrfr{<0.01}
\def\AomegaksSystVtx{0.07}
\def\AomegaksSystFbtg{0.01}
\def\AomegaksSystResol{0.03}
\def\AomegaksSystPhys{<0.01}
\def\AomegaksSystFit{0.02}
\def\AomegaksSystBG{0.10}
\def\AomegaksSystBGdt{0.01}
\def\AomegaksSystTagIntrfr{0.04}
\def\SkspizSystVtx{0.01}
\def\SkspizSystFbtg{0.01}
\def\SkspizSystResol{0.05}
\def\SkspizSystPhys{0.01}
\def\SkspizSystFit{0.01}
\def\SkspizSystBG{0.01}
\def\SkspizSystBGdt{0.06}
\def\SkspizSystTagIntrfr{<0.01}
\def\AkspizSystVtx{0.04}
\def\AkspizSystFbtg{0.01}
\def\AkspizSystResol{<0.01}
\def\AkspizSystPhys{<0.01}
\def\AkspizSystFit{<0.01}
\def\AkspizSystBG{0.01}
\def\AkspizSystBGdt{0.03}
\def\AkspizSystTagIntrfr{0.06}
\def\SksksksSystVtx{0.04}
\def\SksksksSystFbtg{0.01}
\def\SksksksSystResol{0.05}
\def\SksksksSystPhys{<0.01}
\def\SksksksSystFit{0.03}
\def\SksksksSystBG{0.02}
\def\SksksksSystBGdt{0.02}
\def\SksksksSystTagIntrfr{0.01}
\def\AksksksSystVtx{0.04}
\def\AksksksSystFbtg{0.01}
\def\AksksksSystResol{0.02}
\def\AksksksSystPhys{<0.01}
\def\AksksksSystFit{0.02}
\def\AksksksSystBG{0.03}
\def\AksksksSystBGdt{0.01}
\def\AksksksSystTagIntrfr{0.05}
\def\SkpkmksSystVtx{0.02}
\def\SkpkmksSystFbtg{<0.01}
\def\SkpkmksSystResol{0.02}
\def\SkpkmksSystPhys{<0.01}
\def\SkpkmksSystFit{0.01}
\def\SkpkmksSystBG{0.02}
\def\SkpkmksSystBGdt{0.01}
\def\SkpkmksSystTagIntrfr{<0.01}
\def\AkpkmksSystVtx{0.04}
\def\AkpkmksSystFbtg{0.01}
\def\AkpkmksSystResol{0.01}
\def\AkpkmksSystPhys{<0.01}
\def\AkpkmksSystFit{0.01}
\def\AkpkmksSystBG{0.01}
\def\AkpkmksSystBGdt{<0.01}
\def\AkpkmksSystTagIntrfr{0.06}
\def\SjpsikzSystVtx{0.015}
\def\SjpsikzSystFbtg{0.006}
\def\SjpsikzSystResol{0.005}
\def\SjpsikzSystPhys{0.001}
\def\SjpsikzSystFit{0.007}
\def\SjpsikzSystBG{0.005}
\def\SjpsikzSystBGdt{0.006}
\def\SjpsikzSystTagIntrfr{0.003}
\def\AjpsikzSystVtx{0.030}
\def\AjpsikzSystFbtg{0.007}
\def\AjpsikzSystResol{0.001}
\def\AjpsikzSystPhys{0.001}
\def\AjpsikzSystFit{0.005}
\def\AjpsikzSystBG{0.004}
\def\AjpsikzSystBGdt{0.005}
\def\AjpsikzSystTagIntrfr{0.017}
\def\SjpsiksKsvVal{+0.73}
\def\SjpsiksKsvStat{0.08}
\def\AjpsiksKsvVal{+0.01}
\def\AjpsiksKsvStat{0.04}
\def\LifejpsiksKsvVal{1.55}
\def\LifejpsiksKsvStat{0.05}
\def\SjpsiksKsvResultStat{\SjpsiksKsvVal\pm\SjpsiksKsvStat}
\def\AjpsiksKsvResultStat{\AjpsiksKsvVal\pm\AjpsiksKsvStat}
\def\LifejpsiksKsvResultStat{\LifejpsiksKsvVal\pm\LifejpsiksKsvStat}

\def\fKKKsBGinphiKs{2.75\pm 0.14}

%% file: title-sss05.tex
\title{\boldmath Time-Dependent $CP$
       Asymmetries in $b \to s\overline{q}q$ Transitions
       and $\sinbb$ in $\bz\to\jpsi\kz$ Decays
       with 386 Million $B\overline{B}$ Pairs}

\date{\today}


%% file: author-conf2005.tex
\affiliation{Aomori University, Aomori}
\affiliation{Budker Institute of Nuclear Physics, Novosibirsk}
\affiliation{Chiba University, Chiba}
\affiliation{Chonnam National University, Kwangju}
\affiliation{University of Cincinnati, Cincinnati, Ohio 45221}
\affiliation{University of Frankfurt, Frankfurt}
\affiliation{Gyeongsang National University, Chinju}
\affiliation{University of Hawaii, Honolulu, Hawaii 96822}
\affiliation{High Energy Accelerator Research Organization (KEK), Tsukuba}
\affiliation{Hiroshima Institute of Technology, Hiroshima}
\affiliation{Institute of High Energy Physics, Chinese Academy of Sciences, Beijing}
\affiliation{Institute of High Energy Physics, Vienna}
\affiliation{Institute for Theoretical and Experimental Physics, Moscow}
\affiliation{J. Stefan Institute, Ljubljana}
\affiliation{Kanagawa University, Yokohama}
\affiliation{Korea University, Seoul}
\affiliation{Kyoto University, Kyoto}
\affiliation{Kyungpook National University, Taegu}
\affiliation{Swiss Federal Institute of Technology of Lausanne, EPFL, Lausanne}
\affiliation{University of Ljubljana, Ljubljana}
\affiliation{University of Maribor, Maribor}
\affiliation{University of Melbourne, Victoria}
\affiliation{Nagoya University, Nagoya}
\affiliation{Nara Women's University, Nara}
\affiliation{National Central University, Chung-li}
\affiliation{National Kaohsiung Normal University, Kaohsiung}
\affiliation{National United University, Miao Li}
\affiliation{Department of Physics, National Taiwan University, Taipei}
\affiliation{H. Niewodniczanski Institute of Nuclear Physics, Krakow}
\affiliation{Nippon Dental University, Niigata}
\affiliation{Niigata University, Niigata}
\affiliation{Nova Gorica Polytechnic, Nova Gorica}
\affiliation{Osaka City University, Osaka}
\affiliation{Osaka University, Osaka}
\affiliation{Panjab University, Chandigarh}
\affiliation{Peking University, Beijing}
\affiliation{Princeton University, Princeton, New Jersey 08544}
\affiliation{RIKEN BNL Research Center, Upton, New York 11973}
\affiliation{Saga University, Saga}
\affiliation{University of Science and Technology of China, Hefei}
\affiliation{Seoul National University, Seoul}
\affiliation{Shinshu University, Nagano}
\affiliation{Sungkyunkwan University, Suwon}
\affiliation{University of Sydney, Sydney NSW}
\affiliation{Tata Institute of Fundamental Research, Bombay}
\affiliation{Toho University, Funabashi}
\affiliation{Tohoku Gakuin University, Tagajo}
\affiliation{Tohoku University, Sendai}
\affiliation{Department of Physics, University of Tokyo, Tokyo}
\affiliation{Tokyo Institute of Technology, Tokyo}
\affiliation{Tokyo Metropolitan University, Tokyo}
\affiliation{Tokyo University of Agriculture and Technology, Tokyo}
\affiliation{Toyama National College of Maritime Technology, Toyama}
\affiliation{University of Tsukuba, Tsukuba}
\affiliation{Utkal University, Bhubaneswer}
\affiliation{Virginia Polytechnic Institute and State University, Blacksburg, Virginia 24061}
\affiliation{Yonsei University, Seoul}
  \author{K.~Abe}\affiliation{High Energy Accelerator Research Organization (KEK), Tsukuba} 
  \author{K.~Abe}\affiliation{Tohoku Gakuin University, Tagajo} 
  \author{I.~Adachi}\affiliation{High Energy Accelerator Research Organization (KEK), Tsukuba} 
  \author{H.~Aihara}\affiliation{Department of Physics, University of Tokyo, Tokyo} 
  \author{K.~Aoki}\affiliation{Nagoya University, Nagoya} 
  \author{K.~Arinstein}\affiliation{Budker Institute of Nuclear Physics, Novosibirsk} 
  \author{Y.~Asano}\affiliation{University of Tsukuba, Tsukuba} 
  \author{T.~Aso}\affiliation{Toyama National College of Maritime Technology, Toyama} 
  \author{V.~Aulchenko}\affiliation{Budker Institute of Nuclear Physics, Novosibirsk} 
  \author{T.~Aushev}\affiliation{Institute for Theoretical and Experimental Physics, Moscow} 
% ...
\collaboration{The Belle Collaboration}

%% file: abstract-sss05.tex
\begin{abstract}
  We present measurements of time-dependent $CP$ asymmetries in 
  $\bz\to$
  $\phi(1020)\kz$, 
  $\eta'\kz$, 
  $\ks\ks\ks$, 
  $\ks\piz$,
  $\fzero(980)\ks$,
  $\omega(782)\ks$
  and
  $\kp\km\ks$
  decays 
  based on a sample of $\nbb\times 10^6$ $B\bbar$ pairs
  collected at the $\ufs$ resonance with
  the Belle detector at the KEKB energy-asymmetric $e^+e^-$ collider.
  These decays are dominated by the $b \to s$ gluonic penguin
  transition and are sensitive to new $CP$-violating phases
  from physics beyond the standard model.
  One neutral $B$ meson is fully reconstructed in
  one of the specified decay channels,
  and the flavor of the accompanying $B$ meson is identified from
  its decay products.
  $CP$-violation parameters $\sinbbeff$ and $\cala$ for each of the decay 
  modes are obtained from the asymmetries in the distributions of
  the proper-time intervals between the two $B$ decays.
  We also perform an improved measurement of $CP$ asymmetries in 
  $\bz\to\jpsi\kz$ decays using the same data sample. 
  The same analysis procedure mentioned above
  yields $\sinbb = \SjpsikzResultSS$, which serves as a reference point for the standard model,
  and $\cala = \AjpsikzResultSS$.


\end{abstract}


%% file: main-sss05.tex
\section{Introduction}
\label{sec:introduction}

The flavor-changing $b \to s$ transition proceeds through loop penguin
diagrams.
Such loop diagrams play an important role in testing the standard model (SM) 
and new physics because particles beyond the SM can contribute via
additional loop diagrams.
$CP$ violation in the $b \to s$ transition is especially sensitive to physics
at a very high-energy scale~\cite{Akeroyd:2004mj}.
Theoretical studies indicate that
large deviations from the SM expectations
are allowed for time-dependent $CP$ asymmetries in
$\bz$ meson decays~\cite{bib:lucy}.
Experimental investigations have recently been launched
at the two $B$ factories, each of which has produced more than
$10^8$ $B\bbar$ pairs.
The first measurement 
of the $CP$-violating asymmetry
in $\bz\to \phi\ks$ decays~\cite{footnote:CC},
which are dominated by the $\btosss$ transition,
by the Belle collaboration
indicated deviation from the SM expectation~\cite{Abe:2003yt}.
Measurements with a larger data sample are required to 
confirm this difference. It is also essential to examine
additional modes that
are sensitive to the same $b \to s$ penguin amplitude.
In this spirit, experimental results based on a
sample of $\nbblastsummer\times 10^6$ $B\bbar$ pairs
using decay modes
$\bz\to$
$\phi\kz$, 
$\eta'\ks$, 
$\ks\ks\ks$,
$\ks\piz$,
$\fzero\ks$,
$\omega\ks$, 
and
$\kp\km\ks$~\cite{footnote:mesons}
have already been 
reported~\cite{Chen:2005dr,Sumisawa:2005fz}. 
The combined result differs from the SM expectation by
2.4 standard deviations.
Since measurements by the BaBar collaboration
also yield a similar deviation~\cite{bib:HFAG, bib:BaBar_sss},
the present world average 
differs from the SM expectation by 3.7 standard deviations.

In the SM, $CP$ violation arises from a single irreducible phase, 
the Kobayashi-Maskawa (KM) phase~\cite{Kobayashi:1973fv},
in the weak-interaction quark-mixing matrix. In
particular, the SM predicts $CP$ asymmetries in the time-dependent
rates for $\bz$ and
$\bzb$ decays to a common $CP$ eigenstate $\fCP$~\cite{bib:sanda}. 
In the decay chain $\Upsilon(4S)\to \bz\bzb \to f_{CP}f_{\rm tag}$,
where one of the $B$ mesons decays at time $t_{CP}$ to a 
final state $f_{CP}$ 
and the other decays at time $t_{\rm tag}$ to a final state  
$f_{\rm tag}$ that distinguishes between $B^0$ and $\bzb$, 
the decay rate has a time dependence
given by
\begin{equation}
\label{eq:psig}
{\cal P}(\Delta{t}) = 
\frac{e^{-|\Delta{t}|/{\taubz}}}{4{\taubz}}
\biggl\{1 + \fq\cdot 
\Bigl[ \cals\sin(\dmd\Delta{t})
   + \cala\cos(\dmd\Delta{t})
\Bigr]
\biggr\}.
\end{equation}
Here $\cals$ and $\cala$ are $CP$-violation parameters,
$\taubz$ is the $B^0$ lifetime, $\dmd$ is the mass difference 
between the two $B^0$ mass
eigenstates, $\Delta{t}$ = $t_{CP}$ $-$ $t_{\rm tag}$, and
the $b$-flavor charge $\fq$ = +1 ($-1$) when the tagging $B$ meson
is a $B^0$ ($\bzb$).
To a good approximation,
the SM predicts $\cals = -\xi_f\sinbb$, where $\xi_f = +1 (-1)$
corresponds to  $CP$-even (-odd) final states, and $\cala =0$
for both $\btoccs$ and 
$\btosqq$ transitions.
Therefore, a comparison of $CP$-violation parameters between
$\btosqq$ and $\btoccs$ decays is an important test of the SM.

Recent theoretical studies~\cite{bib:b2s_SM_uncertainties}
find that
$\bz\to\phi\kz$, $\etap\kz$ and $\ks\ks\ks$ have
the smallest hadronic uncertainties among the modes listed above.
The effective $\sinbb$ values, $\sinbbeff$, obtained from
these decays are expected to agree with $\sinbb$ from
the $\bz\to\jpsi\kz$ decay within 0.04.
Larger deviations
would indicate a new $CP$-violating phase beyond the SM.
The other modes may be affected by a larger amount
by the $b\to u$ transition
that has a weak phase $\phi_3$.
Correspondingly, the SM predictions for the
$\sinbbeff$ values of these modes suffer larger uncertainties.

Belle's previous measurements of $CP$ violation in
$\bz\to$
$\phi\ks$, 
$\phi\kl$, 
$\eta'\ks$, 
$\ks\ks\ks$,
$\ks\piz$,
$\fzero\ks$,
$\omega\ks$
and
$\kp\km\ks$
decays
were based on a $\lintlastsummer$ fb$^{-1}$ data sample
containing $\nbblastsummer\times 10^6$ $B\bbar$ pairs.
In this report, we describe improved measurements 
for these decays
incorporating an additional $\lintthisyear$ fb$^{-1}$
data sample that contains $\nbbthisyear\times 10^6$ $B\bbar$
pairs for a total of $\nbb\times 10^6$ $B\bbar$ pairs.
We also measure $CP$ asymmetries for $\bz\to\eta'\kl$ and
$\eta'\ks$ followed by $\ks\to\piz\piz$, 
which were not included in the previous analysis.

Recent measurements of time-dependent $CP$ asymmetries in
decay modes governed by the $\btoccs$ transition
by Belle~\cite{bib:CP1_Belle,bib:BELLE-CONF-0436}
and BaBar~\cite{bib:CP1_BaBar}
have determined 
$\sinbb = \sinbbWAResult$~\cite{bib:HFAG}, where
$\bz\to\jpsi\ks$, $\jpsi\kl$, $\psi(2S)\ks$, $\chi_{c1}\ks$
and $\eta_c\ks$ decays are used.
%
In this report, 
we describe improved measurements
of $CP$-violation parameters $\cals$ and $\cala$ in 
$\bz\to\jpsi\ks$ and $\jpsi\kl$ decays,
which are the modes with the largest statistics
and with the smallest theoretical uncertainties~\cite{Boos:2004xp,Atwood:2003tg},
as a firm reference point for the SM.

Among the $b\to s$ modes listed above,
all of the two-body final states are $CP$ eigenstates
with a $CP$ eigenvalue $\xi_f=-1$
($\phi\ks$, $\eta'\ks$, $\ks\piz$ and
$\omega\ks$)
or
$\xi_f=+1$
($\phi\kl$, $\eta'\kl$ and $\fzero\ks$).
While the three body state $\ks\ks\ks$ is a
$CP$ eigenstate with $\xi_f=+1$~\cite{Gershon:2004tk},
the $\kp\km\ks$ state is in general
a mixture of both $CP$-even and -odd final states.
Excluding $\kp\km$ pairs that are consistent 
with a $\phi \to\kp\km$ decay from the $\bz \to K^+K^-\ks$ sample,
we find that the $\kp\km\ks$ state is primarily $CP$-even;
a measurement of the $CP$-even fraction $f_+$
using the isospin relation~\cite{Garmash:2003er}
with a $\lint$~fb$^{-1}$ data sample gives
$f_+ = \FcpkpkmksResultSS$.
The SM expectation for this mode is
$\cals = -(2f_{+}-1)\sin 2\phi_1$.
In this report, we define 
$\xi_f \equiv 2f_+ - 1 = 
\xifkpkmksResultSS$
for the $\bz\to\kp\km\ks$ decay, and measure
$\sinbbeff \equiv -\xi_f^{-1}\cals$. 

The decays $\bz\to\phi\ks$ and $\phi\kl$ are combined
in this analysis by redefining $\cals$ as $-\xi_f\cals$ to take
the opposite $CP$ eigenvalues into account, 
and are collectively called ``$\bz\to\phi\kz$''.
Likewise, $CP$ asymmetries for ``$\bz\to\eta'\kz$'' or ``$\bz\to\jpsi\kz$'' 
are obtained by combining 
the decays $\bz\to\eta'\ks$ and $\eta'\kl$, or
$\bz\to\jpsi\ks$ and $\jpsi\kl$.

At the KEKB energy-asymmetric 
$e^+e^-$ (3.5 on 8.0~GeV) collider~\cite{bib:KEKB},
the $\Upsilon(4S)$ is produced
with a Lorentz boost of $\beta\gamma=0.425$ nearly along
the electron beamline ($z$).
Since the $B^0$ and $\bzb$ mesons are approximately at 
rest in the $\Upsilon(4S)$ center-of-mass system (cms),
$\Delta t$ can be determined from the displacement in $z$ 
between the $f_{CP}$ and $f_{\rm tag}$ decay vertices:
$\Delta t \simeq (z_{CP} - z_{\rm tag})/(\beta\gamma c)
 \equiv \Delta z/(\beta\gamma c)$.

The Belle detector is a large-solid-angle magnetic
spectrometer that
consists of a silicon vertex detector (SVD),
a 50-layer central drift chamber (CDC), an array of
aerogel threshold Cherenkov counters (ACC),
a barrel-like arrangement of time-of-flight
scintillation counters (TOF), and an electromagnetic calorimeter
comprised of CsI(Tl) crystals (ECL) located inside
a superconducting solenoid coil that provides a 1.5~T
magnetic field.  An iron flux-return located outside of
the coil is instrumented to detect $K_L^0$ mesons and to identify
muons (KLM).  The detector
is described in detail elsewhere~\cite{Belle}.
Two inner detector configurations were used. A 2.0 cm radius beampipe
and a 3-layer silicon vertex detector (SVD-I) were used for the
first $\lintsvdone$ fb$^{-1}$ data sample (DS-I) that contains
$\nbbsvdone\times 10^6$ $B\bbar$ pairs,
while a 1.5 cm radius beampipe, a 4-layer
silicon detector (SVD-II)~\cite{Ushiroda}
and a small-cell inner drift chamber were used for
the rest, a $\lintsvdtwo$ fb$^{-1}$ data sample (DS-II)
that contains $\nbbsvdtwo\times 10^6$ $B\bbar$ pairs.

\section{Event Selection, Flavor Tagging and Vertex Reconstruction}
\subsection{Overview}
We reconstruct the following $\bz$ decay modes to
measure $CP$ asymmetries:
$\bz\to$
$\phi\ks$, 
$\phi\kl$, 
$\eta'\ks$,
$\eta'\kl$,
$\ks\ks\ks$,
$\ks\piz$,
$\fzero\ks$,
$\omega\ks$
and
$\kp\km\ks$.
We exclude $K^+K^-$ pairs that are consistent with a $\phi \to K^+K^-$ decay 
from the $\bz \to K^+K^-\ks$ sample.
The intermediate meson states are reconstructed from the following decays:
$\piz\to\gamma\gamma$,
$\ks \to \pip\pim$,
$\eta\to\gamma\gamma$,
$\rhoz\to\pip\pim$,
$\omega\to\pip\pim\piz$,
$\eta'\to\rhoz\gamma$ or $\eta\pip\pim$,
$\fzero\to\pip\pim$,
and $\phi\to K^+K^-$.
In addition, 
$\ks\to\piz\piz$ decays are used
for $\bz\to\phi\ks$ and $\eta'\ks$ decays,
and $\eta\to\pip\pim\piz$ for the case
$\bz\to\eta'\ks$ ($\ks\to\pip\pim$).

\subsection{\boldmath $\bz\to\phi\ks$ and $\kp\km\ks$}
\label{sec:bztophiks}
Charged tracks reconstructed with the CDC
for kaon and pion candidates, except for tracks from $\ks\to\pip\pim$ decays,
are required to originate from the interaction point (IP).
We distinguish charged kaons from pions based on
a kaon (pion) likelihood $\mathcal{L}_{K(\pi)}$
derived from the TOF, ACC and $dE/dx$ measurements in the CDC.

Pairs of oppositely charged tracks that have an invariant mass
within 0.015 GeV/$c^2$
of the nominal $\ks$ mass are used to reconstruct $\ks\to\pip\pim$ decays.
The distance of closest approach of the 
candidate charged tracks to the IP in the plane
perpendicular to the $z$ axis is required to be larger than 0.02~cm
for high momentum ($>1.5$ GeV/$c$) $\ks$ candidates and 
larger than 0.03~cm 
for those with momentum less than 1.5~GeV/$c$.
The $\pip\pim$ vertex is required to be displaced from
the IP by a minimum transverse distance of 0.22~cm for
high-momentum candidates and 0.08~cm for the remaining candidates.
The mismatch in the $z$ direction at the $\ks$ vertex point for the $\pip\pim$
tracks must be less than 2.4~cm for high-momentum candidates and 
less than 1.8~cm
for the remaining candidates.
The direction of the pion pair momentum must also agree with
the direction of the vertex point from the IP
to within 0.03 rad for high-momentum candidates, and 
to within 0.1 rad for the remaining candidates.
The resolution of the reconstructed $\ks$ mass is 0.003~GeV$/c^2$.

Photons are identified as isolated ECL clusters
that are not matched to any charged track.
To select $\ks\to\piz\piz$ decays,
we reconstruct $\piz$ candidates from pairs of photons
with $E_\gamma > 0.05$ GeV, where
$E_\gamma$ is the photon energy measured with the ECL.
Photon pairs with an invariant mass between
0.08 and 0.15 GeV$/c^2$ and a momentum above 0.1 GeV/$c$
are used as $\piz$ candidates.
Initially, the $\piz$ decay vertex is assumed to be the IP.
An asymmetric mass window is used to take into account the lower tail of
the mass distribution due to the distance between the IP and the
true $\piz$ vertex.
Candidate $\ks\to\piz\piz$ decays are required to have
an invariant mass between 0.47 GeV/$c^2$ and 0.52 GeV/$c^2$,
where we perform a fit with constraints on
the $\ks$ vertex and the $\piz$ masses
to improve the $\piz\piz$ invariant mass resolution.
We also require that the distance between the IP and the
reconstructed $\ks$ decay vertex be larger than $-10$~cm,
where the positive direction is defined by the $\ks$ momentum.

Candidate $\phi \to \kp\km$ decays are required to
have an invariant mass that is within 0.01 GeV/$c^2$ 
of the nominal $\phi$ meson mass.
Since the $\phi$ meson selection is effective in reducing background events,
we impose only minimal kaon-identification requirements;
$\rkpi \equiv \mathcal{L}_K /(\mathcal{L}_K + \mathcal{L}_\pi) > 0.1$
is required, 
where the kaon likelihood ratio $\mathcal{R}_{K/\pi}$ has values 
between 0 (likely to be a pion) and 1 (likely to be a kaon).
We use a more stringent kaon-identification requirement,
$\rkpi > 0.6$,
to select non-resonant $\kp\km$ candidates 
for the decay $\bz \to \kp\km\ks$.
We exclude $\kp\km$ pairs
with an invariant mass within 0.015 GeV/$c^2$ of the 
nominal $\phi$ meson mass to reduce the $\phi$ contribution to a negligible
level.
To remove $\chi_{c0}\to K^+K^-$, $J/\psi \to K^+K^-$ and $D^0\to K^+K^-$
decays, $K^+K^-$ pairs with an invariant mass within 0.015~GeV$/c^2$ of the 
nominal masses of $\chi_{c0}$ and $J/\psi$ or within 0.01~GeV$/c^2$ of the nominal 
$D^0$ mass are rejected.
$D^+ \to \ks K^+$ decays are also removed by rejecting $\ks K^+$
pairs with an invariant mass within 0.01~GeV$/c^2$ of the nominal $D^+$ mass.

For reconstructed $B\to\fCP$ candidates, we identify $B$ meson decays using the
energy difference $\dE\equiv E_B^{\rm cms}-E_{\rm beam}^{\rm cms}$ and
the beam-energy constrained mass $\mb\equiv\sqrt{(E_{\rm beam}^{\rm cms})^2-
(\pbstar)^2}$, where $E_{\rm beam}^{\rm cms}$ is
the beam energy in the cms, and
$E_B^{\rm cms}$ and $\pbstar$ are the cms energy and momentum of the 
reconstructed $B$ candidate, respectively.
The resolution of $\mb$ is about 0.003~GeV$/c^2$.
Because of the smallness of $\pbstar$,
the $\mb$ resolution is dominated by the beam-energy spread,
which is common to all decay modes.
The resolution in $\dE$ depends on the reconstructed decay mode.
The $\dE$ resolution is 
0.013~GeV for $\phi\ks~(\ks\to\pip\pim)$ and $\kp\km\ks$.
The $\dE$ distribution for $\phi\ks~(\ks\to\piz\piz)$ has a tail toward lower
$\dE$ 
due to $\gamma$ energy leakage in the ECL. 
The typical $\dE$ resolution for $\phi\ks~(\ks\to\piz\piz)$ is 0.058~GeV 
for the main component
and the typical width of the tail component is about 0.14~GeV.
The $B$ meson signal region is defined as 
$|\dE|<0.06$ GeV for $\bz \to \phi \ks~(\ks\to\pip\pim)$,
$-0.15~{\rm GeV} < \dE < 0.1$ GeV for $\bz \to \phi \ks~(\ks\to\piz\piz)$,
$|\dE|<0.04$ GeV for $\bz \to K^+K^-\ks$,
and $5.27~{\rm GeV}/c^2 <\mb<5.29~{\rm GeV}/c^2$ for all decays.

The dominant background to the 
$\bz\to\phi\ks$ 
and
$\kp\km\ks$
decays comes from
$e^+e^- \rightarrow 
u\overline{u},~d\overline{d},~s\overline{s}$, or $c\overline{c}$
continuum events. Since these tend to be jet-like, while
the signal events tend to be spherical,
we use a set of variables that characterize the event topology
to distinguish between the two.
We combine 
$\sperp$, $\theta_T$ and modified Fox-Wolfram moments~\cite{Abe:2001nq}
into a Fisher discriminant $\calf$, where
$\sperp$ is
the scalar sum of the transverse momenta of 
particles other than the reconstructed $B$ candidate
outside a $45^\circ$ cone around the 
candidate $\phi$ meson direction 
(the thrust axis of the $B$ candidate for $\kp\km\ks$ decays)
divided by the scalar sum of their total momenta, and $\theta_T$ is
the angle between the thrust axis of the $B$ candidate
and that of the other particles in the cms.
We also use the angle of the reconstructed $B$
candidate with respect to the beam direction in the cms
($\theta_B$).
We combine
$\calf$ and $\cos\theta_B$ 
into a signal [background]
likelihood variable, which is defined as 
${\cal L}_{\rm sig[bkg]} \equiv
{\cal L}_{\rm sig[bkg]}(\calf)\times
{\cal L}_{\rm sig[bkg]}(\cos\theta_B)$.
We impose requirements on the likelihood ratio
$\rsigbkg \equiv \lsig/(\lsig+\lbkg)$ to
maximize the figure-of-merit (FoM) defined as
$\nsigmc/\sqrt{\nsigmc+\nbkg}$, where $\nsigmc$ ($\nbkg$) 
represents the expected
number of signal (background) events in the signal region.
We estimate $\nsigmc$ using Monte Carlo (MC) events, while
$\nbkg$ is determined from events outside the signal region.

We define two $\rsigbkg$ regions for the decay
$\bz\to\phi\ks~(\ks\to\pip\pim)$.
We require $\rsigbkg \geq 0.65$ for the high-$\rsigbkg$ region.
The requirement for the low-$\rsigbkg$ region
depends on the flavor-tagging quality, $r$, which is
described in Sec.~\ref{sec:flavor tagging}.
The threshold values range from 0.1 (used for $r>0.875$)
to 0.35 (used for $r<0.25$).
For the $\bz\to\phi\ks~(\ks\to\piz\piz)$ candidates,
the $\rsigbkg$ threshold values depend on $r$
and range from 0.4 to 0.75, which are more stringent than
those for the $\ks\to\pip\pim$ case.
For the $\bz\to\kp\km\ks$ candidates, we require $|\cos\theta_T|<0.9$ 
prior to the $\rsigbkg$ requirement. The $\rsigbkg$ threshold values
range from 0.25 to 0.65.
The $\rsigbkg$ requirement reduces the continuum background
by 65\% for $\bz\to\phi\ks~(\ks\to\pip\pim)$, 
92\% for $\bz\to\kp\km\ks$ and 
93\% for $\bz\to\phi\ks~(\ks\to\piz\piz)$,
retaining 91\% of the signal 
     for $\bz\to\phi\ks~(\ks\to\pip\pim)$,
72\% for $\bz\to\kp\km\ks$ and 
78\% for $\bz\to\phi\ks~(\ks\to\piz\piz)$.

We use events outside the signal region 
as well as a large MC sample to study the background components.
The dominant background is from continuum.
The contributions from $B\overline{B}$ events are small.
We estimate the contamination of $\bz\to\kp\km\ks$ and 
$\bz\to\fzero\ks~(\fzero\to\kp\km)$ decays in the $\bz\to\phi\ks$ sample
from the Dalitz plot for $B\to\kp\km K$ candidates with a method that is
described elsewhere~\cite{Garmash:2003er}.
The contamination of $\bz\to\kp\km\ks$ events in the $\bz\to\phi\ks$ sample 
is $\fKKKsBGinphiKs$\%, which is taken into account in our
signal yield extraction.
The background fraction from the decay $\bz\to\fzero\ks~(\fzero\to\kp\km)$, 
which has a
$CP$ eigenvalue opposite to $\phi\ks$, is found to be 
consistent with zero. 
The influence of the $\fzero\ks$ background is
treated as a source of systematic uncertainty.

\begin{figure}
\includegraphics[width=0.48\textwidth]{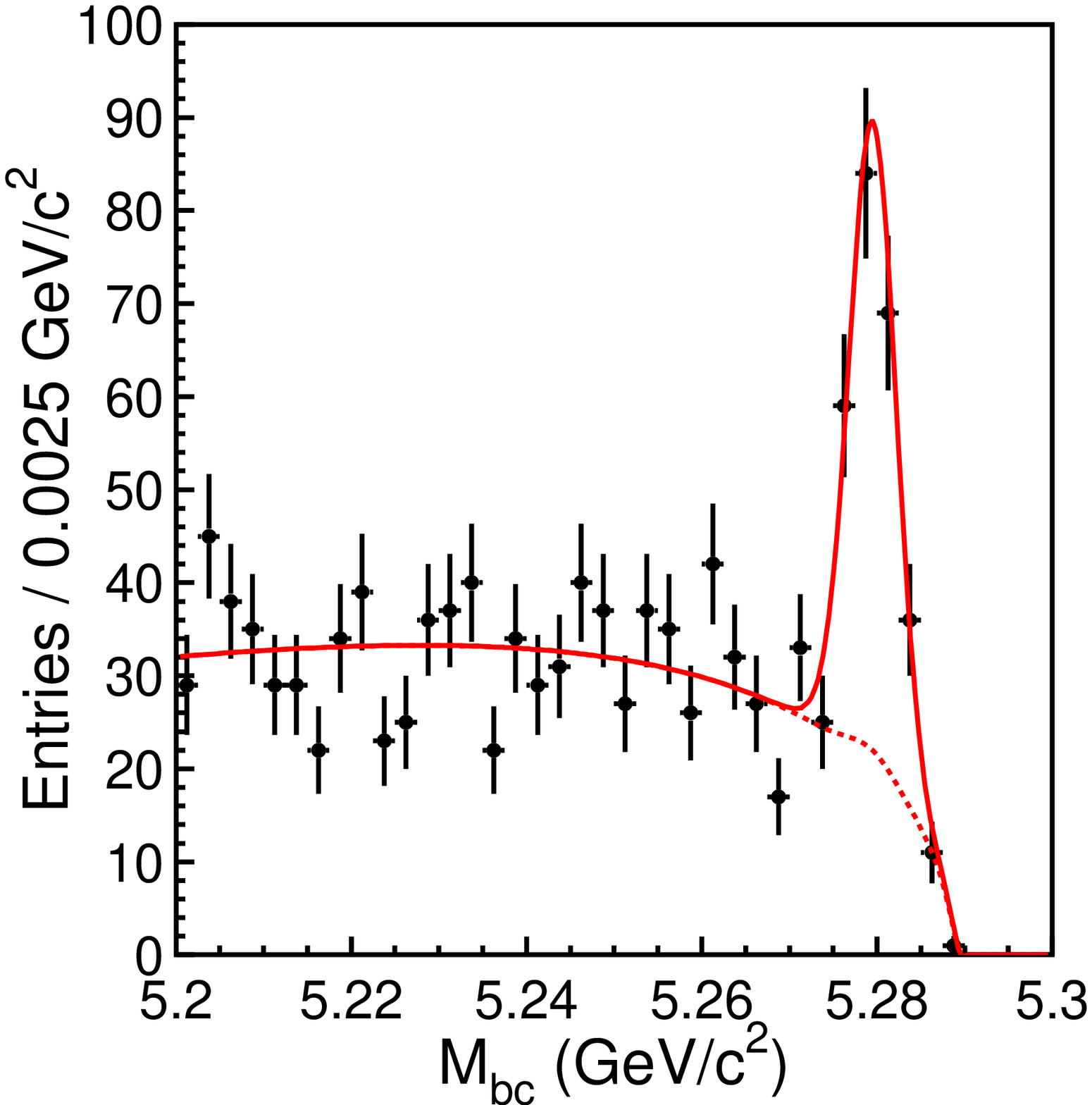}
\includegraphics[width=0.48\textwidth]{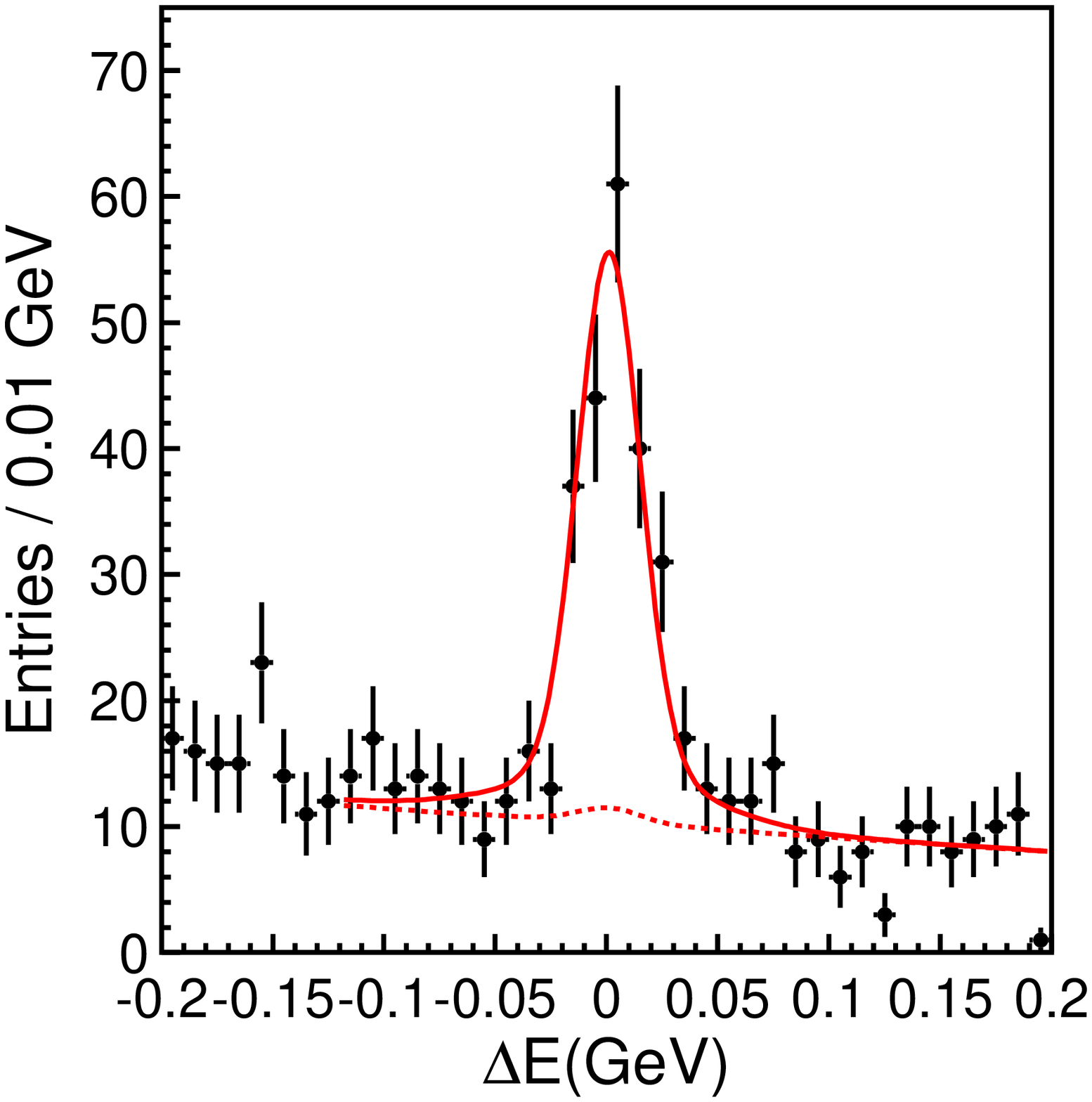}
\includegraphics[width=0.48\textwidth]{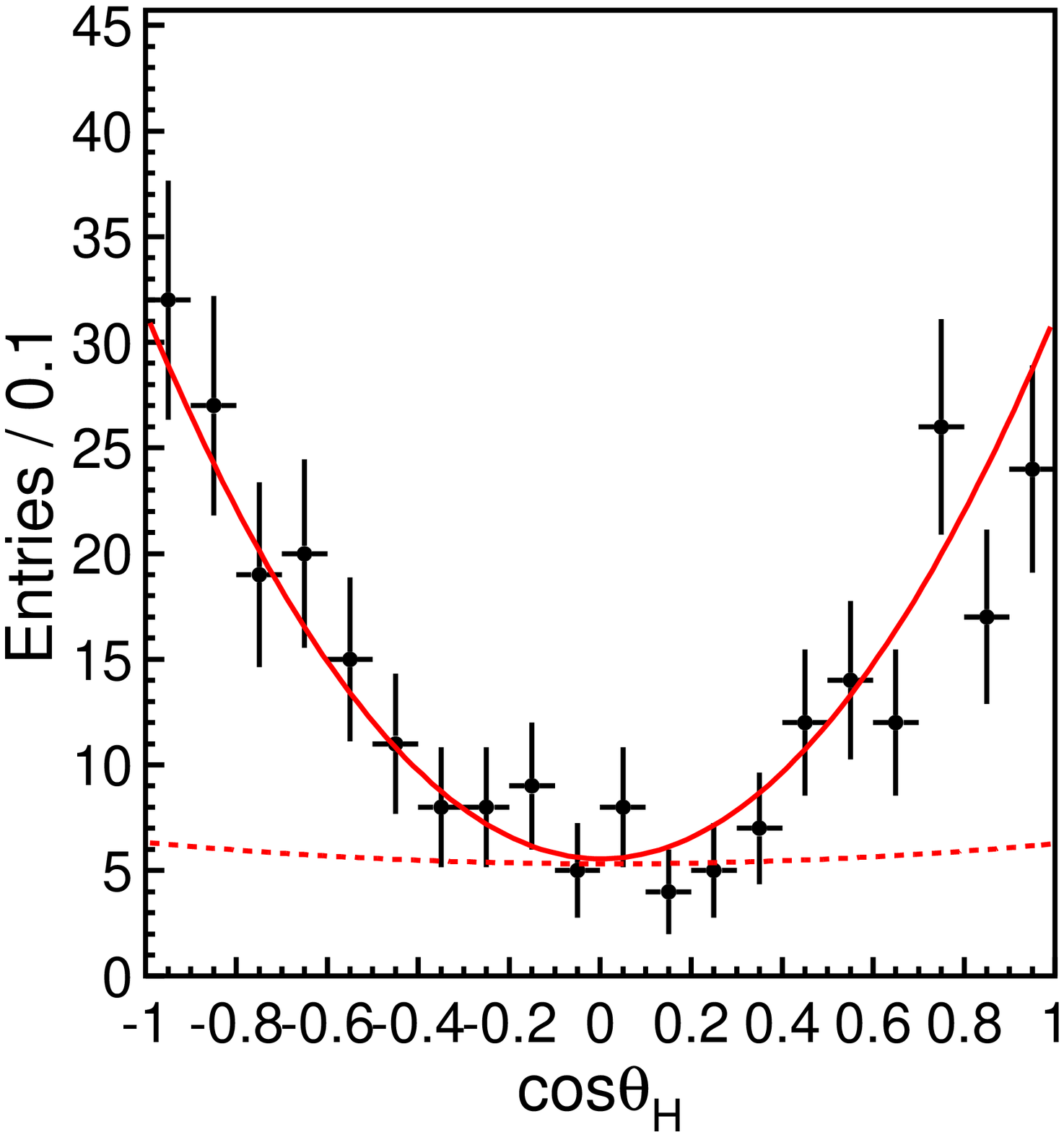}
\caption{Distributions of (a) $\mb$ in the $\dE$ signal region,
(b) $\dE$ in the $\mb$ signal region and
(c) $\cos\theta_H$ in the $\dE$-$\mb$ signal region
for $\bz\to\phi\ks$ candidates.
Solid curves show the fits to signal plus background distributions,
and dashed curves show the background contributions.}
\rput[l]( -6.3,  16.9)  {\Large(a)}
\rput[l](  1.7,  16.9)  {\Large(b)}
\rput[l]( -2.3, 9.0)  {\Large(c)}

\label{fig:phiks}
\end{figure}

Figures~\ref{fig:phiks}(a), (b) and (c) show
the distributions of 
$\mb$ in the $\dE$ signal region,
$\dE$ in the $\mb$ signal region
and 
$\cos\theta_H$ in the $\dE$-$\mb$ signal region
for the reconstructed $\bz\to\phi\ks$ candidates.
Here the helicity angle $\theta_H$ is defined as the
angle between the $B$ meson momentum and the daughter
$\kp$ momentum in the $\phi$ meson rest frame.
%
%
%
%
The signal yield for the $\bz\to\phi\ks$
decay is determined
from an unbinned three-dimensional maximum-likelihood fit
to the $\dE$-$\mb$-$\cos\theta_H$ distribution~\cite{footnote:costhetaH}.
The fit region is
defined as
$-0.12~{\rm GeV} < \dE < 0.25~{\rm GeV}$ for the
$\ks\to\pip\pim$ channel,
$-0.25~{\rm GeV} < \dE < 0.25~{\rm GeV}$ for the
$\ks\to\piz\piz$ channel and
$\mb > 5.2~{\rm GeV/}c^2$
for both cases.
The signal distribution for $\phi\ks~(\ks\to\pip\pim)$ 
is modeled with a Gaussian function (a sum of two Gaussian functions)
for $\mb$ ($\dE$).
The $\phi\ks~(\ks\to\piz\piz)$ signal distribution 
is modeled with a smoothed histogram obtained from MC events.
For the continuum background,
we use the ARGUS parameterization~\cite{bib:ARGUS} 
for $\mb$
and a linear function for $\dE$.
Finally, the $\cos\theta_H$ distribution for the 
$\bz\to\phi\ks$ signal (continuum)
is modeled with a second-order polynomial and is
determined from MC (events in the $\dE$-$\mb$ sideband).
The $\cos\theta_H$ distribution 
for the non-resonant $\bz\to\kp\km\ks$ background is
also determined from MC and is included in the fit,
with a ratio between the non-resonant component and
the $\phi\ks$ signal fixed at the measured value.
The fits yield a total of $\Nsigphiks$(stat) $\bz\to\phi\ks$ events in the signal
region.

\begin{figure}
\includegraphics[width=0.48\textwidth]{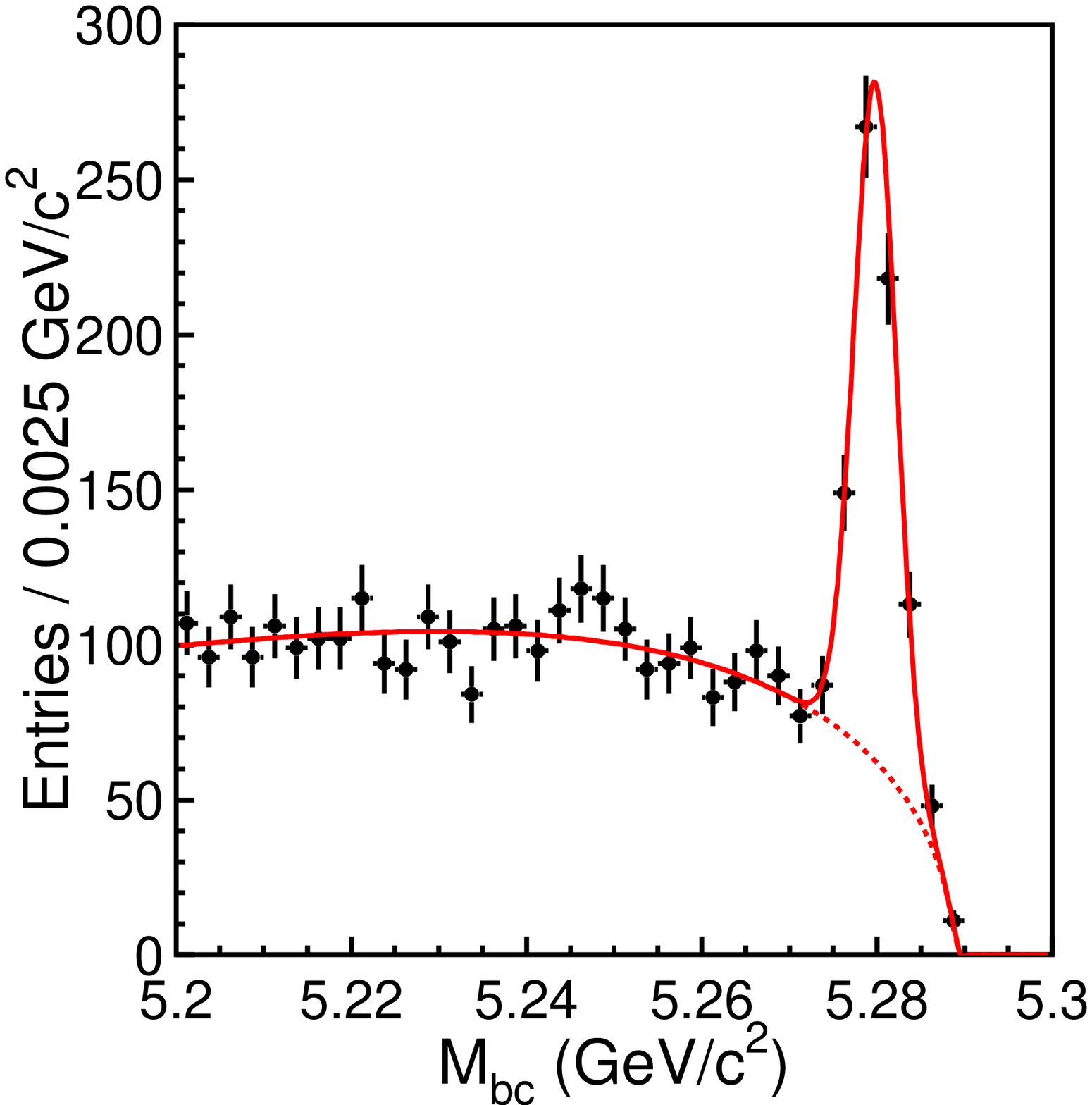}
\includegraphics[width=0.48\textwidth]{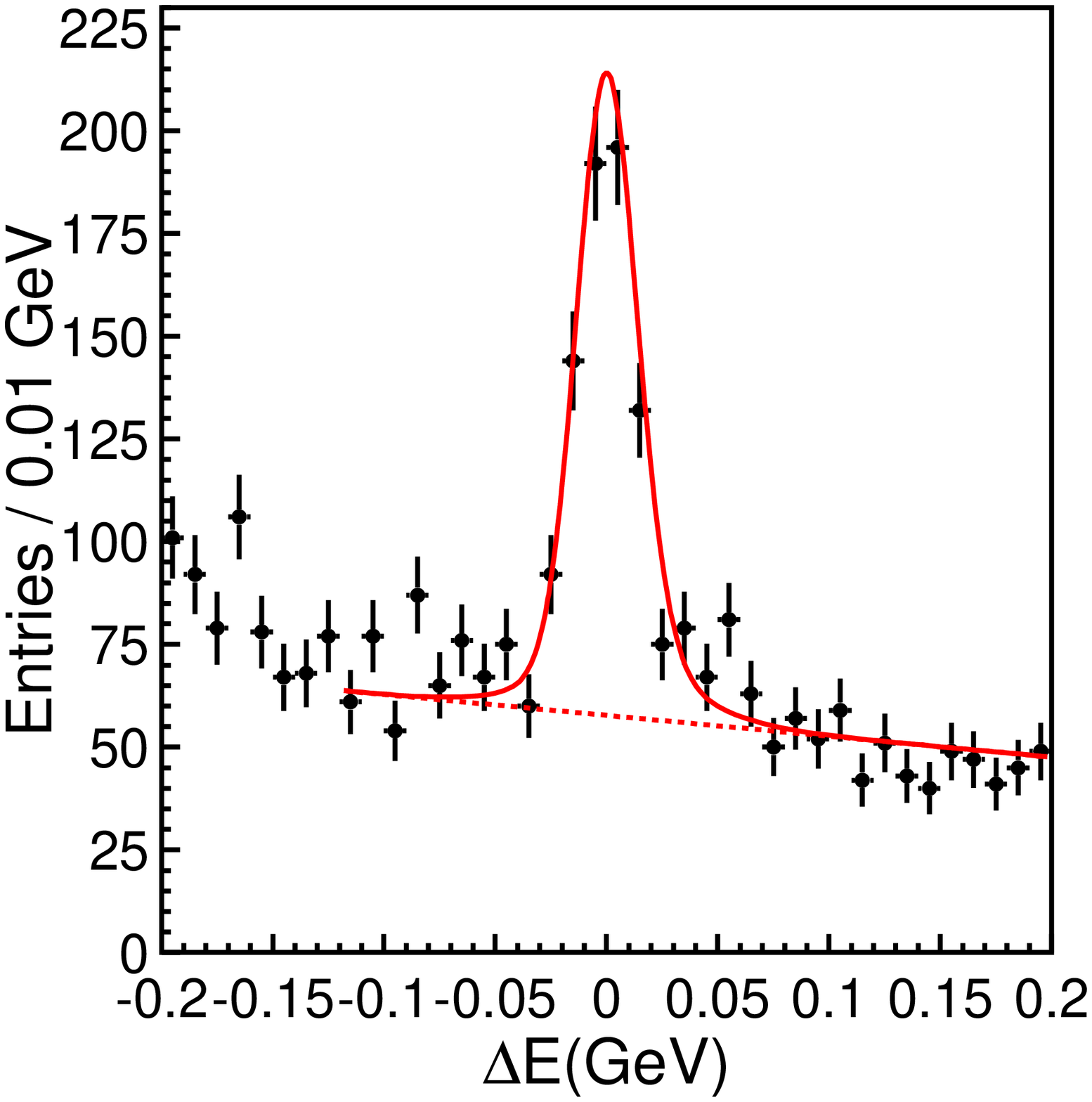}
\caption{Distributions of (a) $\mb$ within the $\dE$ signal region,
(b) $\dE$ within the $\mb$ signal region
for $\bz\to\kp\km\ks$ candidates.
Solid curves show the fits to signal plus background distributions,
and dashed curves show the background contributions.}
\label{fig:kpkmks}

\rput[l]( -6.3,  9.0)  {\Large(a)}
\rput[l](  1.8,  9.0)  {\Large(b)}

\end{figure}

Figures~\ref{fig:kpkmks}(a) and (b) show
distributions of 
$\mb$ in the $\dE$ signal region and
$\dE$ in the $\mb$ signal region
for the reconstructed $\bz\to\kp\km\ks$ candidates
after flavor tagging and vertex reconstruction.
%
%
%
%
The signal yield for the $\bz\to\kp\km\ks$ decay is determined
from an unbinned two-dimensional maximum-likelihood fit
to the $\dE$-$\mb$ distribution in the fit region
defined as
$-0.12~{\rm GeV} < \dE < 0.25~{\rm GeV}$ 
and $\mb > 5.2~{\rm GeV/}c^2$.
The signal and background distributions 
are modeled in the same way as the $\bz\to\phi\ks~(\ks\to\pip\pim)$
case.
The fit yields 
$\Nsigkpkmks$(stat) $\bz\to\kp\km\ks$ events in the signal region.

\subsection{\boldmath $\bz\to\phi\kl$}
\label{sec:bztophikl}
Candidate $\phi \to \kp\km$ decays
are selected with the criteria described above. 
We select $\kl$ candidates based on KLM and ECL
information. There are three classes of $\kl$ candidates, 
which we refer to as KLM, ECL and KLM+ECL candidates.
The KLM candidates are selected from hit clusters in the KLM
that are not associated with either an ECL cluster nor with a charged track.
The requirements for the KLM candidates are the same
as those used in the $\bz\to\jpsi\kl$ selection
for our previous $\sinbb$ measurement~\cite{bib:CP1_Belle}.
ECL candidates are selected from ECL clusters 
if there is no KLM candidate.
We use a $\kl$ likelihood ratio~\cite{bib:CP1_Belle}, which is
calculated from
the following information: the distance between the
ECL cluster and the closest extrapolated charged track position;
the ECL cluster energy; $E_9/E_{25}$, the ratio of energies summed in
$3\times 3$ and $5\times 5$ arrays of CsI(Tl) crystals surrounding
the crystal at the center of the shower; the ECL shower
width and the invariant mass of the shower. 
The likelihood ratio is required to be greater than 0.69.
A KLM+ECL candidate is
an ECL cluster with cluster energy greater than 0.16 GeV
that has an associated KLM cluster.
Here we impose less stringent requirements 
than those for KLM candidates
to select the cluster in the KLM detector.
The $\kl$ likelihood ratio for the ECL cluster
is required to be greater than 0.56.
For all KLM, KLM+ECL and ECL candidates, we also require
that the cosine of the angle between the $\kl$ direction
and the direction of the missing momentum of the event
in the laboratory frame be greater than 0.6.

Since the energy of the $\kl$ is not measured,
$\mb$ and $\dE$ cannot be calculated
in the same way as for the other final states.
Using the four-momentum of a reconstructed
$\phi$ candidate and the $\kl$ flight direction,
we calculate the momentum of the $\kl$ candidate
requiring $\dE=0$. We then calculate
$\pbstar$, the momentum of the $B$ candidate in the
cms, and define the
$B$ meson signal region as 
$0.2~{\rm GeV/}c < \pbstar < 0.5~{\rm GeV/}c$.
We impose the requirement $\rsigbkg > 0.80$,
which rejects 95.7\% of the continuum background
and 67.0\% of backgrounds from $B$ decays, while retaining
65.2\% of signal events.
Here $\rsigbkg$ is based on the discriminating variables
used for the $\bz\to\phi\ks$ decay and the number of
tracks originating from the IP with a momentum above 0.1 GeV/$c$.
%
%
We exclusively reconstruct and reject
$\bz\to\kp\km\ks$ (including $\phi\to\kp\km$ and $\fzero\to\kp\km)$, 
$\phi\kstarz$ $(\kstarz\to \kp\pim$ or $\ks\piz)$,
$\phi\piz$, $\phi\eta$, 
$\bp\to\phi\kp$, and $\phi\kstarp$ $(\kstarp\to\ks\pip$ or $\kp\piz)$
decays.
If there is more than one candidate $\bz\to\phi\kl$ decay
in the signal region,
priority is given to KLM candidates. If there still exist
multiple candidates,
we take the one with the $\kl$ candidate closest to
the expected $\kl$ direction.
%
%

We study the background components
using a large MC sample
as well as data taken with cms energy 60~MeV
below the nominal $\Upsilon(4S)$ mass (off-resonance data).
The dominant background is from continuum.
A MC study shows that background events
from $B$ decays 
are dominated by inclusive $B\to\phi\kl X$ decays
that include $B\to\phi K^*$ decays.
%
%

\begin{figure}
\includegraphics[width=0.48\textwidth]{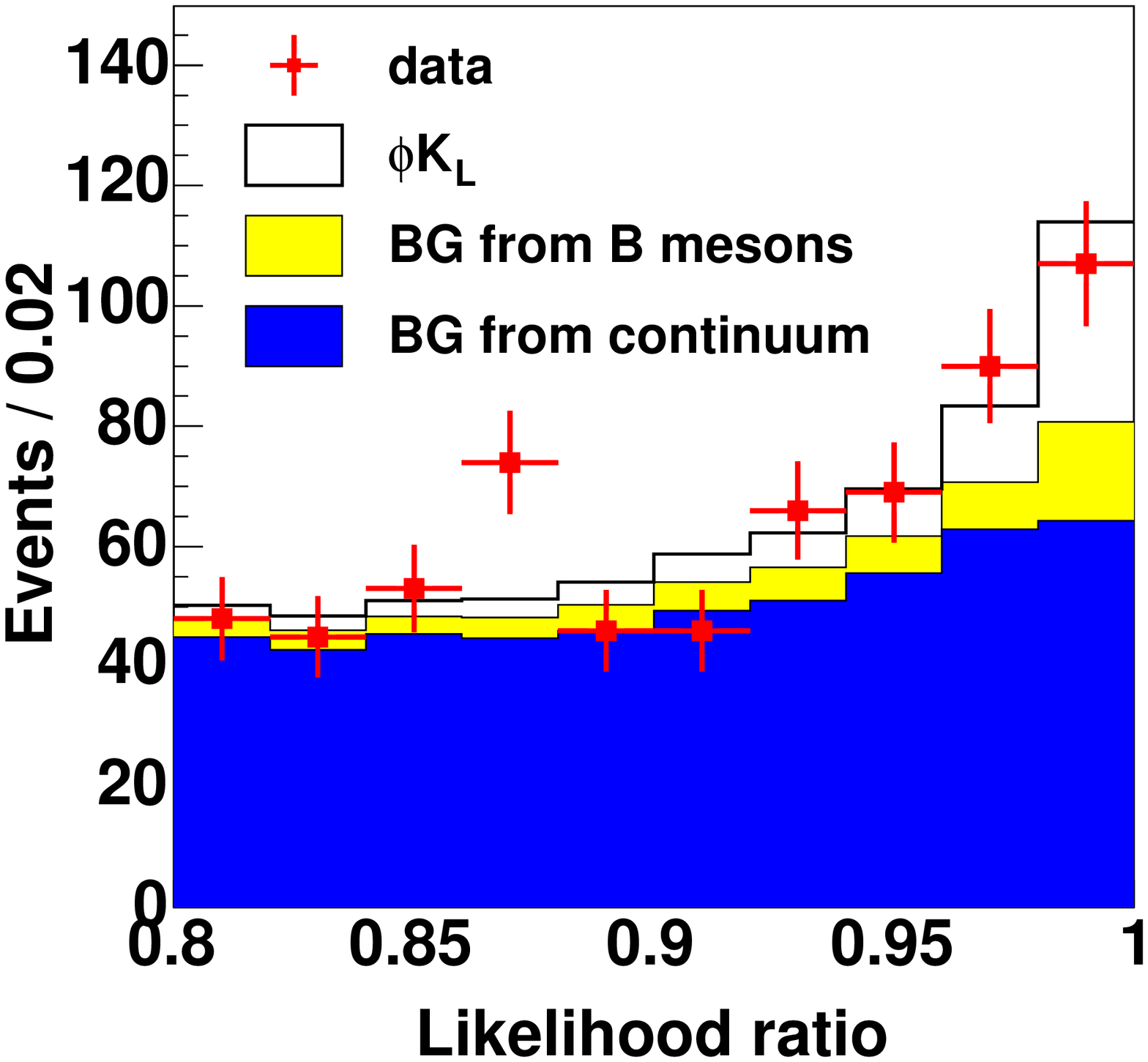}
\includegraphics[width=0.48\textwidth]{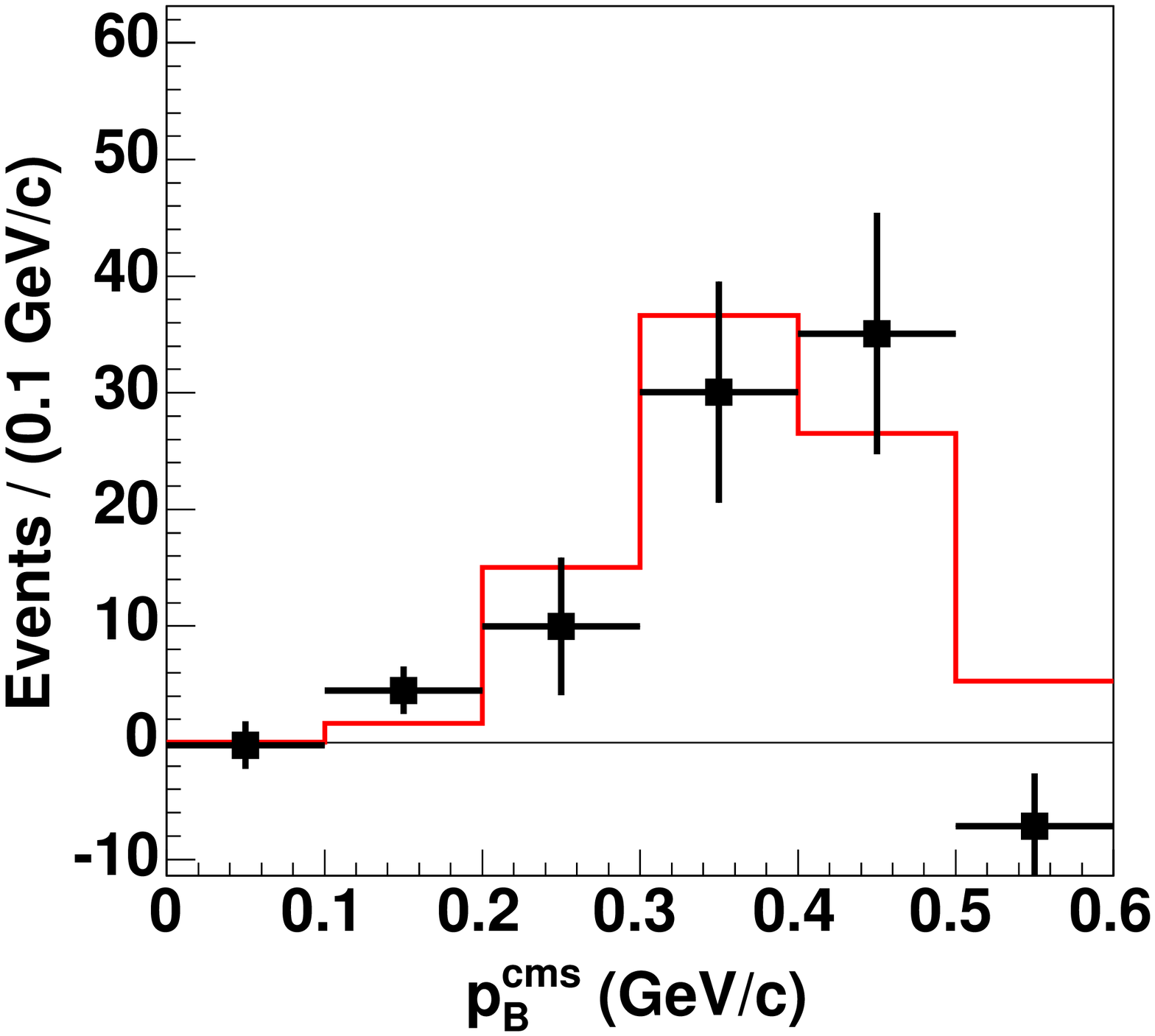}
\caption{(a) Distribution of $\rsigbkg$ in the $\pbstar$ signal
region. The solid histogram shows the fit to the signal plus background distribution,
and the shaded histograms show the background contributions.
(b) Background-subtracted $\pbstar$ distribution
for $\bz\to\phi\kl$ candidates. The solid histogram shows the 
result of the three-dimensional fit to the $\rsigbkg$-$\pbstar$-$r$ distribution.}
\label{fig:phikl}

\rput[l]( -1.4,  9.4)  {\Large(a)}
\rput[l](  1.5,  9.4)  {\Large(b)}

\end{figure}

The signal yield is determined from
an extended three-dimensional binned maximum-likelihood fit 
to the $\rsigbkg$-$\pbstar$-$r$ distribution in the fit region
$0.8 < \rsigbkg \le 1.0$, $0~{\rm GeV/}c < \pbstar \le 0.6~{\rm GeV/}c$
and $r > 0.25$, where the total likelihood is a product of
the likelihood for each of three variables.
The $\bz\to\phi\kl$ signal shape is obtained from MC events.
Background from $B\bbar$ pairs is also modeled with
MC. We fix the ratio between the signal
and the $B\bbar$ background based on
known branching fractions and MC-determined reconstruction
efficiencies with the $\kl$ detection efficiency corrected from
$\bz\to\jpsi\kl$ data. 
The uncertainty in the ratio is treated
as a source of systematic error.
The continuum background distribution is represented
by a histogram obtained from MC events;
we confirm that the function well describes 
both the off-resonance data and the events in
a $\pbstar$ sideband region defined as 
$1.0~{\rm GeV/}c < \pbstar \le 1.6~{\rm GeV/}c$.
The fit yields $\Nsigphikl$ $\bz\to\phi\kl$ events,
where the error is statistical only.
The result is in agreement with
the expected $\bz\to\phi\kl$ signal yield 
(59 events) obtained from MC after applying
the efficiency correction from the
$\bz\to\jpsi\kl$ data.
%
%
Figure~\ref{fig:phikl}(a) shows the $\rsigbkg$ distribution
in the $\pbstar$-$r$ signal region. Figure~\ref{fig:phikl}(b) 
shows signal yields obtained for six $\pbstar$ intervals separately.
The yields agree with the distribution obtained by the three-dimensional fit. 
%
%
%
%
%

\subsection{\boldmath $\bz\to\eta'\ks$}
\label{sec:bztoetapks}
Candidate $\ks \to \pip\pim$ and $\piz\piz$ decays
are selected with the same criteria as those used for
the $\bz\to\phi\ks$ decay.
Charged pions from the $\eta$, $\rhoz$ or $\eta'$ decay
are selected from tracks originating from the IP.
We reject kaon candidates by requiring $\rkpi < 0.9$. 
Candidate photons from
$\piz\to\gamma\gamma$ decays are
required to have $E_\gamma > 0.05$~GeV.
The reconstructed $\piz$ candidate is required to satisfy 
$0.118~{\rm GeV}/c^2 < \mgg < 0.15~{\rm GeV}/c^2$
and
$\ppizcms > 0.1~{\rm GeV}/c$, where
$\mgg$ and $\ppizcms$ are the invariant mass and
the momentum in the cms, respectively. 
Candidate photons from
$\eta\to\gamma\gamma~(\eta'\to\rhoz\gamma)$ decays are
required to have $E_\gamma > 0.05~(0.1)$~GeV.
The invariant mass of the photon pair
is required to be between 0.5 and
0.57~GeV/$c^2$ for the $\eta\to\gamma\gamma$ decay.
The $\pip\pim\piz$ invariant mass is required
to be between 0.535 and 0.558~GeV/$c^2$ for the
$\eta\to\pip\pim\piz$ decay, which is used only for the
reconstruction of the $\bz\to\etap\ks~(\ks\to\pip\pim)$ decay.
A kinematic fit with an $\eta$ mass constraint is
performed using the fitted vertex of the $\pi^+\pi^-$ tracks from
the $\eta^\prime$ as the decay point. 
For $\eta^\prime\to\rhoz\gamma$ decays, candidate $\rhoz$ mesons
are reconstructed from pairs of vertex-constrained $\pi^+\pi^-$
tracks with invariant mass between
0.55 and 0.92~GeV/$c^2$. 
The $\eta^\prime\to\eta\pip\pim$ candidates are required
to have a reconstructed mass
between 0.94 and 0.97~GeV/$c^2$ (0.95 and 0.966~GeV/$c^2$)
for the $\eta\to\gamma\gamma$ ($\eta\to\pip\pim\piz$) decay.
Candidate $\eta^\prime\to\rhoz\gamma$ decays are required to
have a reconstructed mass from 0.935 to 0.975~GeV/$c^2$.

The $B$ meson signal region is defined as 
$|\dE|<0.06$ GeV for 
$\bz \to \eta'\ks~(\eta'\to \rhoz\gamma,~\ks\to\pip\pim)$,
$-0.1$ GeV $< \dE <0.08$ GeV for 
$\bz \to \eta'\ks~(\eta'\to\eta\pip\pim,~\eta\to\gamma\gamma,~\ks\to\pip\pim) $,
$-0.08$ GeV $< \dE <0.06$ GeV for 
$\bz \to \eta'\ks~(\eta'\to\eta\pip\pim,~\eta\to\pip\pim\piz,~\ks\to\pip\pim) $,
$-0.15$ GeV $< \dE < 0.1$ GeV for 
$\bz \to \eta'\ks~(\ks\to\piz\piz)$,
and $5.27~{\rm GeV}/c^2 <\mb<5.29~{\rm GeV}/c^2$ for all decays.
%
%
The continuum suppression is 
based on the
likelihood ratio $\rsigbkg$ obtained from
the same discriminating variables 
used for the $\bz\to\phi\ks$ decay, except for the decay mode
$\eta'\to\rho\gamma~(\rho\to\pip\pim)$
where $\cos\theta_H$ is included.
Here $\theta_H$ is defined as
the angle between the $\eta'$ meson momentum and the daughter
$\pip$ momentum in the $\rho$ meson rest frame.
The minimum $\rsigbkg$ requirement
depends both on the decay mode and on the
flavor-tagging quality, and
ranges from 0 (i.e., no requirement) to $0.4$ for the
decay $\bz\to\etap\ks~(\ks\to\pip\pim)$ and 
from 0.2 to 0.9 for the 
decay $\bz\to\etap\ks~(\ks\to\piz\piz)$.
For the $\eta'\to\rhoz\gamma$ mode, we also require $|\cos\theta_T|<0.9$ prior
to the $\rsigbkg$ requirement.
With these requirements, the continuum background in
the $\bz\to\etap\ks~(\ks\to\pip\pim)$ mode
is reduced by  
87\% for $\eta'\to\rhoz\gamma$,
58\% for $\eta'\to\eta\pip\pim~(\eta\to\gamma\gamma)$ and
31\% for $\eta'\to\eta\pip\pim~(\eta\to\pip\pim\piz)$,
while retaining 78\% of the signal
     for $\eta'\to\rhoz\gamma$,
94\% for $\eta'\to\eta\pip\pim~(\eta\to\gamma\gamma)$ and
97\% for $\eta'\to\eta\pip\pim~(\eta\to\pip\pim\piz)$.
The continuum background for
the $\bz\to\etap\ks~(\ks\to\piz\piz)$ candidates
is reduced by
90\% (97\%) while retaining 81\% (54\%) of signal events
for $\etap\to\eta\pip\pim~(\rho\gamma)$.

We use events outside the signal region 
as well as a large MC sample to study the background components
in $\bz\to\etap\ks$.
The dominant background is from continuum.
In addition, according to MC simulation, there is  a small ($\sim 3\%$) 
combinatorial background from $B\overline{B}$ events
in $\bz\to\eta'\ks~(\eta'\to\rhoz\gamma)$. 
The contributions from $B\overline{B}$ events are smaller for other modes.
The influence of these backgrounds
is treated as a source of systematic uncertainty.

\begin{figure}
\includegraphics[width=0.48\textwidth]{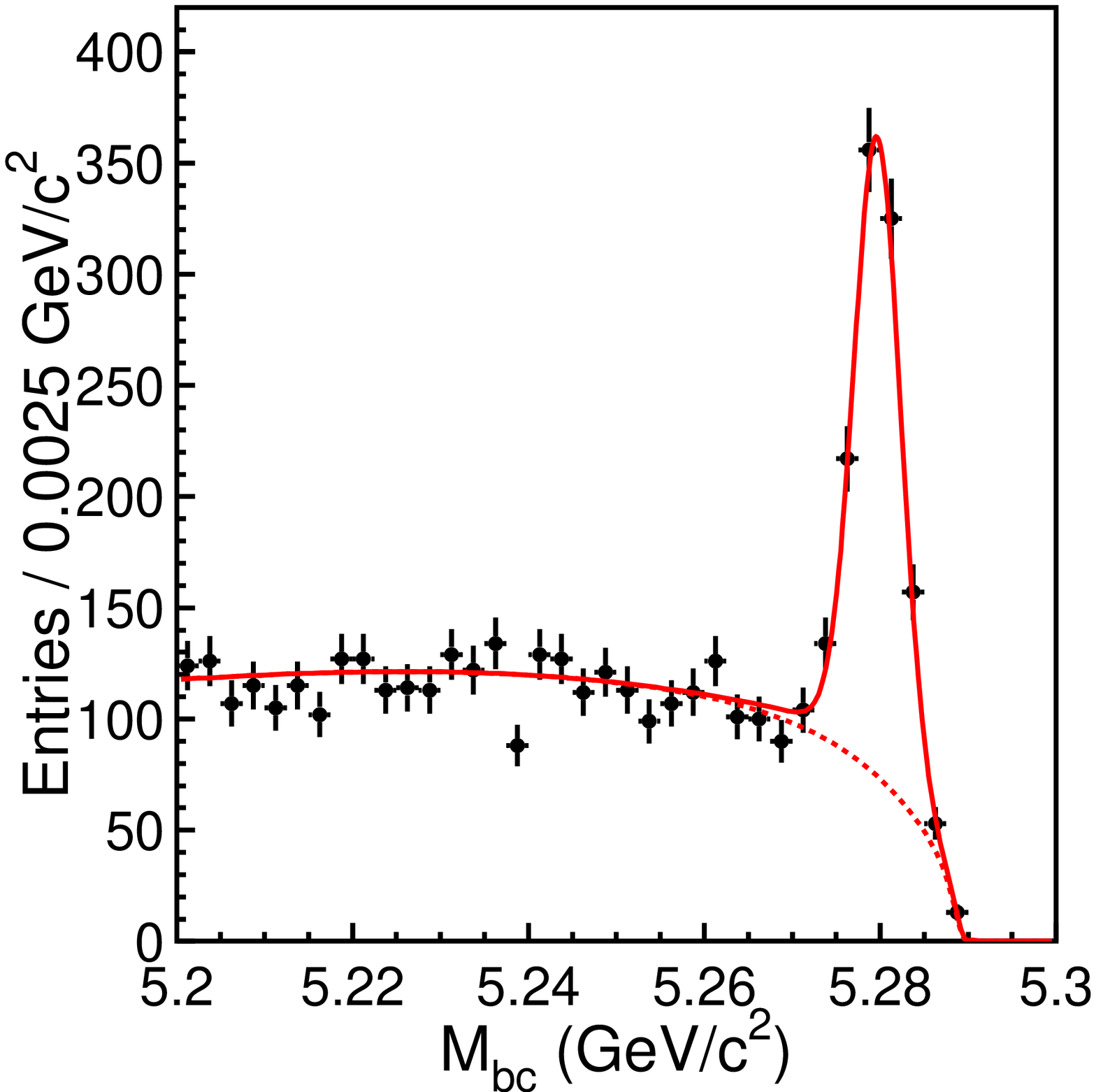}
\includegraphics[width=0.48\textwidth]{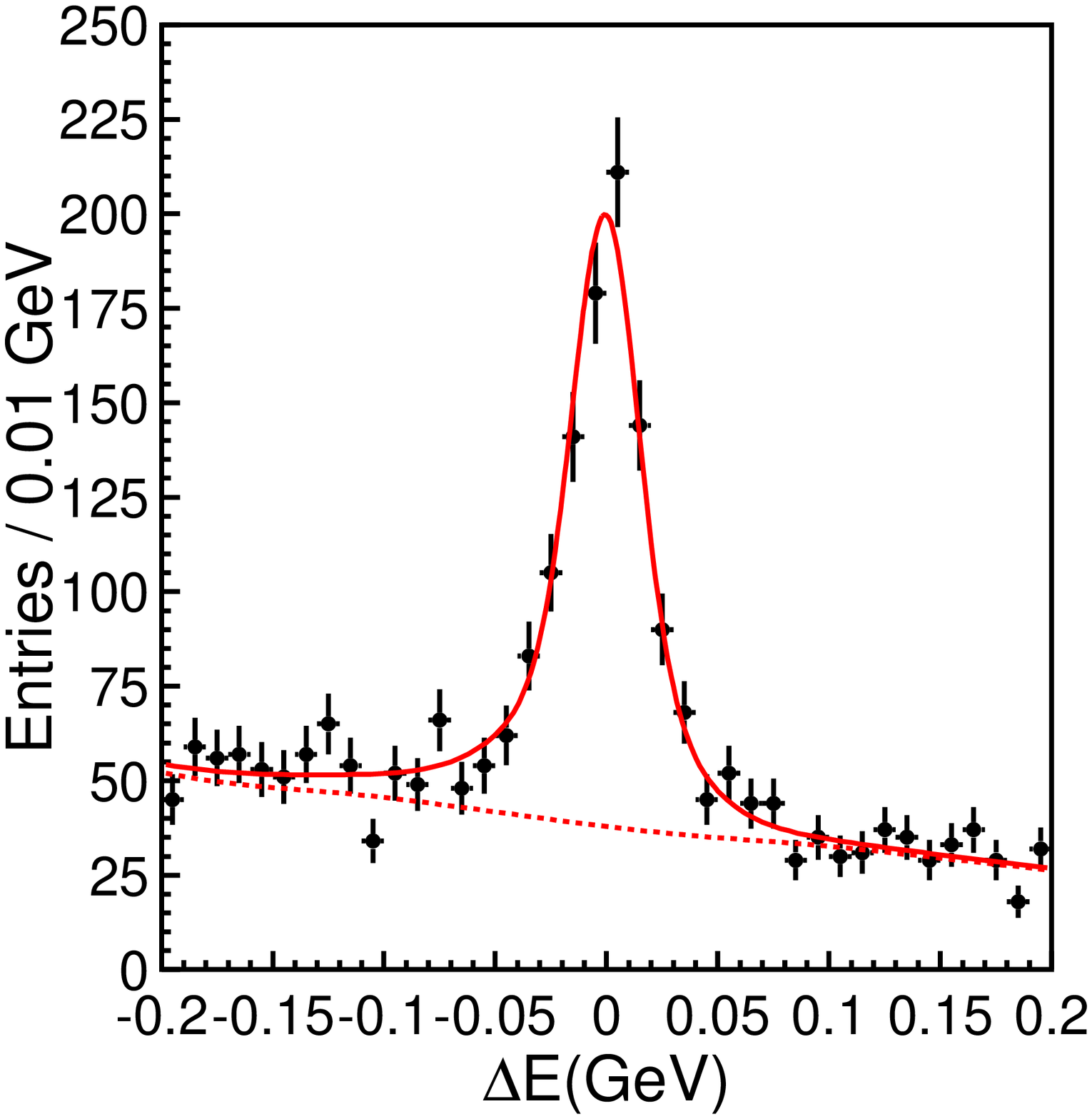}
\caption{Distributions of (a) $\mb$ within the $\dE$ signal region,
(b) $\dE$ within the $\mb$ signal region
for $\bz\to\etap\ks$ candidates.
Solid curves show the fits to signal plus background distributions,
and dashed curves show the background contributions.}
\label{fig:etapks}
\rput[l]( -6.3,  9.0)  {\Large(a)}
\rput[l](  1.8,  9.0)  {\Large(b)}

\end{figure}

Figure~\ref{fig:etapks}(a) shows
the $\mb$ distribution for the reconstructed $\bz\to\eta'\ks$ candidates
within the $\dE$ signal region
after flavor tagging and vertex reconstruction,
where all subdecay modes are combined.
The $\dE$ distribution for the $\bz\to\eta'\ks$ candidates 
within the $\mb$ signal region is shown in Fig.~\ref{fig:etapks}(b).
%
%
%
%
The signal yields are determined
from unbinned two-dimensional maximum-likelihood fits
to the $\dE$-$\mb$ distributions.
The fit region is defined as
$-0.25~{\rm GeV} < \dE < 0.25~{\rm GeV}$
and $\mb > 5.2~{\rm GeV/}c^2$.
We perform the fit for each final state separately.
The $\eta'\ks~(\ks\to\pip\pim)$ signal distribution 
is modeled with a sum of two (three) Gaussian functions
for $\mb$ ($\dE$).
The $\eta'\ks~(\ks\to\piz\piz)$ signal distribution 
is modeled with a smoothed histogram.
For the continuum background, 
we use the ARGUS parameterization for $\mb$
and a linear function for $\dE$.
For the $\eta'\to\rho\gamma$ mode, we include
in the fits the $B\bbar$ background shape obtained from MC.
The fits yield a total of $\Nsigetapks$ $\bz\to\eta'\ks$ events
in the signal region,
where the error is statistical only.

\subsection{\boldmath $\bz\to\eta'\kl$}
\label{sec:bztoetapkl}
Candidate $\etap\to\eta\pip\pim$ $(\eta\to\gamma\gamma)$
decays are selected with the same criteria as those used
for the $\bz\to\etap\ks$ analysis. 
The $\kl$ selection is adopted from
the $\phi\kl$ analysis, with a
likelihood ratio optimized for the $\bz\to\etap\kl$ decay
that is required to be greater than
0.50 (0.40) for KLM+ECL (ECL) candidates.
The best candidate is formed from the $\etap$ candidate with
the smallest $\chi^2$ value in its mass-constrained fit and the $\kl$
candidate whose measured direction is closest to the expected direction.
The following exclusive modes are reconstructed and are rejected:
$B\to \etap\piz$, $\etap\pi^\pm$,
$\etap\eta$, $\etap\ks$, $\etap K^\pm$, $\etap K^{*0}$ $(\to \ks\piz$
or $K^\pm \pi^\mp)$, $\etap K^{*\pm}$, $\etap\rho^0$ and $\etap\rho^\pm$.
The $B$ meson signal region is defined as
$\rsigbkg > 0.8$,
$0.2~{\rm GeV/}c < \pbstar < 0.5~{\rm GeV/}c$
and $r > 0.25$ (0.5 for ECL candidates).

\begin{figure}
\includegraphics[width=0.95\textwidth]{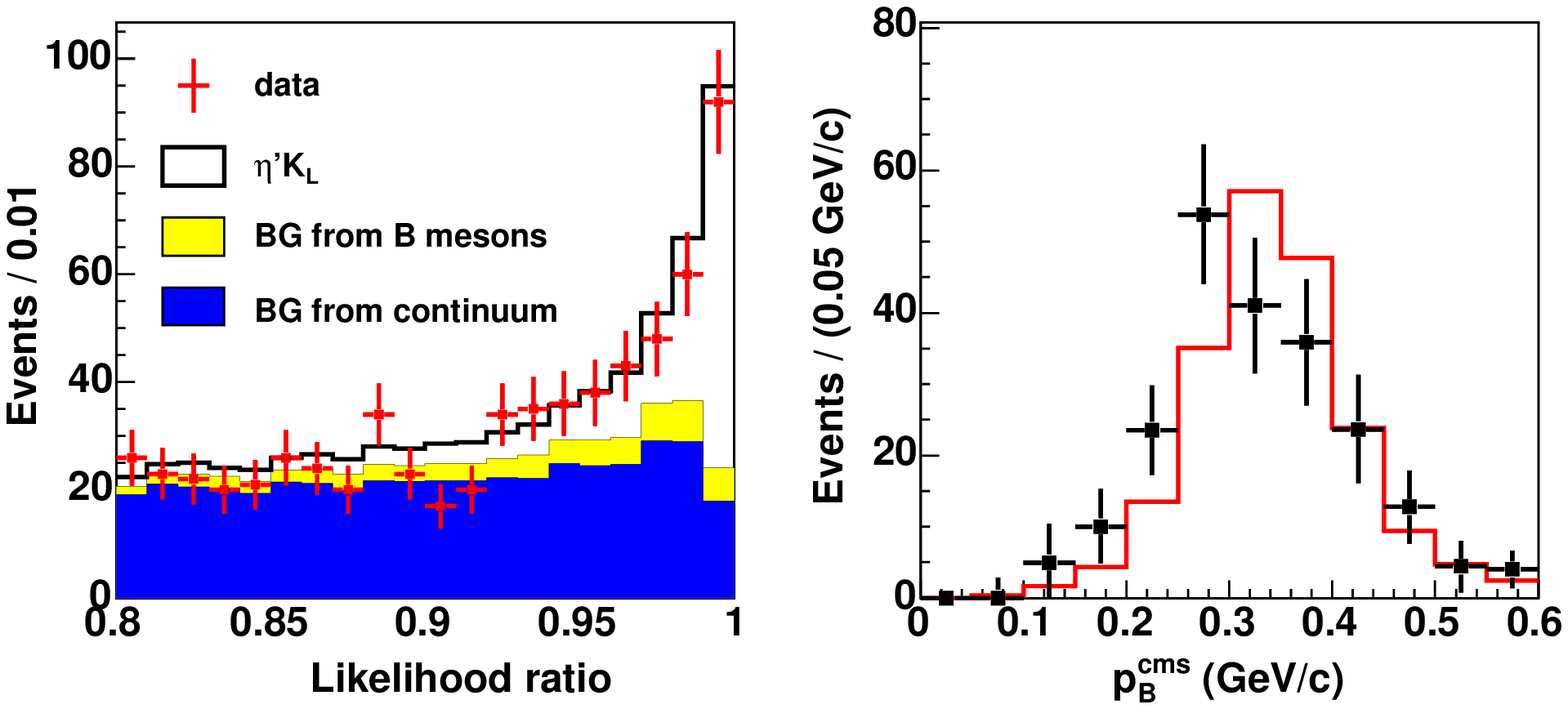}
\caption{Distributions of (a) $\rsigbkg$ in the $\pbstar$ signal
region and 
(b) background-subtracted $\pbstar$
for $\bz\to\etap\kl$ candidates.
Solid histograms show the fits to signal plus background distributions,
and the shaded histogram shows the background contributions.}
\label{fig:etapkl}
\rput[l]( -1.8,  8.6)  {\Large(a)}
\rput[l](  1.5,  8.6)  {\Large(b)}

\end{figure}

The signal yield is determined from
an extended three-dimensional maximum-likelihood fit 
to the $\rsigbkg$-$\pbstar$-$r$ distribution.
The procedure to determine the signal and background 
distributions is the same as that for the $\bz\to\phi\kl$ decay.
The fit yields $\Nsigetapkl$ $\bz\to\etap\kl$ events,
where the error is statistical only.
The result is in good agreement with
the expected $\bz\to\etap\kl$ signal yield 
(180 events) obtained from MC after applying
the efficiency correction from the
$\bz\to\jpsi\kl$ data.
Figure~\ref{fig:etapkl}(a) shows the $\rsigbkg$ distribution
in the $\pbstar$-$r$ signal region. Figure~\ref{fig:etapkl}(b) 
shows signal yields obtained for twelve $\pbstar$ intervals separately.
The yields agree with the distribution obtained by the three-dimensional fit. 
%
%
%
%

\subsection{\boldmath $\bz\to\ks\ks\ks$}
\label{sec:bztoksksks}
We reconstruct the $\bz\to\ks\ks\ks$ decay
in the $\kspm\kspm\kspm$ or $\kspm\kspm\kszz$ final state,
where the $\pip\pim$ ($\piz\piz$) 
state from a $\ks$ decay is denoted as $\kspm$ ($\kszz$).
Pairs of oppositely charged tracks with $\pip\pim$ invariant mass
within 0.012~GeV/$c^2$ ($\simeq 3 \sigma$)
of the nominal $\ks$ mass are
used to reconstruct $\kspm$ candidates.
The $\pip\pim$ vertex is required to be displaced from
the interaction point (IP) by a minimum transverse distance of 0.22~cm
for $\ks$ candidates with momentum greater than 1.5 GeV/$c$ and 0.08~cm for
those with momentum less than 1.5 GeV/$c$.
The angle in the transverse plane between the $\ks$ momentum vector and the
direction defined by the $\ks$ vertex and the IP should be less than
0.03~rad (0.1~rad) for the high (low) momentum candidates.
The mismatch in the $z$ direction at the $\ks$ vertex point
for the two charged pion tracks should be less than 2.4~cm (1.8~cm)
for the high (low) momentum candidates.
After two good $\kspm$ candidates have been found that satisfy the criteria
given above, looser requirements are applied for the third $\kspm$ candidate.
The requirement on the transverse direction matching is relaxed to
0.2~rad (0.4~rad for low momentum candidates),
and the mismatch of the two charged pions in the $z$ direction
is required to be less than 5~cm (1~cm if both
pions have hits in the SVD).
We also require that the $\ks$ flight length in the plane perpendicular
to the beam axis be less than 0.5 mm and the $\ks$ momentum be
greater than 0.5 GeV/$c$.
   
To select $\kszz$ candidates,
we reconstruct $\piz$ candidates from pairs of photons
with $E_\gamma > 0.05$~GeV.
The reconstructed $\piz$ candidate is required to have an invariant mass
between 0.08 and 0.15~GeV/$c^2$ and momentum above 0.1~GeV/$c$.
$\kszz $ candidates are required to have
an invariant mass between 0.47 and 0.52~GeV/$c^2$,
and a fit is performed with constraints on the
$\ks$ vertex and $\piz$ masses to improve the $\piz\piz$ invariant mass
resolution.
The $\kszz$ candidate is combined with two good $\kspm$ candidates
to reconstruct a $\bz$ meson.

The $\bz$ meson signal region is defined as 
$|\dE|<0.10$ GeV for $\bz \to \kspm\kspm\kspm$,
$-0.15~{\rm GeV} < \dE < 0.10$ GeV for $\bz \to \kspm\kspm\kszz$,
and $5.27~{\rm GeV}/c^2 < \mb < 5.29~{\rm GeV}/c^2$ for both decays.
To suppress the $e^+e^- \rightarrow q\overline{q}$
continuum background ($q = u,~d,~s,~c$),
we form the likelihood ratio $\rsigbkg$ by
combining likelihoods for two quantities; a Fisher discriminant of
modified Fox-Wolfram moments, and the cosine of the cms $\bz$
flight direction.
The requirement for $\rsigbkg$ depends both on the decay mode
and on the flavor-tagging quality;
after applying all other cuts,
this requirement rejects 94\% of the $q\bar{q}$ background
while retaining 75\% of the signal.

If both $\bz\to\kspm\kspm\kspm$ and $\kspm\kspm\kszz$ candidates are found
in the same event, we choose the $\bz\to\kspm\kspm\kspm$ candidate.
If more than one $\bz\to\kspm\kspm\kspm$ candidate is found,
we check for each of them the quality of the third $\kspm$ candidate, 
which is selected with looser requirements as described above.
We choose the $\bz\to\kspm\kspm\kspm$ candidate in which
the third $\kspm$ candidate satisfies the tight $\kspm$ selection requirements.
If no $\bz$ candidate is found with the tight requirements or
more than one $\bz$ candidate still remain,
we select the one with the smallest value for
$\sum(\Delta M_{\kspm})^2$, where $\Delta M_{\kspm}$ is the
difference between the reconstructed and nominal mass of $\kspm$.
For multiple $\bz\to\kspm\kspm\kszz$ candidates,
we select the $\kspm\kspm$ pair that has the smallest $\sum(\Delta
M_{\kspm})^2$ value and the $\kszz$ candidate with the minimum $\chi^2$
of the constrained fit.

We reject $\ks\ks\ks$ candidates if they are consistent with
$\bz\to\chi_{c0}\ks\to(\ks\ks)\ks$ or $\bz\to D^0\ks\to(\ks\ks)\ks$
decays, i.e. if one of the $\ks$ pairs has an invariant mass within
$\pm 2 \sigma$ of the $\chi_{c0}$ mass or $D^0$ mass, where $\sigma$ is
the $\ks\ks$ mass resolution.

We use events outside the signal region 
as well as a large MC sample to study the background components.
The dominant background is from continuum.
The contamination of $\bz\to\chi_{c0}\ks$ events in the 
$\bz\to\ks\ks\ks$ sample is small.
The contributions from other $B\overline{B}$ events are negligibly small.
The influence of these backgrounds
is treated as a source of systematic uncertainty
in the $CP$ asymmetry measurement.
Backgrounds from the decay $\bz\to D^0\ks$
are found to be negligible.

\begin{figure}
\includegraphics[width=0.48\textwidth]{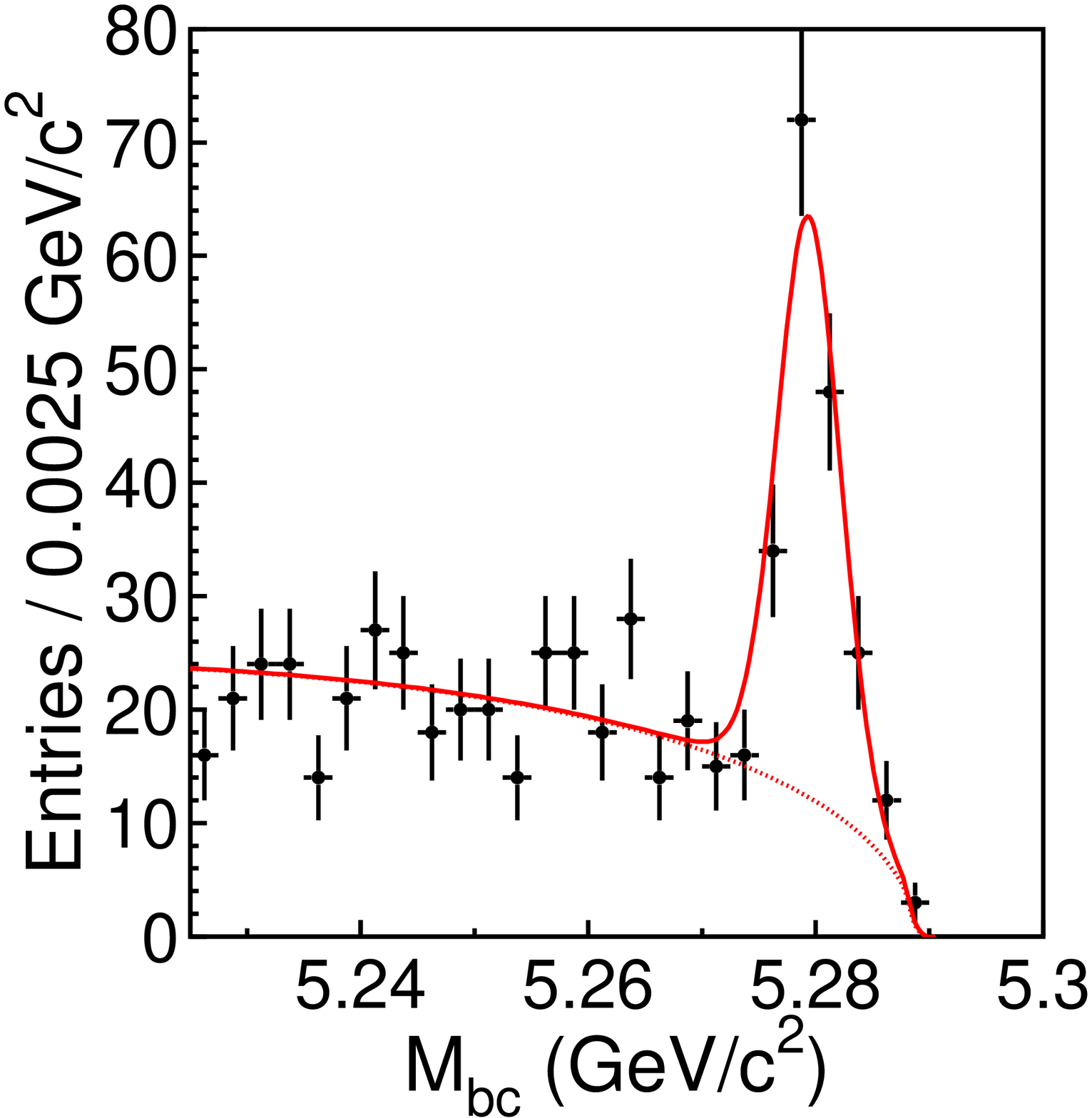}
\includegraphics[width=0.48\textwidth]{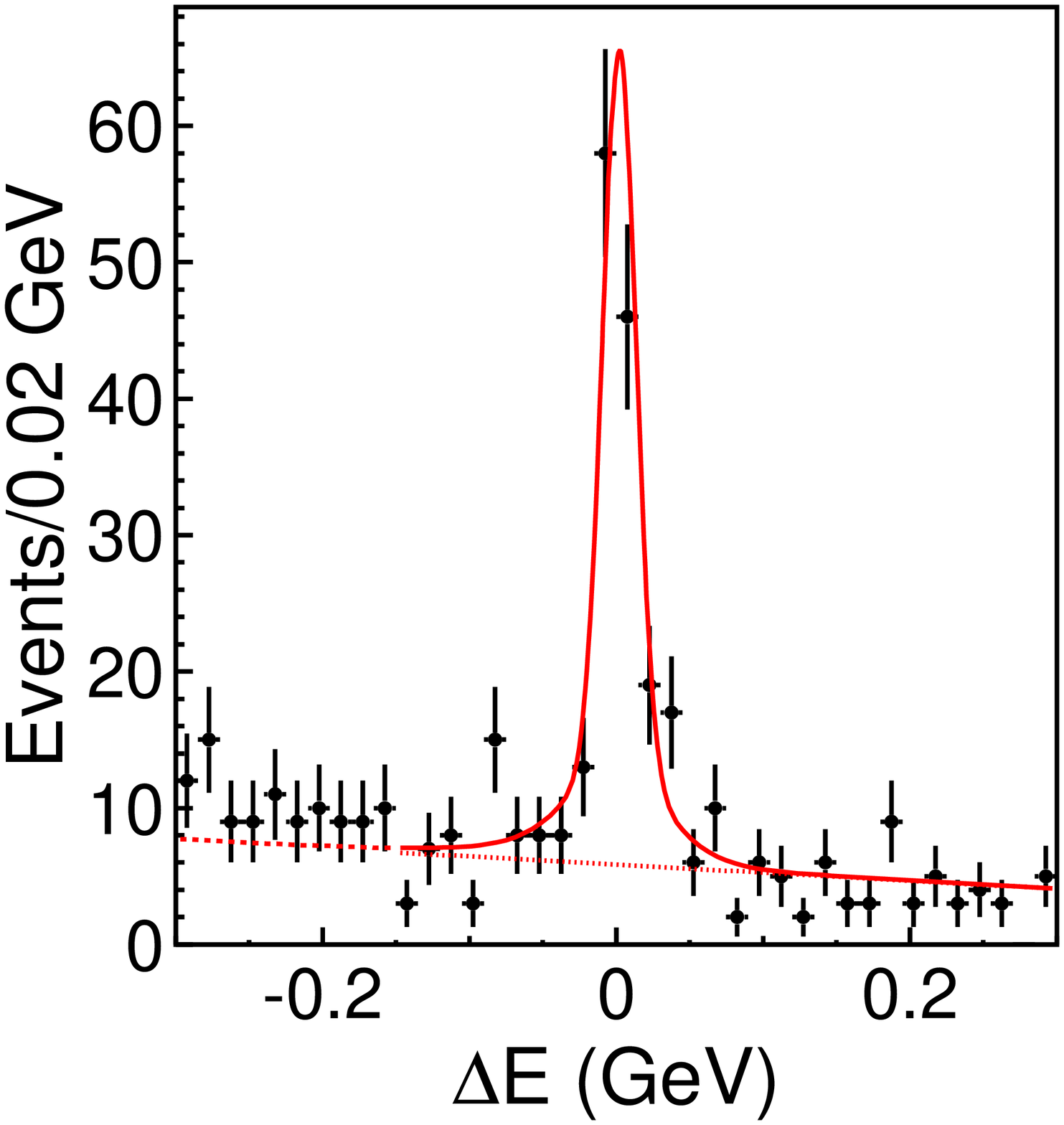}
\caption{Distributions of (a) $\mb$ within the $\dE$ signal region,
(b) $\dE$ within the $\mb$ signal region
for $\bz\to\ks\ks\ks$ candidates.
Solid curves show the fits to signal plus background distributions,
and dashed curves show the background contributions.}
\label{fig:ksksks}
\rput[l]( -6.2,  9.7)  {\Large(a)}
\rput[l](  1.8,  9.7)  {\Large(b)}

\end{figure}

Figure~\ref{fig:ksksks} shows the $\mb$ and $\dE$ distributions for the
reconstructed $\bz\to\ks\ks\ks$ candidates
after flavor tagging and vertex reconstruction.
The signal yield is determined
from an unbinned two-dimensional maximum-likelihood fit
to the $\dE$-$\mb$ distribution.
The $\kspm\kspm\kspm$ signal distribution 
is modeled with a Gaussian function (a sum of two Gaussian functions)
for $\mb$ ($\dE$).
For $\bz\to\kspm\kspm\kszz$ decay,
the signal is modeled with a two-dimensional
smoothed histogram obtained from MC events.
For the continuum background,
we use the ARGUS parameterization
for $\mb$ and a linear function for $\dE$.
The fits after flavor tagging and vertex reconstruction yield 
$\Nsigkspmkspmkspm$ $\bz\to\kspm\kspm\kspm$ events and
$\Nsigkspmkspmkszz$ $\bz\to\kspm\kspm\kszz$ events
for a total of $\Nsigksksks$ $\bz\to\ks\ks\ks$ events 
in the signal region,
where the errors are statistical only.
The obtained purity
 is $\Pkspmkspmkspm$ for the $\kspm\kspm\kspm$
and $\Pkspmkspmkszz$ for the $\kspm\kspm\kszz$ channels.
Here the purity is defined as $\nsig/\nev$,
where $\nsig$ is the number of signal events 
in the signal region obtained by the fit, 
and $\nev$ is the total number of events in the signal region.
%
%
%
%

\subsection{\boldmath $\bz\to\ks\piz$}
\label{sec:bztopizks}
Candidate $\ks \to \pip\pim$ decays
are selected with the same criteria as those used for
the $\bz\to\phi\ks$ decay, except that we use
pairs of oppositely charged pions that have an invariant mass
within 0.018 GeV/$c^2$
of the nominal $\ks$ mass.
The $\piz$ selection criteria are the same as
those used for the $\bz\to\eta'\ks$ decay.

The $B$ meson signal region is defined as 
$-0.15$ GeV $< \dE <0.1$ GeV
and $5.27~{\rm GeV}/c^2 <\mb<5.29~{\rm GeV}/c^2$.
The $\dE$ distribution for $\ks\piz$ has a tail toward lower $\dE$.
The $\dE$ resolution is 0.047~GeV for the main component.
The width of the tail is about 0.1~GeV.
The dominant background is from continuum.
In addition, according to MC simulation, there is a small ($\sim 2\%$)
contamination from other charmless rare $B$ decays.
We use extended modified Fox-Wolfram moments,
which were applied for the selection of
the $\bz\to\piz\piz$ decay~\cite{Chen:2005dr},
to form a Fisher discriminant $\calf$.
We then combine likelihoods for $\calf$ and $\cos\theta_B$
to obtain the event likelihood ratio $\rsigbkg$ for
continuum suppression.

As described below, we include events that do not have
$B$ decay vertex information in our fit
to obtain a better sensitivity for
the $CP$-violation parameter $\cala$.
For events with and without vertex information,
the high-$\rsigbkg$ region is defined as
$\rsigbkg > 0.8$
and the low-$\rsigbkg$ region as
$0.45 < \rsigbkg \le  0.8$ for both DS-I and DS-II.
After applying the high-$\rsigbkg$ requirement, 95\% of the continuum background
is rejected and 62\% of signal events remain.
In the low-$\rsigbkg$ region, 84\% of the continuum background is rejected 
and 24\% of the signal remains.

\begin{figure}
\includegraphics[width=0.48\textwidth]{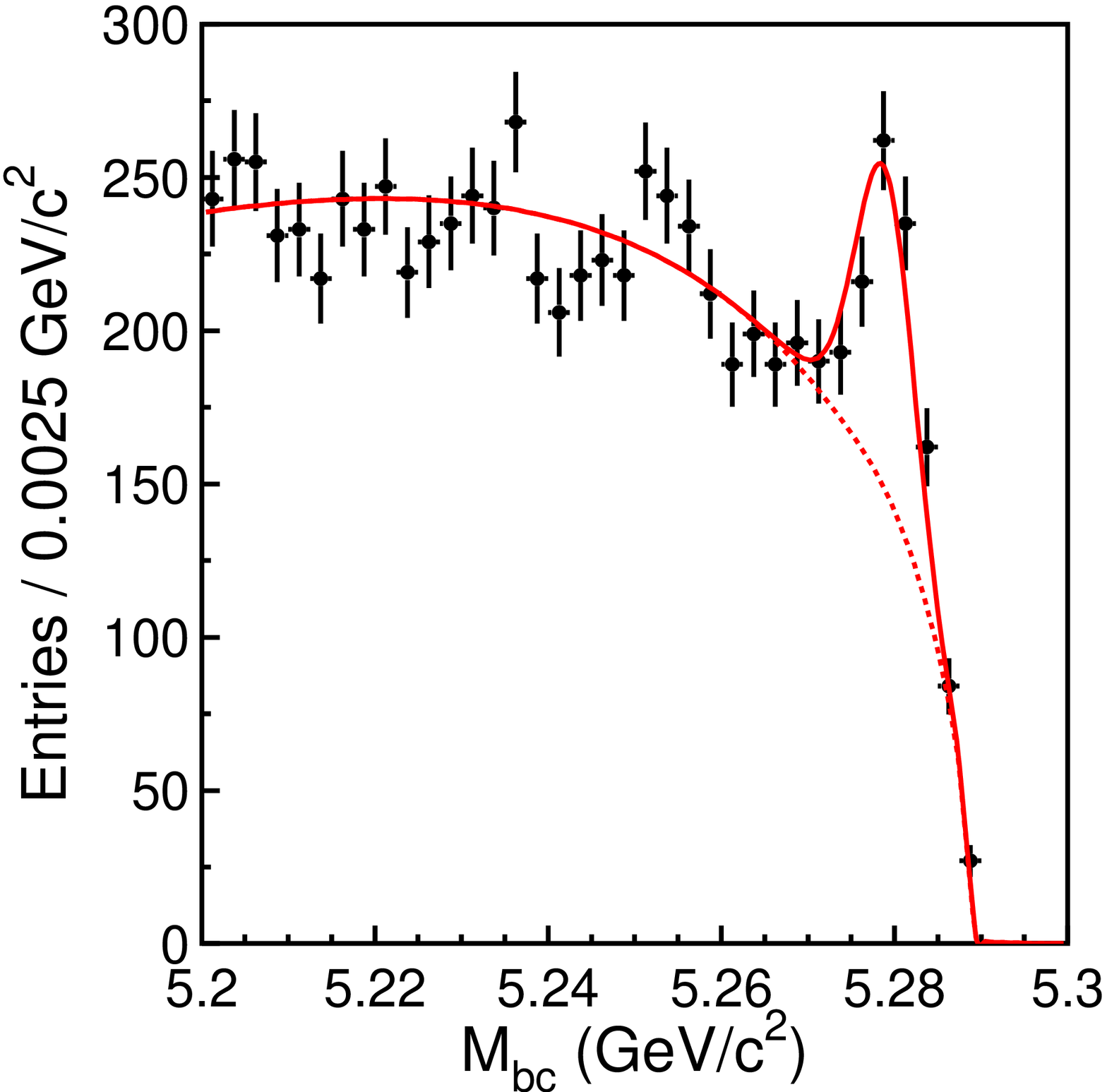}
\includegraphics[width=0.48\textwidth]{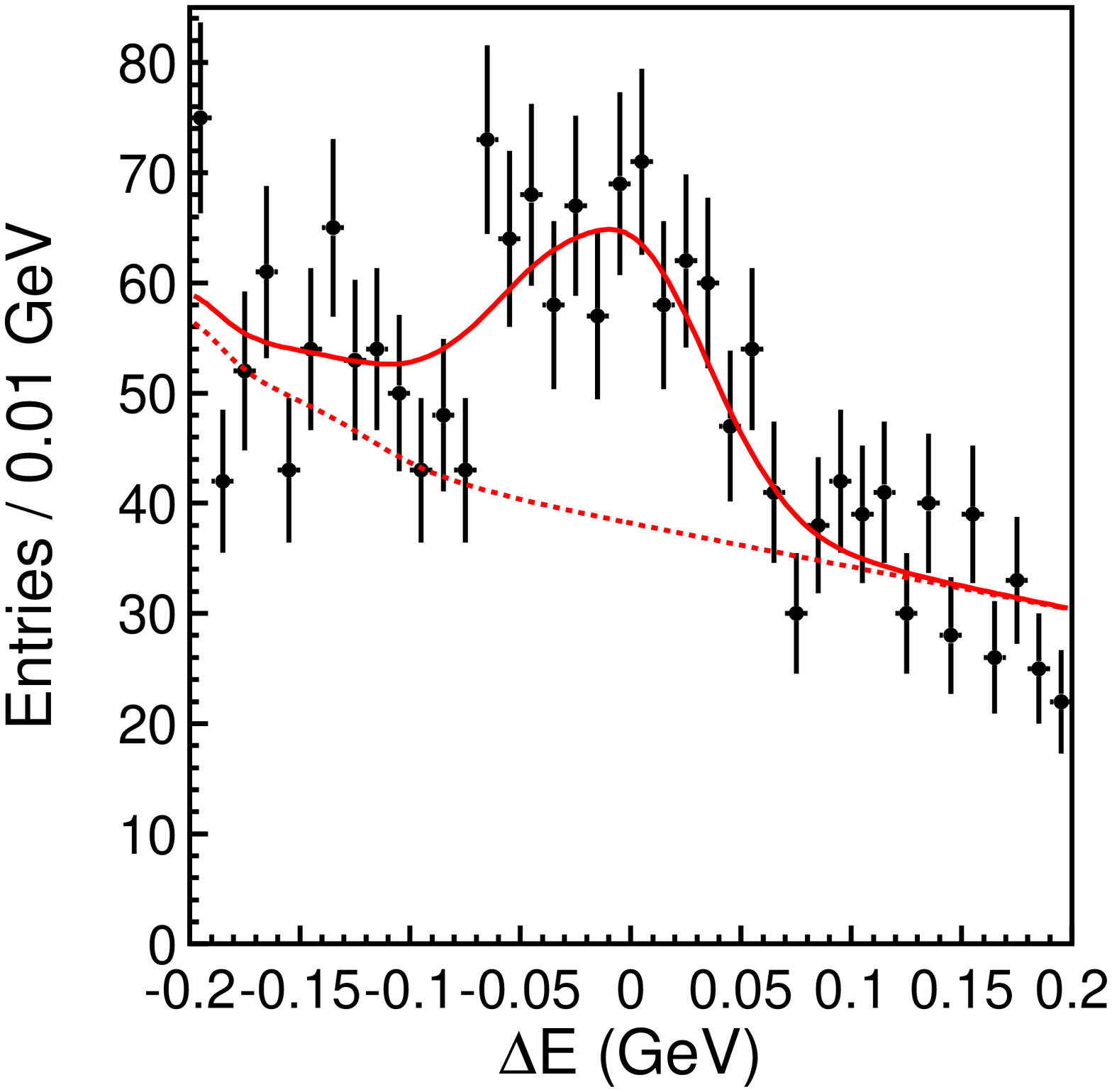}
\caption{Distributions of (a) $\mb$ within the $\dE$ signal region,
(b) $\dE$ within the $\mb$ signal region
for $\bz\to\ks\piz$ candidates.
Solid curves show the fits to signal plus background distributions,
and dashed curves show the background contributions.}
\label{fig:kspiz}
\rput[l]( -1.8,  9.0)  {\Large(a)}
\rput[l](  6.2,  9.0)  {\Large(b)}

\end{figure}

Figure~\ref{fig:kspiz}(a) shows
the $\mb$ distribution for the $\bz\to\ks\piz$ candidates
within the $\dE$ signal region
after flavor tagging and before vertex reconstruction.
Also shown in Fig.~\ref{fig:kspiz}(b) is the $\dE$ distribution
within the $\mb$ signal region.
%
%
%
%
The signal yield is determined
from an unbinned two-dimensional maximum-likelihood fit
to the $\dE$-$\mb$ distribution in the fit region defined as
$ 5.2~{\rm GeV/}c^2 < \mb < 5.29~{\rm GeV/}c^2$ 
and
$-0.3~{\rm GeV} < \dE < 0.3~{\rm GeV}$.
The $\bz\to\ks\piz$ signal distribution is modeled with
a smoothed histogram obtained from MC and calibrated with
data using $\bm\to D^0\pim$ $(D^0\to \km\pip\piz)$.
For the continuum background, 
we use the ARGUS parameterization for $\mb$
and a linear function for $\dE$.
The $B$ decay background distribution is represented by
a smoothed histogram obtained from MC simulation. 
The fits yield $\NsigkspizH$ and $\NsigkspizL$ $\bz\to\ks\piz$ events
in the high-$\rsigbkg$ and 
low-$\rsigbkg$ signal regions, respectively, where the errors are
statistical only.
The same procedure after the vertex reconstruction yields
a total of $106\pm 14$ $\ks\piz$ events.

\subsection{\boldmath $\bz\to\fzero\ks$}
\label{sec:bztofzeroks}

Candidate $\ks\to\pip\pim$ decays are selected
with criteria that are slightly different from
those used for the $\bz\to\phi\ks$ decay
so as to obtain the best sensitivity to $CP$ violation in the $\bz\to\fzero\ks$ decay.
Pairs of oppositely charged tracks that have an invariant mass
between 0.484 GeV/$c^2$ and 0.513 GeV/$c^2$ are used to reconstruct
$K_S \rightarrow \pi^+ \pi^-$ decays. The distance of closest approach of
the candidate charged tracks to the IP in the plane perpendicular to $z$
axis is required to be larger than 0.008 cm. The $\pi^+ \pi^-$ vertex is
required to be displaced from the IP by a minimum transverse distance
distance of 0.1 cm. The direction of the pion pair momentum must also
agree with the direction of the vertex point from the IP to within 0.03
rad.

Pairs of oppositely charged pions that have invariant masses
between 0.890 and 1.088 GeV/$c^2$ are used to reconstruct
$\fzero\to\pip\pim$ decays.
Tracks that are identified as kaons ($\rkpi > 0.7$) or
electrons are not used.
We reject both $\ks\pip$ and $\ks\pim$
combinations with an invariant mass within
0.02 GeV/$c^2$ of the nominal charged
$D$ meson mass to remove background from $D^\pm\to\ks\pi^\pm$.

The $B$ meson signal region is defined as
$-0.03~{\rm GeV} < \dE <0.06$ GeV and $5.27~{\rm GeV}/c^2 <\mb<5.29~{\rm GeV}/c^2$.
The $\dE$ resolution is about 20 MeV.
The dominant background is from continuum. 
The likelihood ratio $\rsigbkg$ is obtained from
$\cos\theta_B$, $\calf$ and $\cos \theta_{H}$, where
the helicity angle $\theta_H$ is defined as the angle
between the $\bz$ meson momentum and the $\pip$ momentum in the $\fzero$
meson rest frame. The requirement for
$\rsigbkg$ depends on the flavor tagging $r$, and the threshold values
range from 0.3 (used for $r > 0.875$) to 0.8 (used for $r < 0.25$).
The continuum background is reduced by 93\%, while retaining 72\% of signal
events with the requirement on $\rsigbkg$.

\begin{figure}
\includegraphics[width=0.48\textwidth]{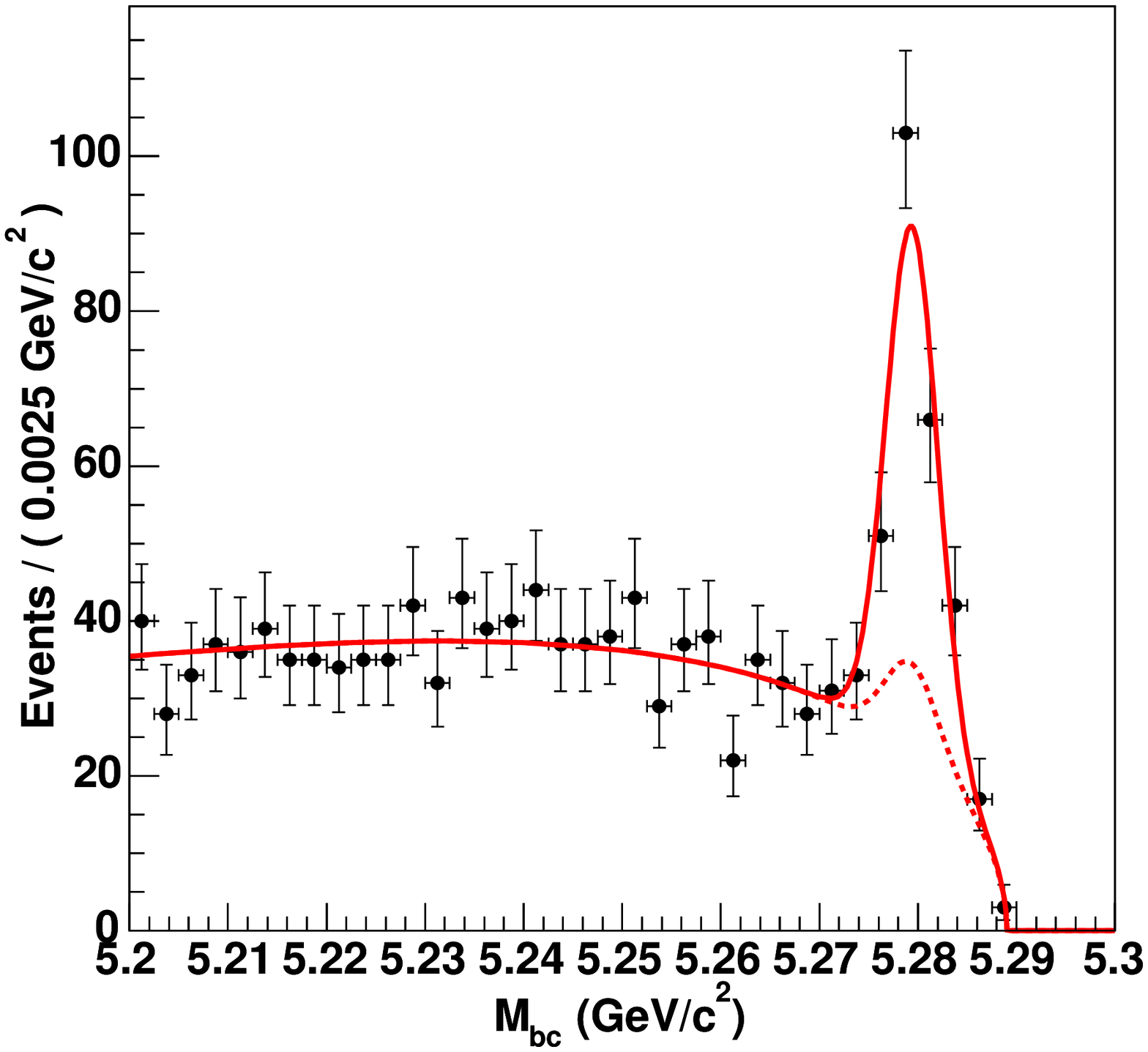}
\includegraphics[width=0.48\textwidth]{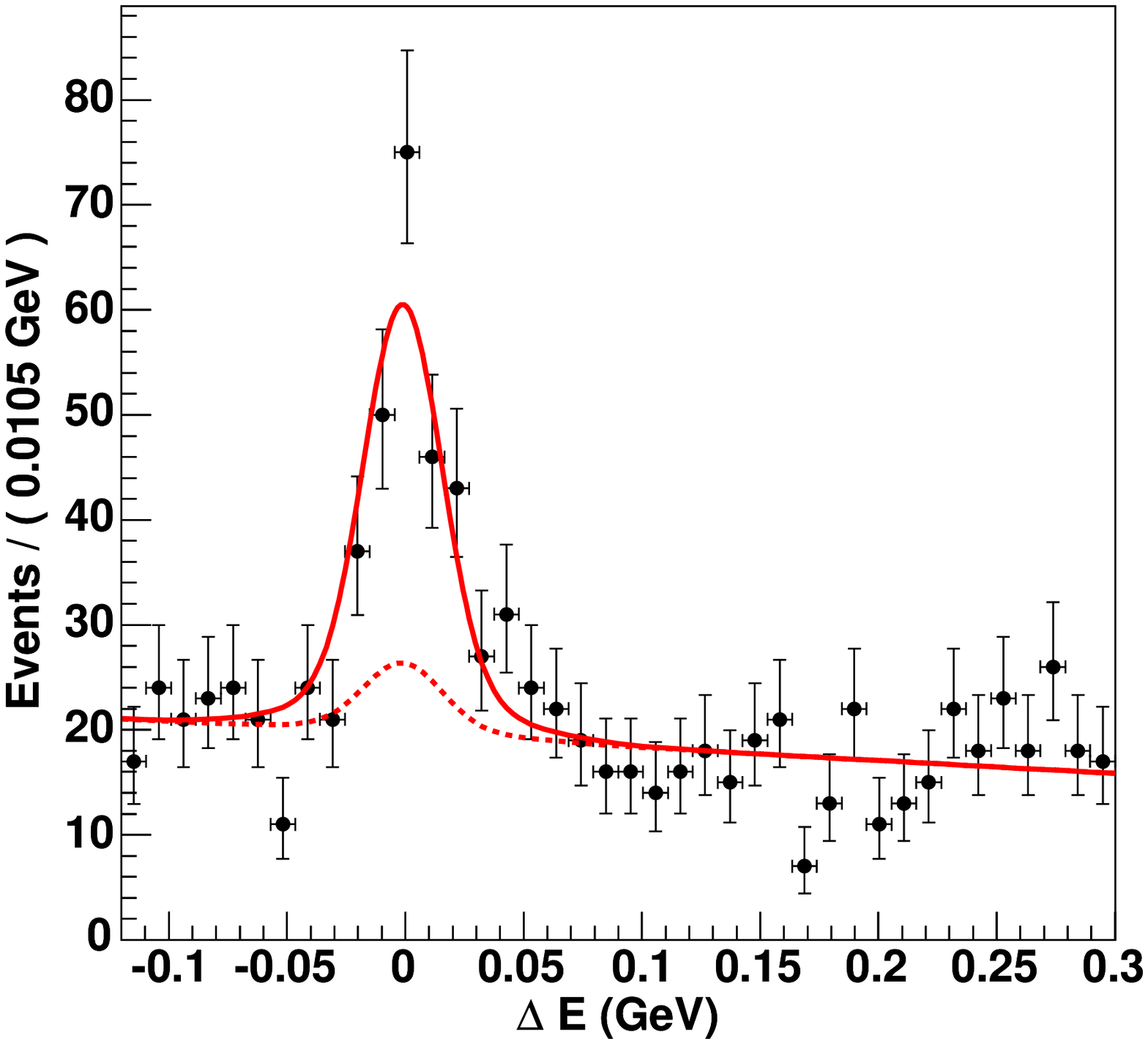}
\vskip 1cm
\includegraphics[width=0.48\textwidth]{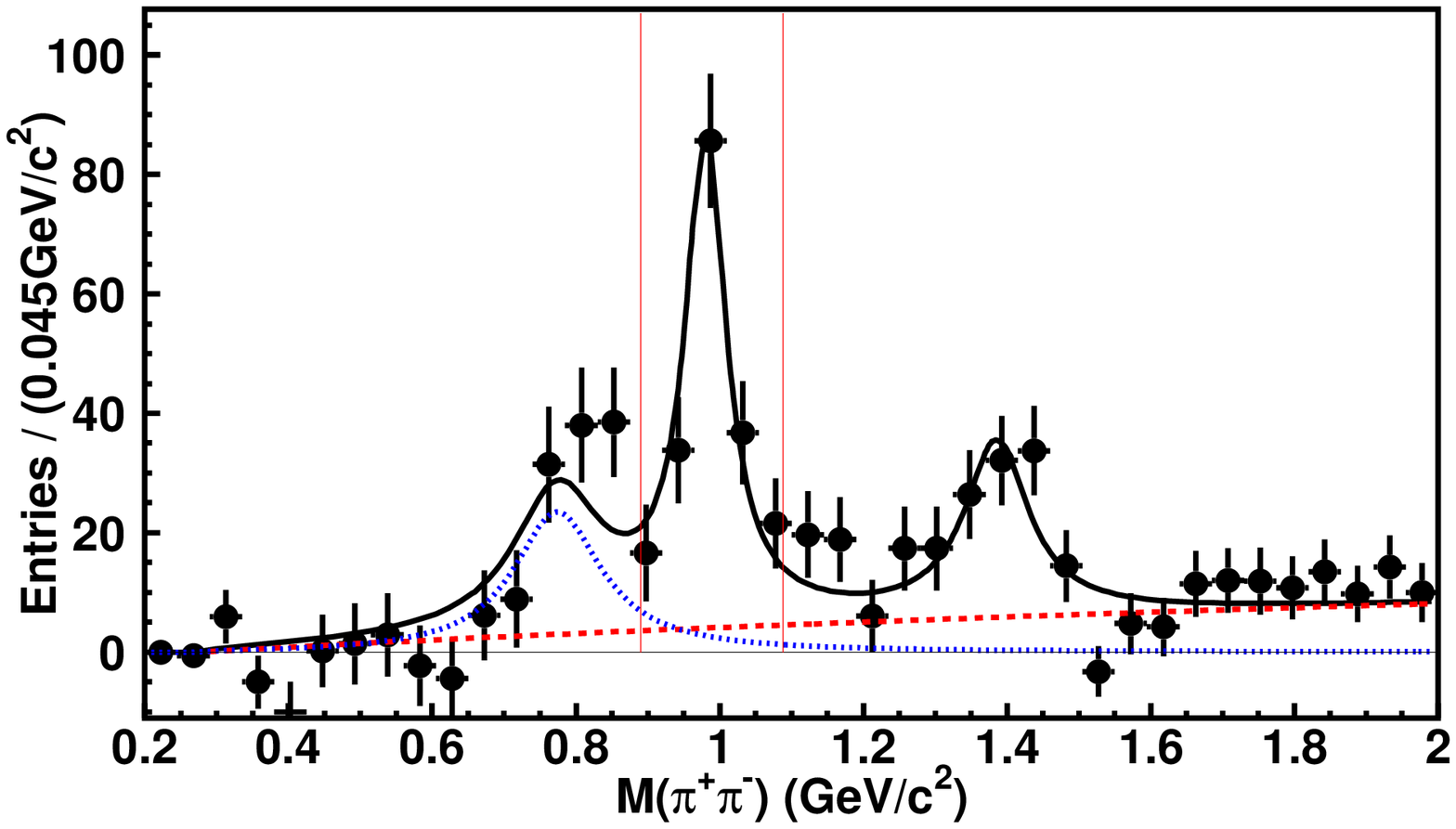}
\caption{Distributions of (a) $\mb$ in the $\dE$ signal region,
(b) $\dE$ in the $\mb$ signal region
and (c) $m_{\pip\pim}$ in the $\dE$-$\mb$ signal region
for $\bz\to\fzero\ks$ candidates.
Solid curves show the fits to signal plus background distributions.
Dashed curves in (a) and (b) show the background contributions.
In (c), 
each point shows a yield for $\bz\to\pip\pim\ks$ including $\fzero\ks$
obtained from a fit to the $\dE$-$\mb$ distribution,
the dotted line shows the $\bz\to\rho\ks$, and
the dashed line shows
other quasi two-body decays as well as
three-body $\bz\to\pip\pim\ks$ decays.
}
\label{fig:fzeroks}
\rput[l]( -6.8,  15.7)  {\Large(a)}
\rput[l](  1.2,  15.7)  {\Large(b)}
\rput[l]( -3.0,  8.0)  {\Large(c)}
\end{figure}

Figure~\ref{fig:fzeroks}(a) shows
the $\mb$ distribution for the reconstructed $\bz\to\fzero\ks$ candidates
within the $\dE$ signal region
after flavor tagging and vertex reconstruction.
The $\dE$ distribution for the $\bz\to\fzero\ks$ candidates
within the $\mb$ signal region is shown in Fig.~\ref{fig:fzeroks}(b).
%
%
%
%
For the signal yield extraction,
we first perform
an unbinned two-dimensional maximum-likelihood fit
to the $\dE$-$\mb$ distribution in the fit region defined as
$\mb > 5.2~{\rm GeV/}c^2$
and
$-0.12~{\rm GeV} < \dE < 0.3~{\rm GeV}$.
The signal is modeled with a Gaussian function
(a sum of two Gaussian functions) for $\mb$ ($\dE$).
For the continuum background,
we use the ARGUS parameterization for $\mb$
and a linear function for $\dE$.
The fit yields the number of
$\bz\to\pip\pim\ks$ events that have
$\pip\pim$ invariant masses within
the $\fzero$ resonance region, which
may include contributions from
$\bz\to\rhoz\ks$ as well as non-resonant three-body
$\bz\to\pip\pim\ks$ decays.
To separate these peaking backgrounds from
the $\bz\to\fzero\ks$ decay,
we perform another fit to the $\pip\pim$ invariant
mass distribution for the events inside the
$\dE$-$\mb$ signal region. We use Breit-Wigner
functions for the $\bz\to\fzero\ks$ signal,
for the $\bz\to\rho\ks$ background and for a possible resonance
above the $\fzero$ mass region, which is referred to as $\fx$. 
The contributions from other resonant or non-resonant $\bz\to\pip\pim\ks$ decays are modeled with
a threshold function.
The combinatorial background is represented by the $\mb$-$\dE$ sideband
and subtracted from the signal region distribution.
The $\pip\pim$ invariant mass distribution with the fit result
is shown in Fig.~\ref{fig:fzeroks}(c).
The fit yields $\Nsigfzeroks$ $\bz\to\fzero\ks$ events.

\subsection{\boldmath $\bz\to\omega\ks$}
\label{sec:bztoomegaks}
Candidate $\ks \to \pip\pim$ decays
are selected with criteria that
are identical to those used for the $\bz\to\phi\ks$ decay.
Pions for the $\omega\to\pip\pim\piz$ decay
are selected with the same criteria used
for the $\eta\to\pip\pim\piz$ decay, except that
we require $\ppizcms > 0.35~{\rm GeV}/c$.
The $\pip\pim\piz$ invariant mass $M_{3\pi}$ is required to be
between 0.73 GeV/$c^2$ and 0.84 GeV/$c^2$.
The $B$ meson signal region is defined as 
$-0.10~{\rm GeV}<\dE<0.08~{\rm GeV}$
and $5.27~{\rm GeV}/c^2 <\mb<5.29~{\rm GeV}/c^2$.
The $\dE$ resolution is 0.028~GeV.
The dominant background is from continuum.
The continuum suppression is based on the
likelihood ratio $\rsigbkg$ obtained from
the same discriminating variables 
used for the $\bz\to\phi\ks$ decay plus
the helicity angle $\theta_H$ defined as the
angle between the $\bz$ meson momentum and the cross product
of the $\pip$ and $\pim$ momenta
in the $\omega$ meson rest frame.
We also require $|\cos\theta_T|<0.9$ prior to the $\rsigbkg$ requirement.
We define two $\rsigbkg$ regions.
The $\rsigbkg$ requirements
depend on the flavor-tagging quality. 
The boundary between the high-$\rsigbkg$ regions
and the low-$\rsigbkg$ regions
is 0.85 for all $r$ values.
The minimum $\rsigbkg$ requirements range
from 0.1 to 0.6 for the low-$\rsigbkg$ regions.
The $\rsigbkg$ and $|\cos\theta_T|$ requirements reject 85\% of
the continuum background while retaining 84\% of the signal.
The contribution from $B\overline{B}$ events is negligibly small.

\begin{figure}
\includegraphics[width=0.48\textwidth]{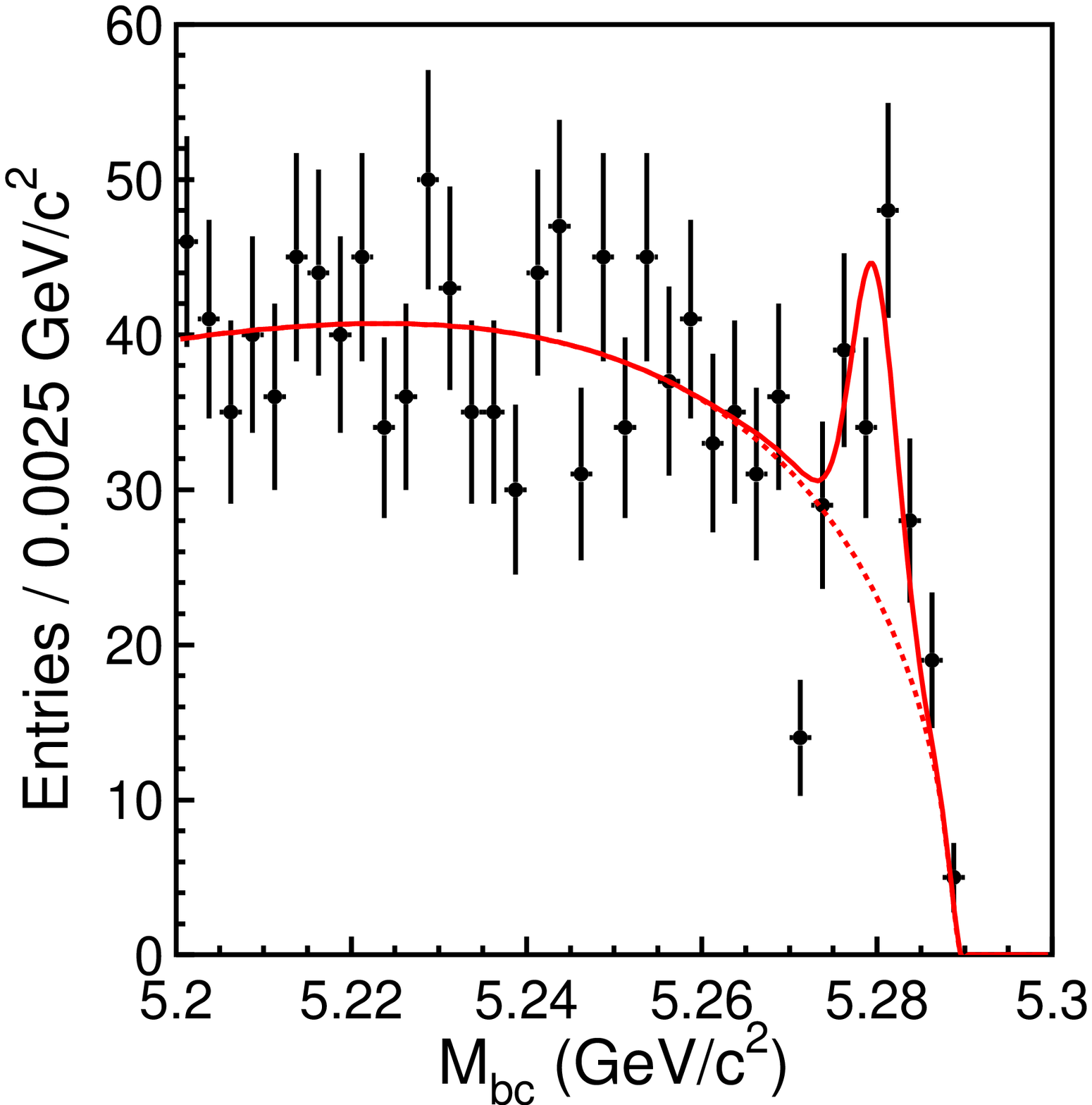}
\includegraphics[width=0.48\textwidth]{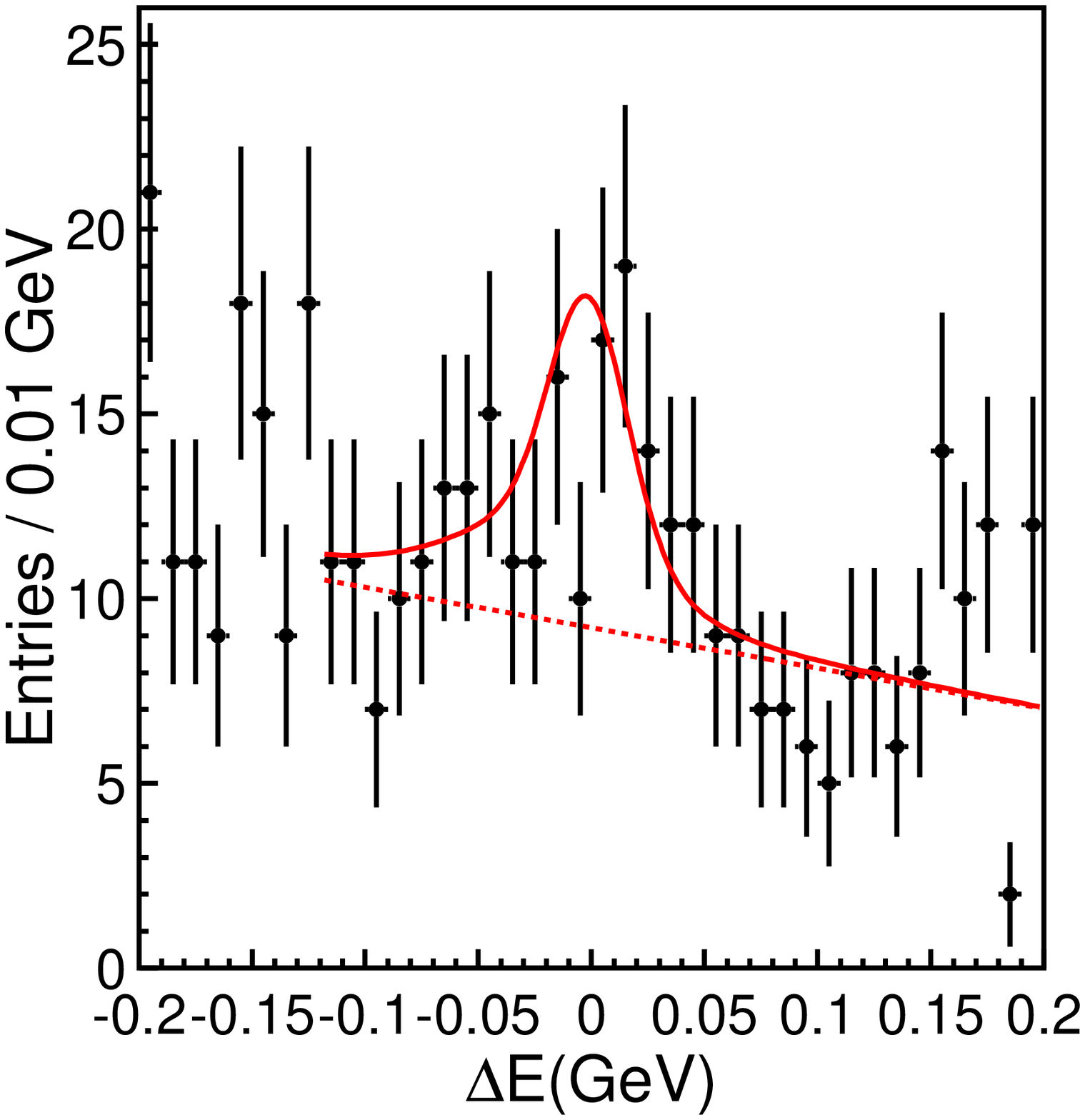}
\includegraphics[width=0.48\textwidth]{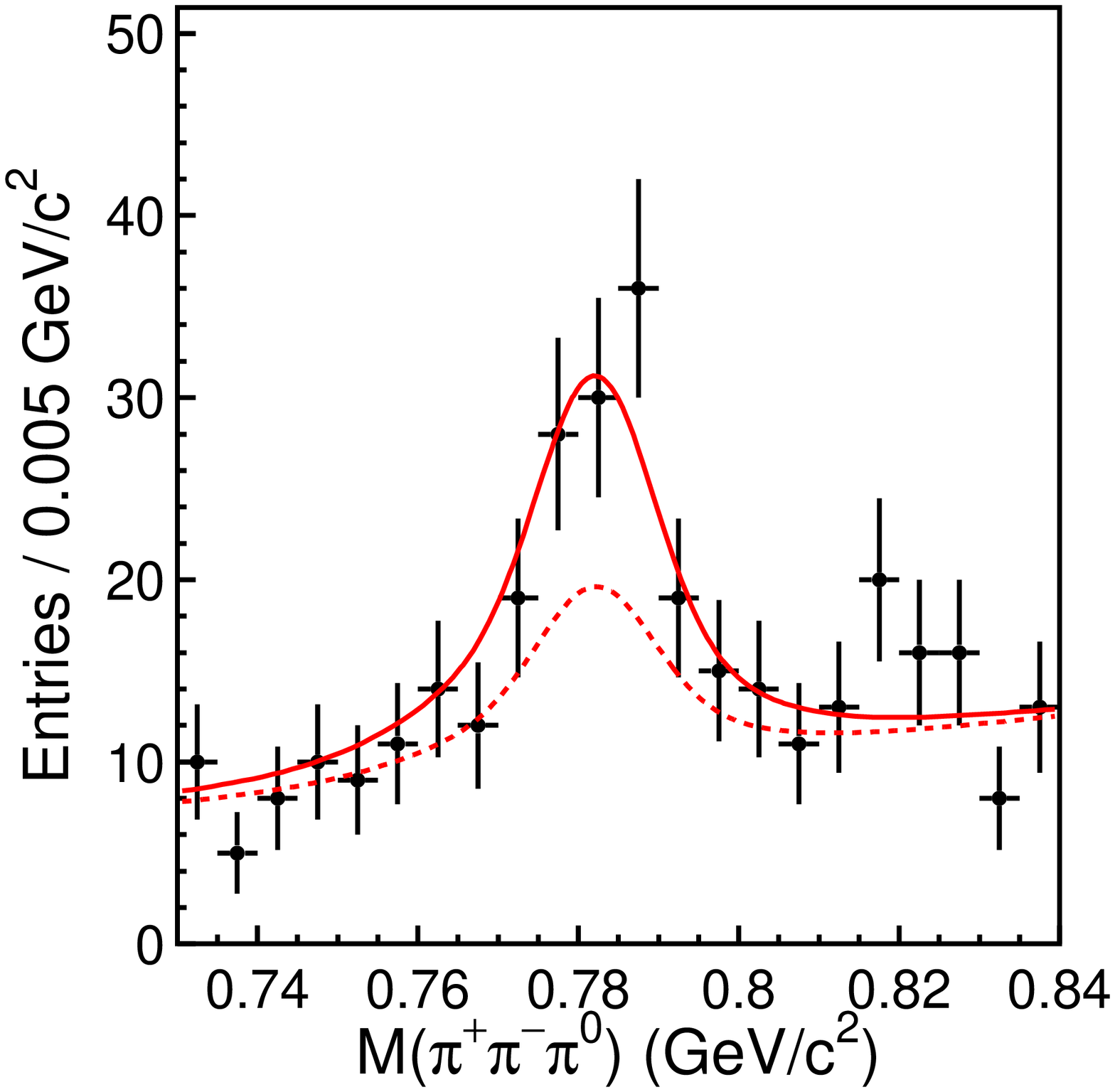}
\caption{Distributions of (a) $\mb$ within the $\dE$ signal region,
(b) $\dE$ within the $\mb$ signal region and
(c) $m_{\pip\pim\piz}$ within the $\dE$-$\mb$ signal region
for $\bz\to\omega\ks$ candidates.
Solid curves show the fits to signal plus background distributions,
and dashed curves show the background contributions.}
\label{fig:omegaks}
\rput[l]( -6.3,  17.5)  {\Large(a)}
\rput[l](  1.7,  17.5)  {\Large(b)}
\rput[l]( -2.3, 9.5)  {\Large(c)}

\end{figure}

Figures~\ref{fig:omegaks}(a-c) show
the $\mb$ distribution for the reconstructed $\bz\to\omega\ks$ candidates
within the $\dE$ signal region,
the $\dE$ distribution within the $\mb$ signal region 
and the $M_{3\pi}$ distribution within the $\dE$-$\mb$
signal region, respectively,
after flavor tagging and vertex reconstruction.
%
%
%
%
The signal yield is determined
from an unbinned three-dimensional maximum-likelihood fit
to the $\dE$-$\mb$-$M_{3\pi}$ distribution in the fit region
defined as
$\mb > 5.2~{\rm GeV/}c^2$,
$-0.12~{\rm GeV} < \dE < 0.25~{\rm GeV}$
and
$0.73~{\rm GeV}/c^2 < M_{3\pi} < 0.84~{\rm GeV}/c^2$.
The $\bz\to\omega\ks$ signal distribution
is modeled with a sum of two (three) Gaussian
functions for $\mb$ ($\dE$ and $M_{3\pi}$).
For the continuum background, 
we use the ARGUS parameterization for $\mb$,
a linear function for $\dE$ and
a second-order polynomial function plus
three Gaussian functions for $M_{3\pi}$.
The fit yields $\Nsigomegaks$ $\bz\to\omega\ks$ events
in the signal region.

\subsection{\boldmath $\bz\to\jpsi\ks$ and $\jpsi\kl$}
\label{sec:bztojpsikzero}
The reconstruction and selection criteria for 
$\bz\to\jpsi\ks$ decays
used in this measurement are the same as those
in the previous publication, which are
described in detail elsewhere~\cite{bib:CP1_Belle}.
We reconstruct $\jpsi$ candidates
via their decays to $\ell^+\ell^-$ ($\ell = \mu,e$), and
$\ks$ candidates via $\ks \to \pip\pim$ decays.
The $B$ meson signal region is defined as 
$|\dE|<0.04$ GeV 
and $5.27~{\rm GeV}/c^2 <\mb<5.29~{\rm GeV}/c^2$.

Candidate $\jpsi\to\mu^+\mu^-$ or $e^+e^-$ decays for
the $\bz \to \jpsi\kl$ mode are
selected by requiring
3.05 GeV/$c^2 < M_{\mu\mu} < 3.13$ GeV/$c^2$ or
2.95 GeV/$c^2 < M_{ee} < 3.13$ GeV/$c^2$, where
$M_{\mu\mu}~(M_{ee})$ is the invariant mass of
the $\mu^+\mu^-$ $(e^+e^-)$ pair.
The momentum of the reconstructed $\jpsi$ candidate
is required to be between 1.38 GeV/$c$ and 2.00 GeV/$c$.
The selection criteria for $\kl$ candidates are identical to
those in the $\bz\to\phi\kl$ analysis, except that
the $\kl$ likelihood ratio for the ECL cluster
is required to be greater than 0.25 
for both KLM+ECL and ECL candidates.
The $B$ signal region is defined as
$0.2~{\rm GeV}/c < \pbstar < 0.45~{\rm GeV}/c$.

\begin{figure}
\includegraphics[width=0.48\textwidth]{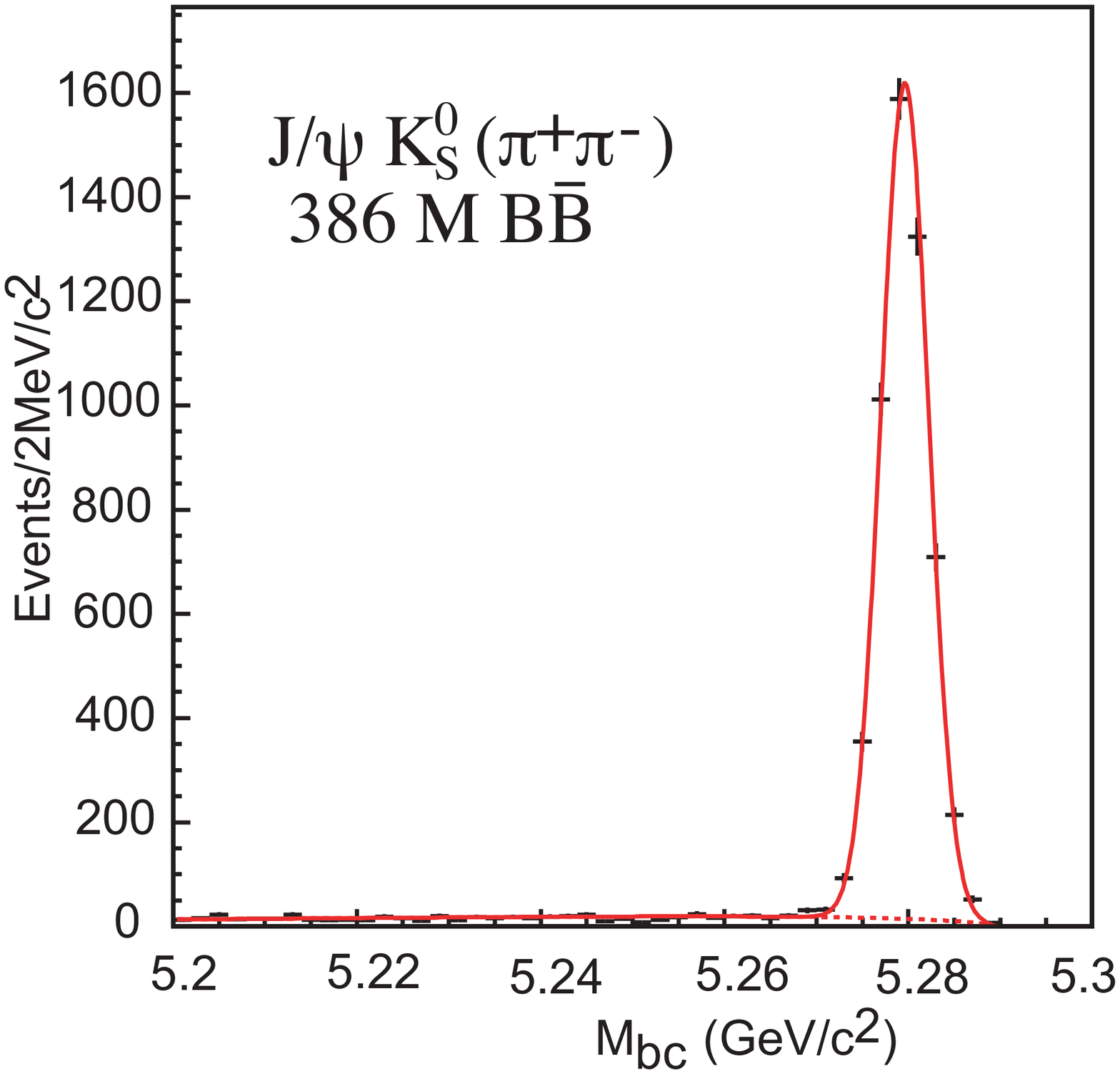}
\includegraphics[width=0.49\textwidth]{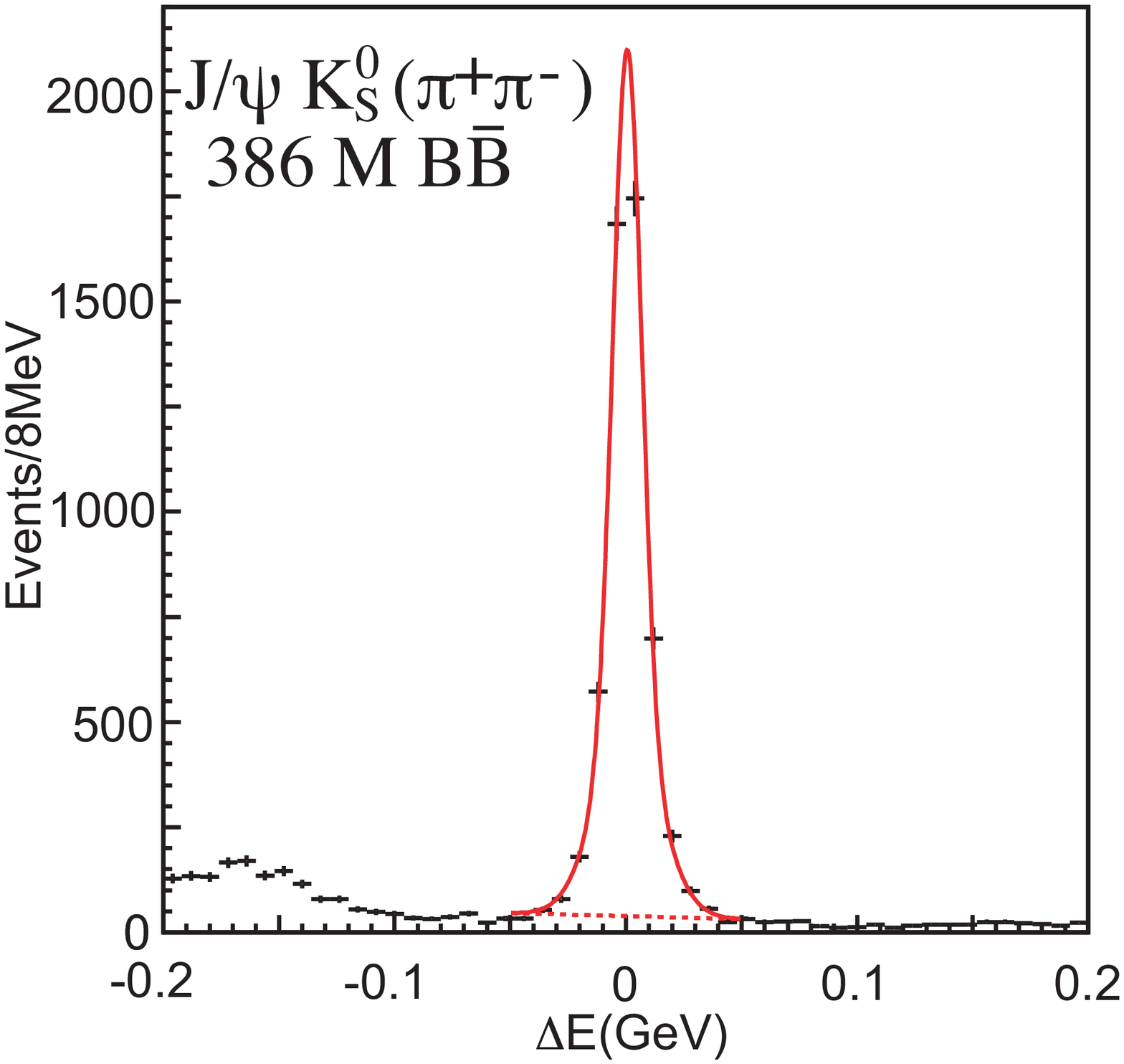}
\includegraphics[width=0.48\textwidth]{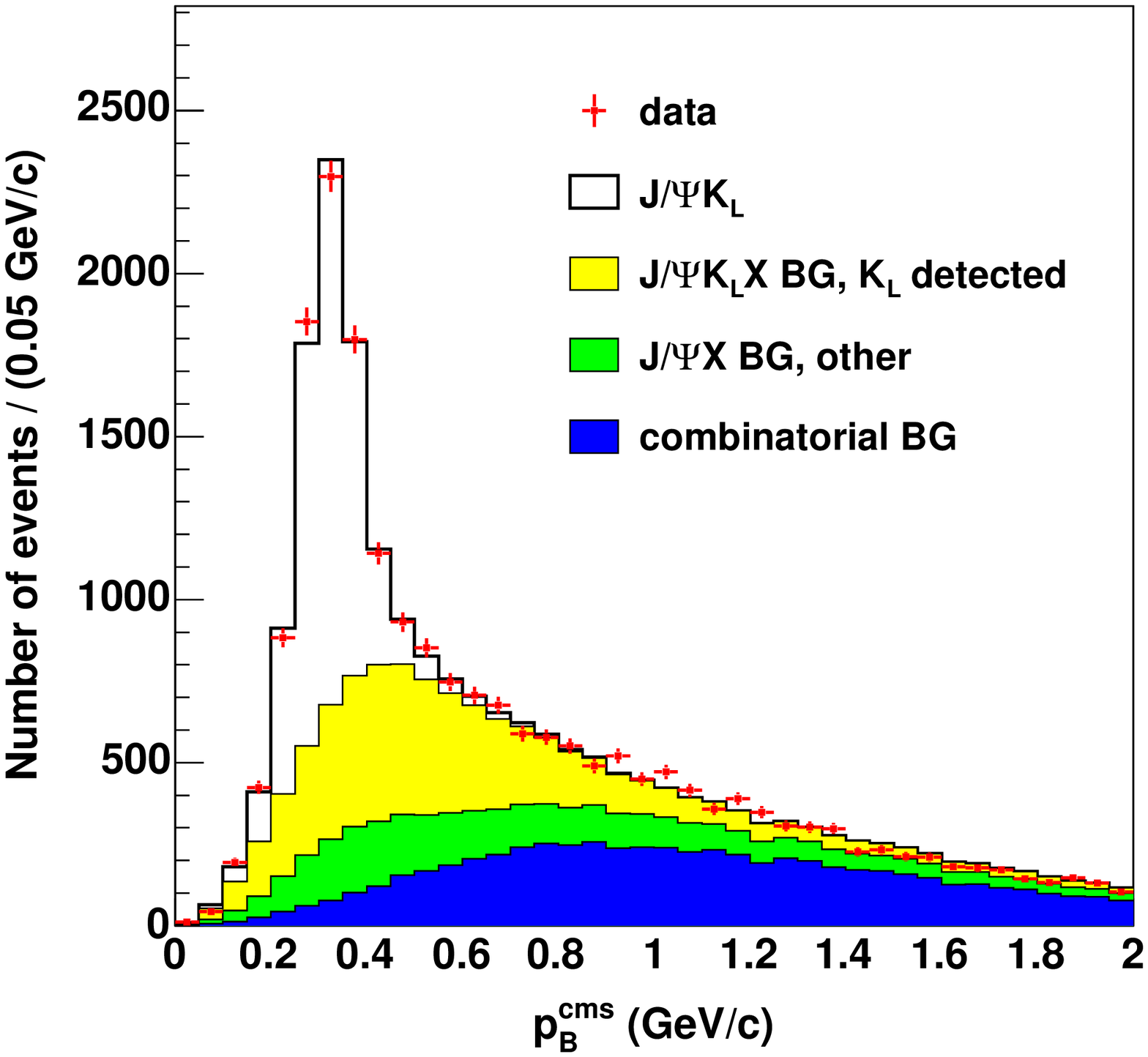}
\caption{Distributions of (a) $\mb$ within the $\dE$ signal region
and
(b) $\dE$ within the $\mb$ signal region for
$\bz\to\jpsi\ks$ candidates,
and
(c) $\pbstar$
for $\bz\to\jpsi\kl$ candidates.}
\label{fig:jpsikz}
\rput[l]( -6.7,  16.7)  {\Large(a)}
\rput[l]( 7.0,  16.7)  {\Large(b)}
\rput[l]( -2.5, 8.8)  {\Large(c)}

\end{figure}

Figure~\ref{fig:jpsikz}(a) shows
the $\mb$ distribution for the reconstructed $\bz\to\jpsi\ks$ candidates
within the $\dE$ signal region
after flavor tagging and vertex reconstruction.
The $\dE$ distribution for the $\bz\to\jpsi\ks$ candidates 
within the $\mb$ signal region is shown in Fig.~\ref{fig:jpsikz}(b).
The signal yield for the $\bz\to\jpsi\ks$ decay is determined
from an unbinned two-dimensional maximum-likelihood fit
to the $\dE$-$\mb$ distribution.
The fit region is defined as
$|\dE| < 0.05~{\rm GeV}$ and $\mb > 5.2~{\rm GeV/}c^2$.
The signal distribution
is modeled with a Gaussian function (a sum of two Gaussian functions)
for $\mb$ ($\dE$).
For the background,
we use the ARGUS parameterization for $\mb$
and a linear function for $\dE$.
Figure~\ref{fig:jpsikz}(c) shows
the $\pbstar$ distribution for the reconstructed 
$\bz\to\jpsi\kl$ candidates.
The signal yield for the $\bz\to\jpsi\kl$ decay is determined
from a binned maximum-likelihood fit to the $\pbstar$ distribution
for each of KLM, KLM+ECL and ECL candidates separately.
The fit region is defined as
$0~{\rm GeV}/c < \pbstar < 2~{\rm GeV}/c$.
The shapes of the signal and background with $J/\psi$ 
are determined from the $J/\psi$ inclusive MC sample. Here
background distributions with $\kl$ and without $\kl$ are treated
separately to minimize the effect of an uncertainty in the $\kl$
detection efficiency in the MC simulation.
The background shape for the case with a fake $J/\psi$ meson is obtained
from events in the sideband of the $\ell^+\ell^-$ mass distribution.

%
%
%
%

\subsection{Flavor Tagging}
\label{sec:flavor tagging}
The $b$-flavor of the accompanying $B$ meson is identified
from inclusive properties of particles
that are not associated with the reconstructed $\bz \to \fCP$ 
decay. We use the same procedure as for our previous
$\sinbb$ measurement~\cite{bib:BELLE-CONF-0436}.
The algorithm for flavor tagging is described in detail
elsewhere~\cite{bib:fbtg_nim}.
We use two parameters, $\fq$ and $r$, to represent the tagging information.
The first, $\fq$, is defined in Eq.~(\ref{eq:psig}).
The parameter $r$ is an event-by-event,
MC-determined flavor-tagging dilution factor
that ranges from $r=0$ for no flavor
discrimination to $r=1$ for unambiguous flavor assignment.
It is used only to sort data into six $r$ intervals listed in
Table~\ref{tab:wtag}.
The wrong tag fractions for the six $r$ intervals, 
$w_l~(l=1,6)$, and differences 
between $\bz$ and $\bzb$ decays, $\dwl$,
are determined from the data;
we use the same values
that were used for the $\sin 2\phi_1$ measurement~\cite{bib:BELLE-CONF-0436}
for DS-I.
Wrong tag fractions for DS-II are separately obtained 
with the same procedure 
and are listed in Table~\ref{tab:wtag}.
%
%
\begin{table*}
  \caption{Event fractions $\epsilon_l$,
    wrong-tag fractions $w_l$, wrong-tag fraction differences $\dwl$,
    and average effective tagging efficiencies
    $\eeff^l = \epsilon_l(1-2w_l)^2$ for each $r$ interval for DS-II.
    Errors for $w_l$ and $\dwl$
    include both statistical and systematic uncertainties.
    The event fractions are obtained from $\jpsi\ks$ data.}
  \begin{ruledtabular}
    \begin{tabular}{ccclll}
      $l$ & $r$ interval & $\epsilon_l$ &\multicolumn{1}{c}{$w_l$} 
          & \multicolumn{1}{c}{$\dwl$}  &\multicolumn{1}{c}{$\eeff^l$} \\
      \hline
%
 1 & 0.000 -- 0.250 & $0.384\pm 0.011$ & $0.467\pm 0.006$ & $+0.005\pm 0.007$ & $0.002\pm 0.001$ \\
 2 & 0.250 -- 0.500 & $0.165\pm 0.007$ & $0.324\pm 0.007$ & $-0.029\pm 0.009$ & $0.021\pm 0.002$ \\
 3 & 0.500 -- 0.625 & $0.105\pm 0.006$ & $0.223\pm 0.010$ & $+0.019\pm 0.011$ & $0.032\pm 0.003$\\
 4 & 0.625 -- 0.750 & $0.112\pm 0.006$ & $0.160\pm 0.011$ & $+0.008\pm 0.011$ & $0.052\pm 0.004$\\
 5 & 0.750 -- 0.875 & $0.089\pm 0.005$ & $0.101\pm 0.009$ & $-0.022\pm 0.010$ & $0.057\pm 0.004$\\
 6 & 0.875 -- 1.000 & $0.144\pm 0.007$ & $0.020\pm 0.006$ & $+0.003\pm 0.006$ & $0.133\pm 0.007$ \\
    \end{tabular}
  \end{ruledtabular}
\label{tab:wtag}
\end{table*}
%
%
The total effective tagging efficiency for DS-II
is determined to be
$\eeff \equiv \sum_{l=1}^6 \epsilon_l(1-2w_l)^2 = \efftot$,
where 
$\epsilon_l$ is the event fraction for each $r$ interval
determined from the $\jpsi\ks$ data and is listed in
Table~\ref{tab:wtag}.
The error includes both statistical and systematic uncertainties.
We find that the wrong tag fractions for DS-II
are slightly smaller than those for DS-I. As a result,
the $\eeff$ value for DS-II is slightly larger than that for DS-I
($\eeff = \efftotdsone$).

\subsection{Vertex Reconstruction}
The vertex position for the $\fCP$ decay 
is reconstructed using charged tracks that have enough SVD hits:
at least one layer with hits on both sides 
and at least one additional $z$ hit in other layers for SVD-I,
and at least two layers with hits on both sides for SVD-II.
A constraint on the IP is also used with the selected tracks;
the IP profile is convolved with the finite $B$ flight length in the plane
perpendicular to the $z$ axis.
The pions from $\ks$ decays are not used
except in the analysis of $\bz\to\ks\piz$ and $\ks\ks\ks$ decays.
The typical vertex reconstruction efficiency and $z$ resolution 
for $\bz\to\phi\ks$ decays
are 95\% and 78 $\mu$m, respectively.
Similar values are obtained for other $\fCP$ decays except for
$\bz\to\ks\piz$ and $\ks\ks\ks$ decays.

The vertex
for $\bz\to\ks\piz$ decays is
reconstructed using
the $\ks$ trajectory and the IP constraint, where
both pions from the $\ks$ decay are required to
have enough SVD hits in the same way as for other $\fCP$ decays.
The reconstruction efficiency depends both on
the $\ks$ momentum and on the SVD geometry;
the efficiency with SVD-II (32\%) is significantly higher than
that with SVD-I (23\%)
because of the larger outer radius and the additional layer.
The typical $z$ resolution of the vertex reconstructed with the $\ks$ is
93~$\mu$m for SVD-I and 110~$\mu$m for SVD-II.

The vertex position for $\bz\to\ks\ks\ks$ decays is
also obtained using $\kspm$ trajectories and a constraint on the IP.
The reconstruction efficiency depends both on
the $\kspm$ momentum and on the SVD geometry.
The vertex efficiencies with SVD-I (SVD-II) are
79\% (86\%) for $\kspm\kspm\kspm$ and 62\% (74\%) for $\kspm\kspm\kszz$.
The typical vertex resolution is
about 97~$\mu$m (113~$\mu$m) for SVD-I (SVD-II) when two or three
$\kspm$ candidates can be used.
The resolution is worse when only one $\kspm$ can be used;
the typical value is 152~$\mu$m (168~$\mu$m) for SVD-I (SVD-II),
which is comparable to the $\ftag$ vertex resolution.

The $\ftag$ vertex determination with SVD-I
remains unchanged from the
previous publication~\cite{Chen:2005dr},
and is described in detail elsewhere~\cite{bib:resol_nim};
to minimize the effect of long-lived particles, 
secondary vertices from charmed hadrons and a small fraction of
poorly reconstructed tracks, we adopt an iterative procedure
in which the track that gives the largest contribution to the
vertex $\chi^2$ is removed at each step 
until a good $\chi^2$ is obtained.
The reconstruction efficiency was measured to be 93\%.
The typical $z$ resolution is $140~\mu$m~\cite{bib:CP1_Belle}.

For SVD-II, we find that
the same vertex reconstruction algorithm results in
a larger outlier fraction when only
one track remains after the iteration procedure.
Therefore, in this case, we repeat the
iteration procedure with
a more stringent requirement on the SVD-II hit pattern;
at least two of the three outer layers have hits on both sides.
The resulting outlier fraction, which is described in
Sec.~\ref{sec:results}, is comparable to
that for SVD-I, while 
the inefficiency caused by this change is small (2.5\%).

\subsection{Summary of Signal Yields}
The signal yields that contribute to the determination 
of $CP$-violation parameters, $\nsig$, for $\bz\to\fCP$ decays
are summarized in Table~\ref{tab:num}.
These yields are obtained after flavor tagging and vertex reconstruction
for all modes except $\bz\to\ks\piz$.
As events with no vertex information reduce the statistical error
on ${\cal A}_{\ks\piz}$ significantly, we include them
in the fit for the $\bz\to\ks\piz$ decay.
The signal purities are also listed in the table.
The signal yields are all consistent with
expected values that are obtained from
previously measured branching fractions~\cite{bib:HFAG}
and reconstruction efficiencies estimated
from MC simulation studies.
%
%
\begin{table}
\caption{
Estimated signal purities and
signal yields $\nsig$ in the signal region for each $\fCP$ mode
that is used to measure $CP$ asymmetries.
The purity is defined as $\nsig/\nev$,
where $\nev$ is the total number of events in the signal region.
Results for $\bz\to\ks\piz$ decays are obtained
with samples after flavor tagging but
before vertex reconstruction.
Results for other decays are obtained
after flavor tagging and vertex reconstruction.}
\label{tab:num}
\begin{ruledtabular}
\begin{tabular}{llllr}
\multicolumn{1}{c}
{Mode}    &         &$\xi_f$        
                           &\multicolumn{1}{c}{purity} 
                                          & \multicolumn{1}{c}{$\nsig$} \\
\hline
$\phi\ks$         & & $-1$ & $\Pphiks$    & $\Nsigphiks$ \\
$\phi\kl$         & & $+1$ & $\Pphikl$    & $\Nsigphikl$ \\
$\eta'\ks$        & & $-1$ & $\Petapks$   & $\Nsigetapks$ \\
$\eta'\kl$        & & $+1$ & $\Petapkl$   & $\Nsigetapkl$ \\
$\ks\ks\ks$       & & $+1$ & $\Pksksks$   & $\Nsigksksks$ \\
$\ks\piz$         & & $-1$ & $\Pkspiz$    & $\Nsigkspiz$ \\
$\fzero\ks$       & & $+1$ & $\Pfzeroks$  & $\Nsigfzeroks$ \\
$\omega\ks$       & & $-1$ & $\Pomegaks$  & $\Nsigomegaks$ \\
$\kp\km\ks$       & & $\xifkpkmksResult$ & $\Pkpkmks$   & $\Nsigkpkmks$ \\
\hline
$\jpsi\ks$        & & $-1$ & $\Pjpsiks$ & $\Nsigjpsiks$ \\
$\jpsi\kl$        & & $+1$ & $\Pjpsikl$   & $\Nsigjpsikl$ \\
\end{tabular}
\end{ruledtabular}
\end{table}
%
%

\section{Results of {\boldmath $CP$} Asymmetry Measurements}
\label{sec:results}
We determine $\cals$ and $\cala$ for each mode by performing an unbinned
maximum-likelihood fit to the observed $\Dt$ distribution.
The probability density function (PDF) expected for the signal
distribution, ${\cal P}_{\rm sig}(\Dt;\cals,\cala,\fq,w_l,\dwl)$, 
is given by Eq.~(\ref{eq:psig}) incorporating
the effect of incorrect flavor assignment. The distribution is
convolved with the
proper-time interval resolution function 
$\rsig$,
which takes into account the finite vertex resolution. 

For the decays 
$\bz\to$
$\phi\ks$, 
$\phi\kl$, 
$\eta'\ks$, 
$\etap\kl$,
$\fzero\ks$, 
$\omega\ks$, 
$\kp\km\ks$, 
$\jpsi\ks$
and
$\jpsi\kl$,
we use flavor-specific $B$ decays governed by
semileptonic or hadronic $b\to c$ transitions
to determine the resolution function.
We perform a simultaneous multiparameter fit to these high-statistics
control samples to obtain the resolution function parameters,
wrong-tag fractions (Section~\ref{sec:flavor tagging}), 
$\dmd$, $\taubp$ and $\taubz$.
We use the same resolution function used for
the $\sinbb$ measurement for DS-I~\cite{bib:BELLE-CONF-0436}.
For DS-II, the following modifications are introduced:
a sum of two Gaussian functions is used
to model the resolution of the $\fCP$ vertex
while a single Gaussian function is used for DS-I;
a sum of two Gaussian functions is used to model
the resolution of the tag-side vertex
obtained with one track and the IP constraint,
while a single Gaussian function is used for DS-I.
These modifications are needed to account for
differences between SVD-I and SVD-II, as well as
different background conditions in DS-I and DS-II.
We test the resolution parameterization using
MC events on which we overlay
beam-related background taken from data.
A fit to the MC sample yields correct values
for all parameters. 

For the $\bz\to\ks\piz$ decay,
we use the resolution function described above
with additional parameters that rescale vertex
errors. The rescaling function depends on
the detector configuration (SVD-I or SVD-II), 
SVD hit patterns of charged pions from the $\ks$ decay,
and $\ks$ decay vertex position in the plane 
perpendicular to the beam axis.
The parameters in the rescaling function are determined from a fit to the 
$\Dt$ distribution of $\bz\to\jpsi\ks$ data.
Here only the $\ks$ and the IP constraint are used for the
vertex reconstruction, the $\bz$ lifetime is
fixed at the world average value, and $b$-flavor tagging
information is not used so that the 
expected PDF is
an exponential function convolved with
the resolution function.

We check the resulting resolution function
by also reconstructing the vertex with
leptons from $\jpsi$ decays and the IP constraint.
We find that the distribution of the
distance between the vertex positions obtained with
the two methods is well represented by
the resolution function convolved with
the well-known resolution for the $\jpsi$ vertex.
Finally, we also perform a fit to the $\bz\to\jpsi\ks$ sample
with $b$-flavor information and obtain
${\cal S}_{\jpsi\ks} = \SjpsiksKsvResultStat$(stat) and
${\cal A}_{\jpsi\ks} = \AjpsiksKsvResultStat$(stat),
which
are in good agreement with our measurement using
leptons from $\jpsi$ decays, which will be described later.
A separate fit to the same sample with $\taubz$ as a free
parameter yields
$\taubz = \LifejpsiksKsvResultStat$(stat) ps,
which is consistent with the world average value.
Thus, we conclude that 
the vertex resolution for the $\bz\to\ks\piz$ decay
is well understood.

For $\bz\to\ks\ks\ks$ candidates,
we use the same resolution function that is used for the $\bz\to\ks\piz$
decay if only one $\kspm$ is available for the
vertex reconstruction.
For events with $n$ (= 2 or 3) $\kspm$ trajectories used in the vertexing,
we adopt a function defined as
$[{\cal V}(\sigma_z)]^{1/n}$ to rescale the vertex error $\sigma_z$.
Here ${\cal V}(\sigma_z)$ is the aforementioned rescaling function for
the case that only one $\kspm$ is available. 
We find from MC simulation that the resolution is well described by
this form for the rescaling function.

We determine the following likelihood for each
event:
\begin{eqnarray}
P_i
&=& (1-\fol)\int \biggl[
\fsig{\cal P}_{\rm sig}(\Dt')R_{\rm sig}(\Dt_i-\Dt') \nonumber \\
&+&(1-\fsig){\cal P}_{\rm bkg}(\Dt')R_{\rm bkg}(\Dt_i-\Dt')\biggr]
d(\Dt')  \nonumber \\
&+&\fol P_{\rm ol}(\Dt_i),
\label{eq:likelihood}
\end{eqnarray}
where $P_{\rm ol}(\Dt)$ is a broad Gaussian function that represents
an outlier component with a small fraction $\fol$~\cite{bib:BELLE-CONF-0436}.
The width of the outlier component for DS-I is determined to be
$(39^{+2}_{-13})$ ps; the fractions of the outlier components are
$(2.1^{+1.2}_{-0.8})\times 10^{-4}$ for events with the $\ftag$ vertex reconstructed
with more than one track, and $(3.1^{+0.3}_{-0.6})\times 10^{-2}$
for the case only one track is used.
Here the errors include both statistical and systematic errors.
Corresponding values for DS-II are
$(35^{+8}_{-11})$ ps,
$(3.6^{+2.0}_{-1.1})\times 10^{-4}$ and
$(1.8^{+0.2}_{-0.3})\times 10^{-2}$.
The signal probability $\fsig$ depends on the $r$ region and
is calculated on an event-by-event basis
as a function of $\pbstar$ for the $\bz\to\jpsi\kl$ decay,
$\pbstar$ and $\rsigbkg$
for the $\bz\to\phi\kl$ and $\etap\kl$ decays, 
$\dE$, $\mb$ and $\cos\theta_H$
for the $\bz\to\phi\ks$ decay,
$\dE$, $\mb$ and $M_{3\pi}$
for the $\omega\ks$ decays,
and
$\dE$ and $\mb$ for the other modes.
A PDF for background events, ${\cal P}_{\rm bkg}(\Dt)$,
is modeled as a sum of exponential and prompt components, and
is convolved with a sum of two Gaussians $R_{\rm bkg}$.
Parameters in ${\cal P}_{\rm bkg} (\Dt)$ and $R_{\rm bkg}$ 
for continuum background are determined by a fit to the $\Dt$ distribution
for events outside the $\dE$-$\mb$ signal region except for the 
$\bz\to\phi\kl$ and $\etap\kl$ decays.
For the $\bz\to\phi\kl$ and $\etap\kl$
decays, we use $\pbstar$ sideband events to obtain the
parameters.
Parameters in ${\cal P}_{\rm bkg}(\Dt)$ and $R_{\rm bkg}$
for $B\overline{B}$ background events in $\bz\to\eta'\ks$,
$\ks\piz$, $\phi\kl$ and $\etap\kl$
decays are determined from MC simulation. 

We fix $\tau_\bz$ and $\dmd$ at
their world average values~\cite{bib:PDG2005}.
We assume no $CP$ asymmetry in the background $\Delta t$ distributions and 
possible $CP$ asymmetries in the $B$ decay backgrounds are treated as
sources of systematic error.
In order to reduce the statistical error on $\cala$,
we include events without vertex information
in the analysis of $\bz\to\ks\piz$.
The likelihood in this case is obtained by integrating 
Eq.~(\ref{eq:likelihood}) over $\Dt_i$.

\begin{figure}
\includegraphics[width=0.48\textwidth]{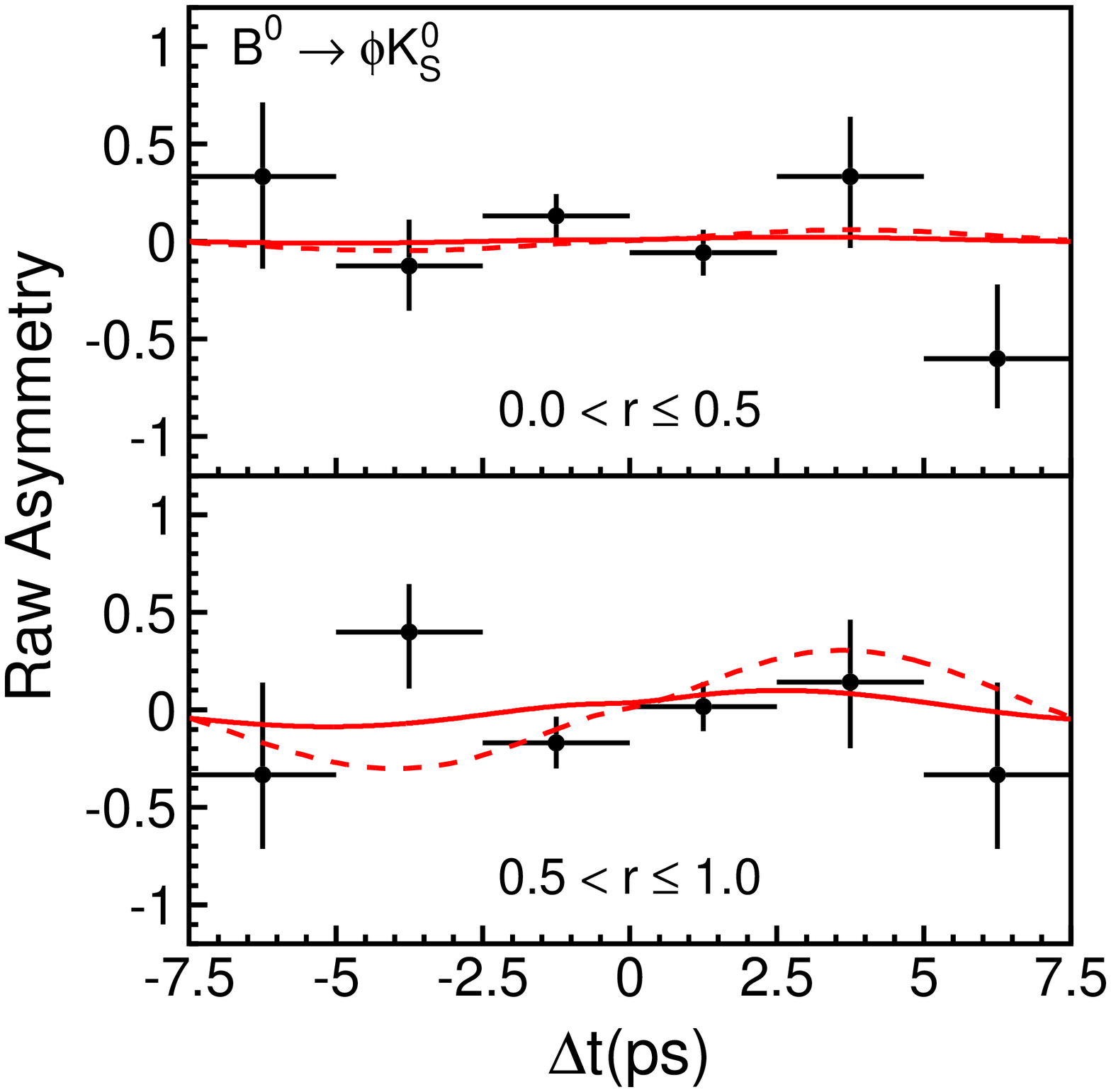} 
\includegraphics[width=0.48\textwidth]{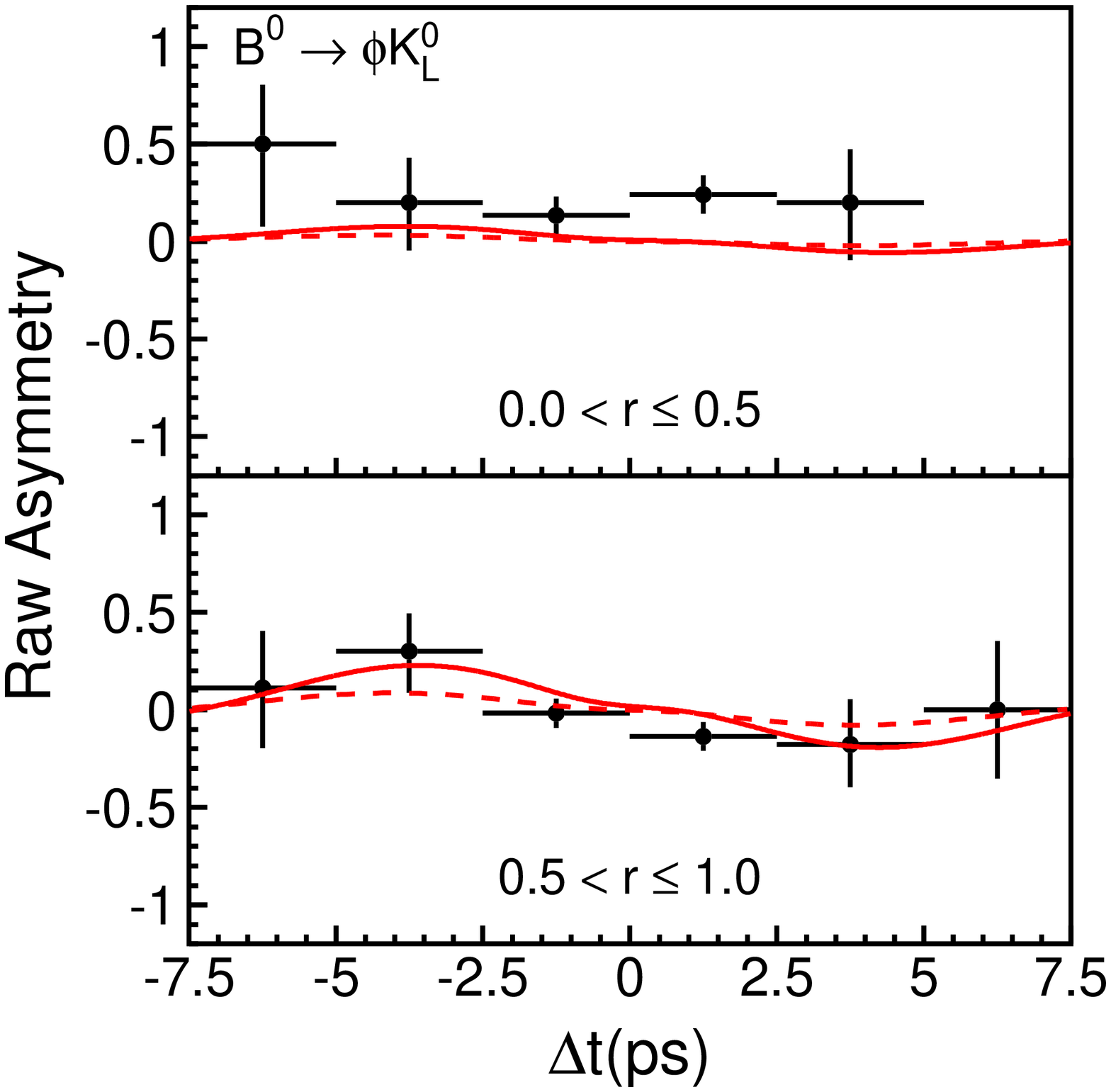} 
\includegraphics[width=0.48\textwidth]{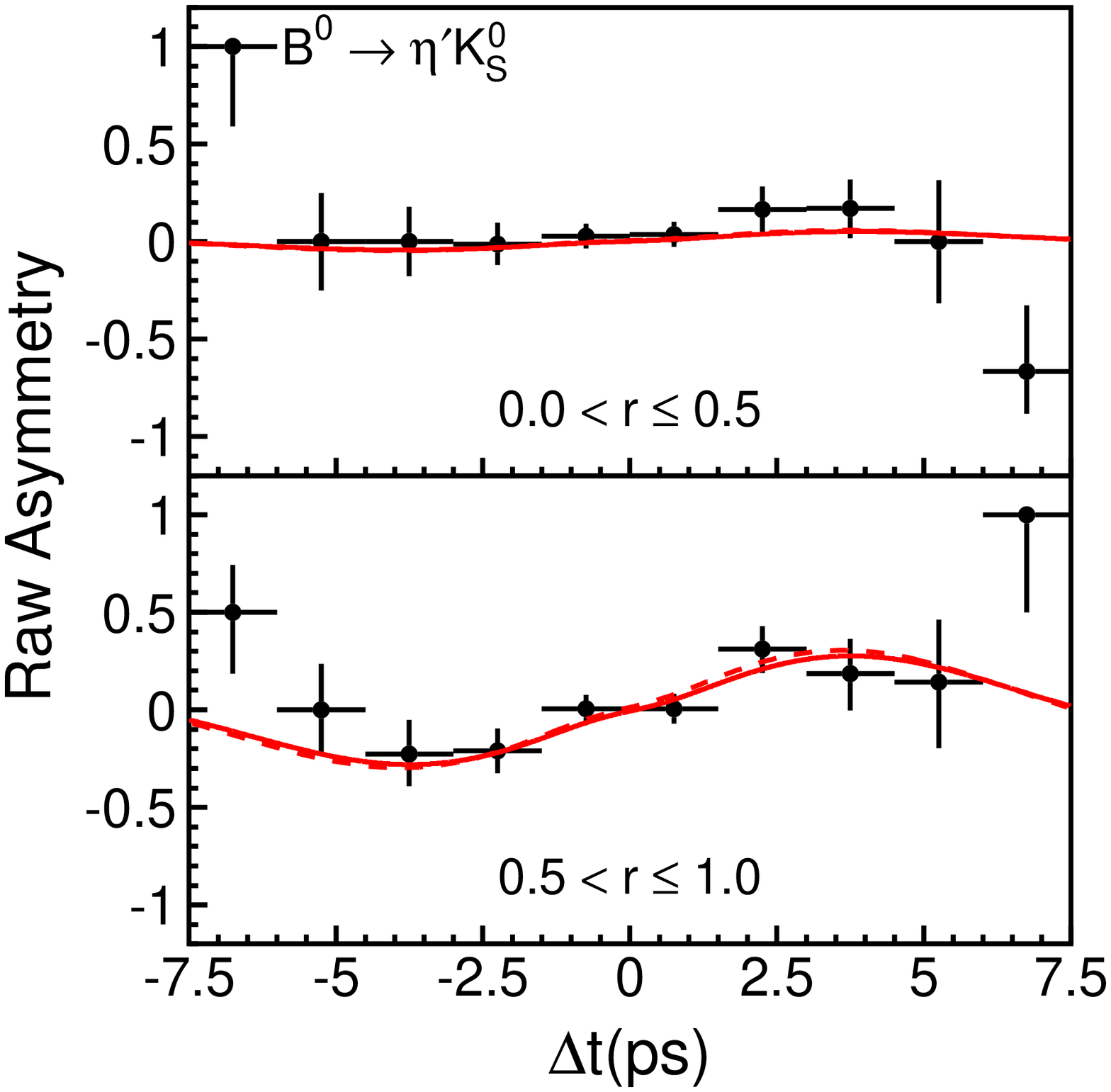} 
\includegraphics[width=0.48\textwidth]{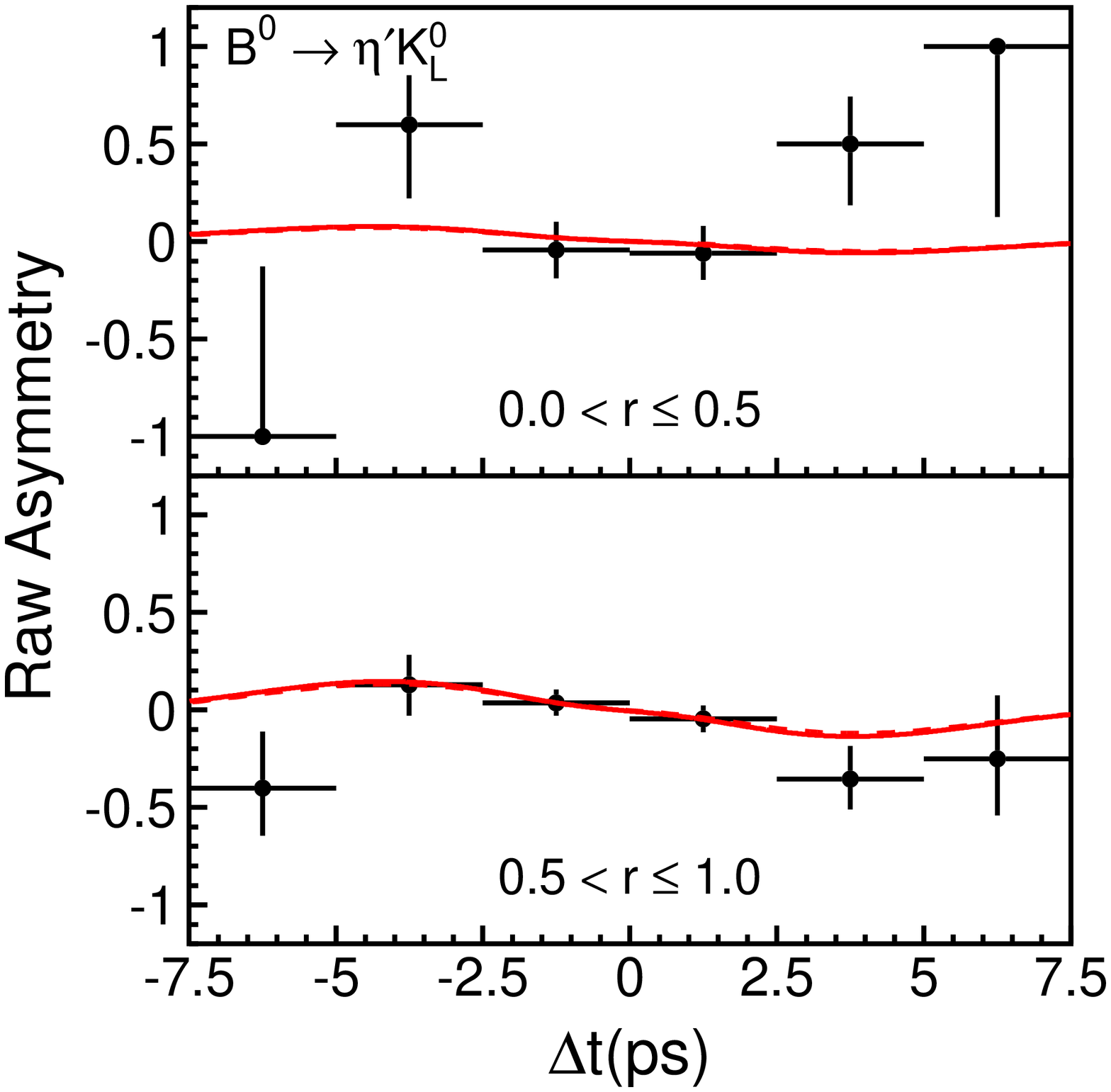} 
\caption{
Raw asymmetry in each $\Dt$ bin with $0 < r \le 0.5$ (top)
and with $0.5 < r \le 1.0$ (bottom) for 
(a) $\bz\to\phi\ks$, 
(b) $\bz\to\phi\kl$, 
(c) $\bz\to\eta'\ks$
and
(d) $\bz\to\eta'\kl$
decays.
The solid curves show the results of the 
unbinned maximum-likelihood fits.
The dashed curves show the
SM expectation with
our measurement of $CP$-violation parameters for the 
$\bz\to\jpsi\kz$ mode
($\sinbb = \SjpsikzVal$ and $\cala = \AjpsikzVal$).}
\label{fig:asym}
\rput[l]( -1.8,  18.0)  {\Large(a)}
\rput[l]( 6.3,  18.0)  {\Large(b)}
\rput[l]( -1.8, 10.1)  {\Large(c)}
\rput[l]( 3.8,  10.1)  {\Large(d)}
\end{figure}

\begin{figure}
\includegraphics[width=0.32\textwidth]{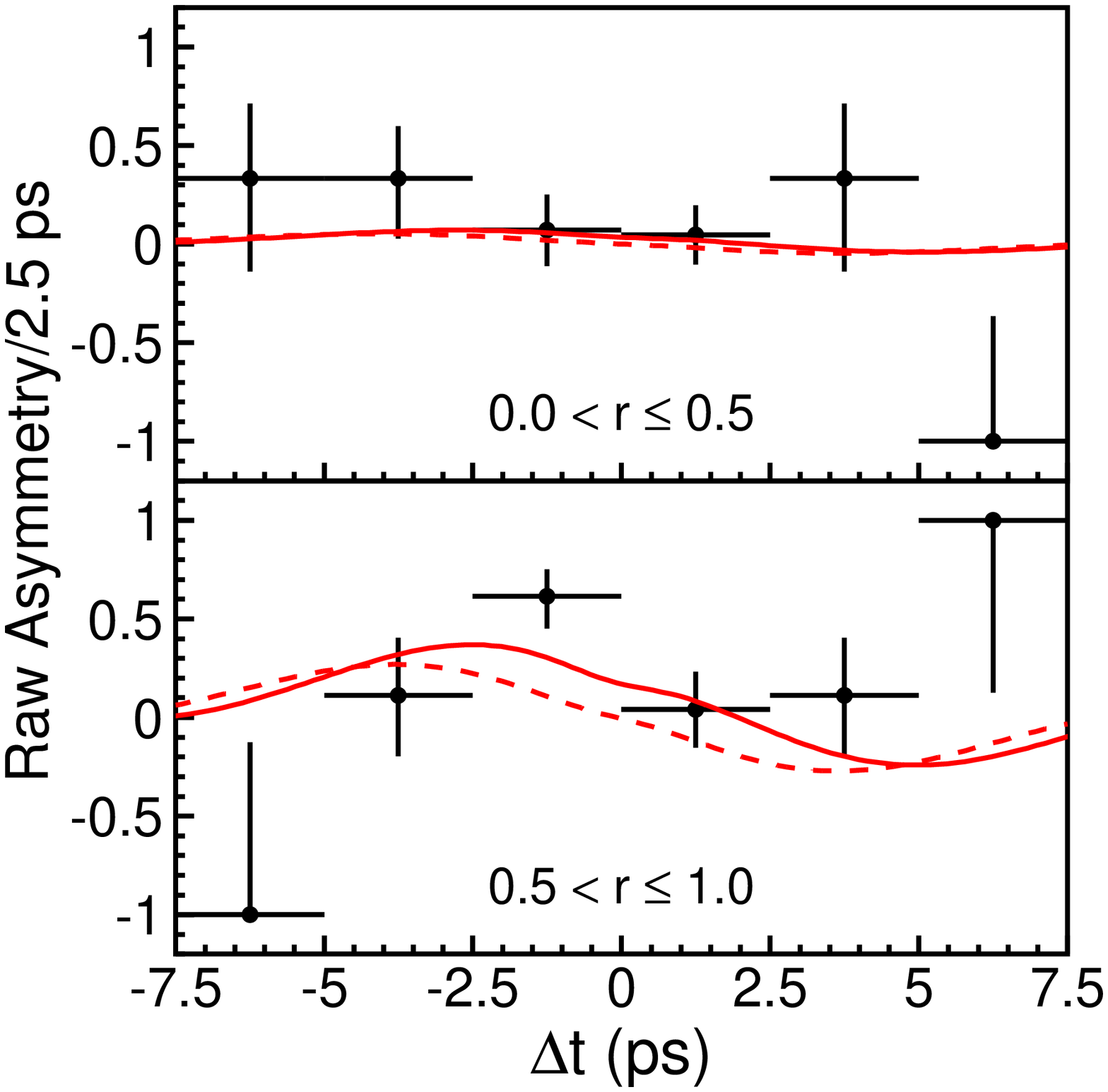} 
\includegraphics[width=0.32\textwidth]{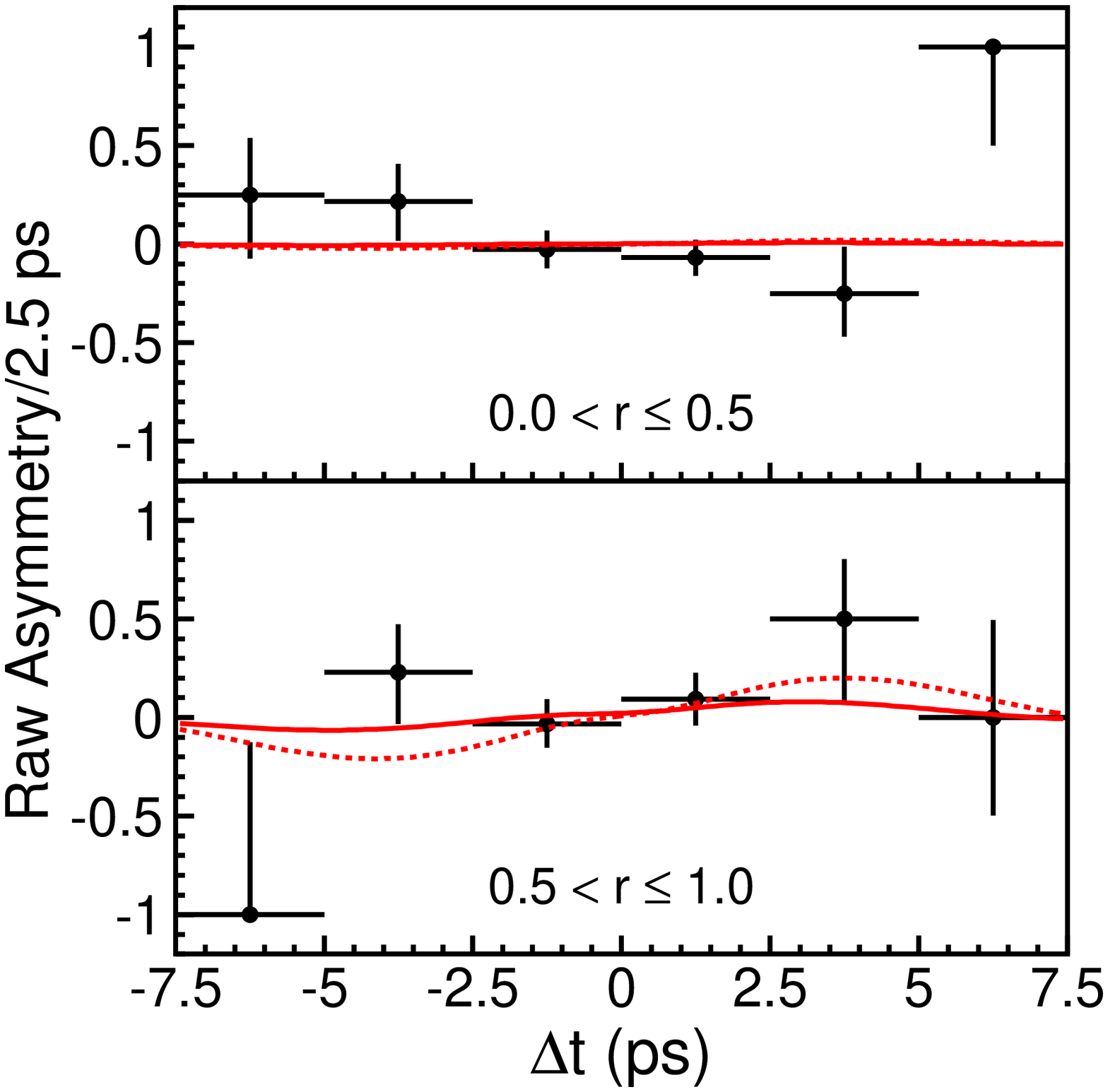} 
\includegraphics[width=0.32\textwidth]{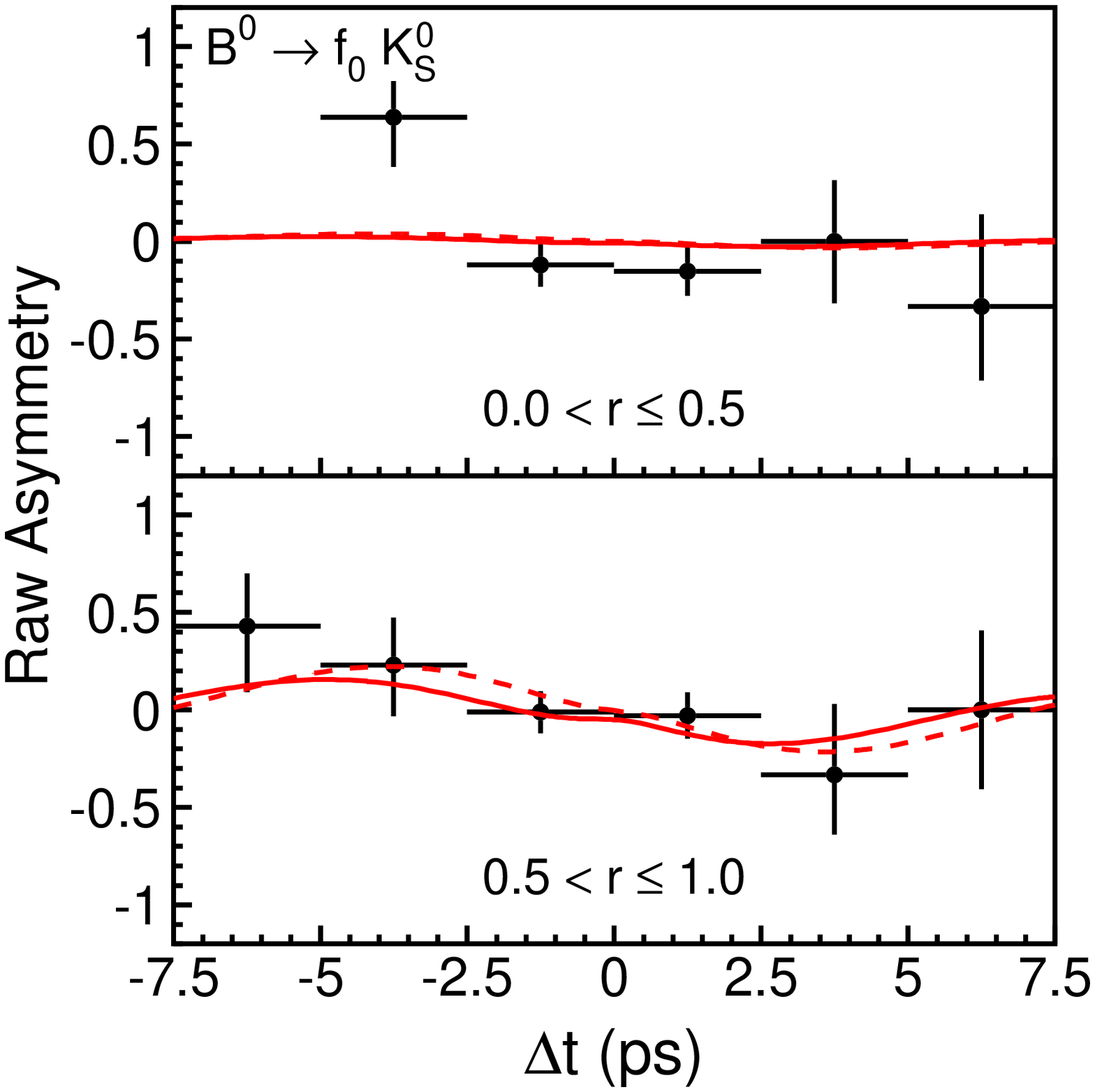} 
\includegraphics[width=0.32\textwidth]{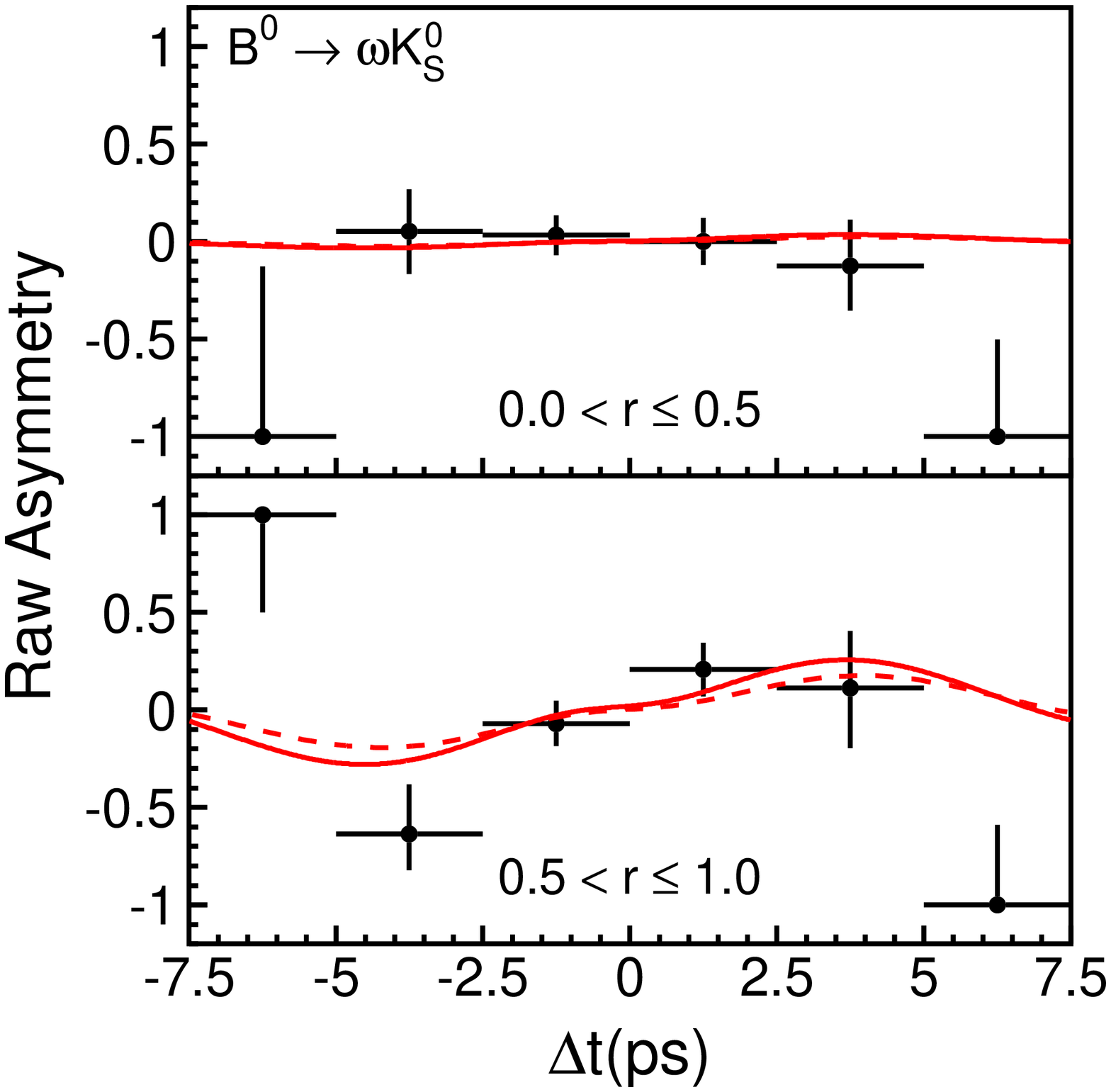} 
\includegraphics[width=0.32\textwidth]{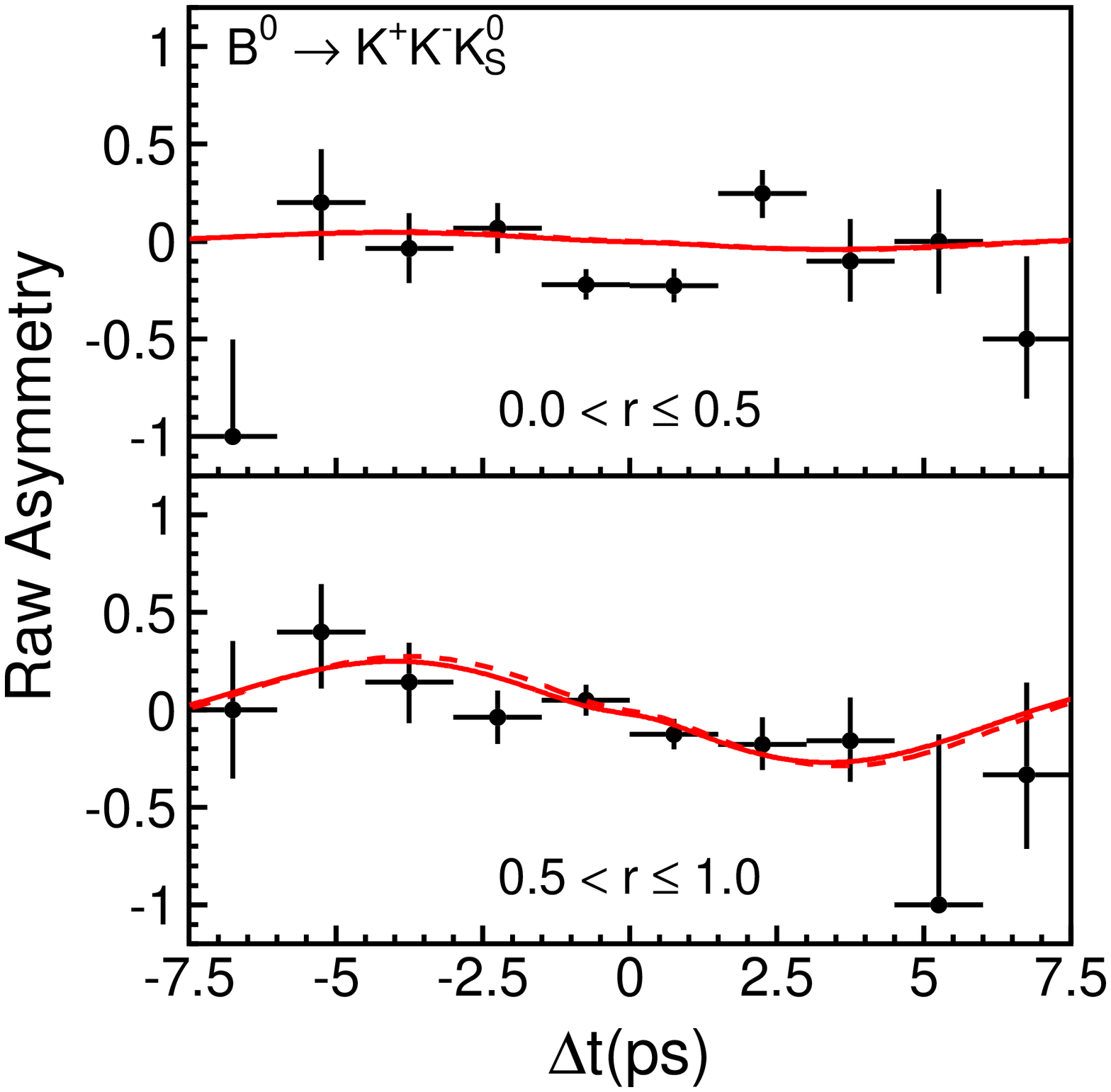} 
\caption{
Raw asymmetry in each $\Dt$ bin with $0 < r \le 0.5$ (top)
and with $0.5 < r \le 1.0$ (bottom) for 
(a) $\bz\to\ks\ks\ks$
(b) $\bz\to\ks\piz$,
(c) $\bz\to\fzero\ks$,
(d) $\bz\to\omega\ks$
and
(e) $\bz\to\kp\km\ks$
decays.
The solid curves show the results of the 
unbinned maximum-likelihood fits.
The dashed curves show the
SM expectation with
our measurement of $CP$-violation parameters for the 
$\bz\to\jpsi\kz$ mode
($\sinbb = \SjpsikzVal$ and $\cala = \AjpsikzVal$).}
\label{fig:asym2}
\rput[l]( -3.9,  13.2)  {(a)}
\rput[l]( 0.0,  13.2)  {(b)}
\rput[l]( 6.9, 13.2)  {(c)}
\rput[l]( -1.2,  7.9)  {(d)}
\rput[l]( 4.3,  7.9)  {(e)}
\end{figure}

\begin{figure}
\includegraphics[width=0.32\textwidth]{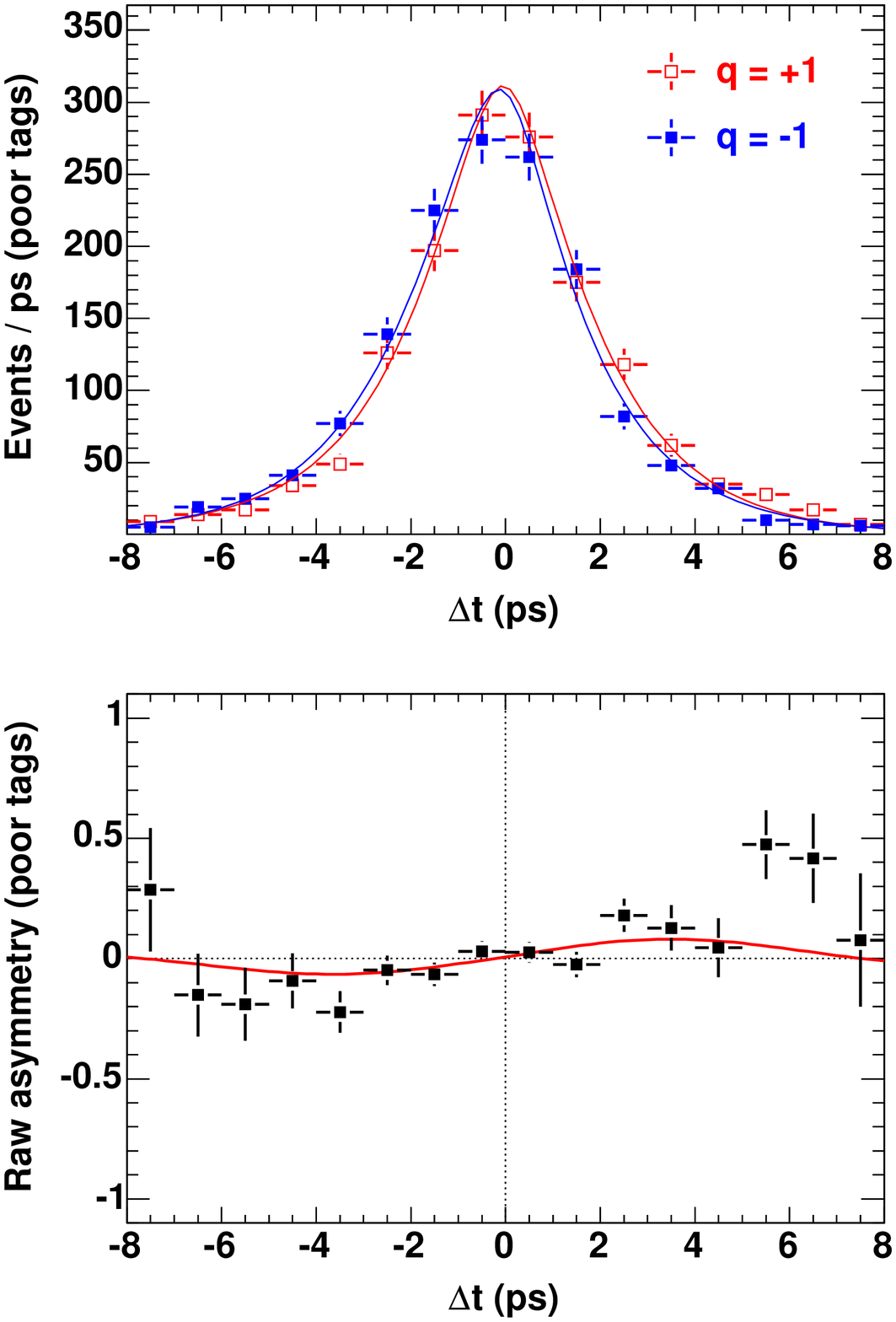} 
\includegraphics[width=0.32\textwidth]{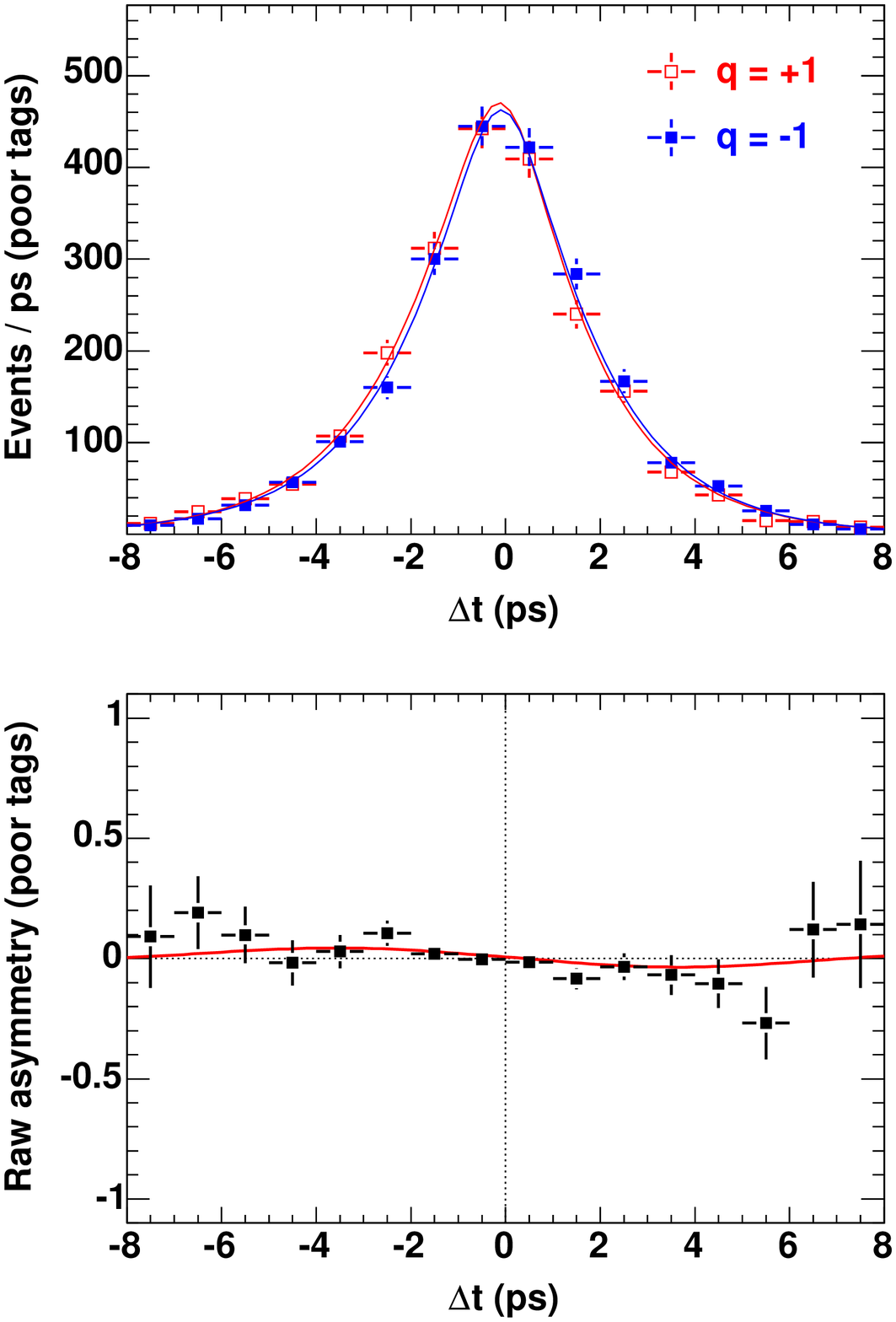} 
\caption{
$\Dt$ distributions in (a) $\bz\to\jpsi\ks$, (b) $\bz\to\jpsi\kl$,
and raw asymmetries in (c) $\bz\to\jpsi\ks$ and (d) $\bz\to\jpsi\kl$
with $0 < r \le 0.5$. 
The curves show the results of the 
unbinned maximum-likelihood fits.}
\label{fig:asym-jpsikz-poot}
\rput[l]( -4.4,  9.5)  {(a) $\jpsi\ks$}
\rput[l](  1.0,  9.5)  {(b) $\jpsi\kl$}
\rput[l]( -4.4,  5.7)  {(c) $\jpsi\ks$}
\rput[l](  1.0,  5.7)  {(d) $\jpsi\kl$}
\end{figure}

\begin{figure}
\includegraphics[width=0.48\textwidth]{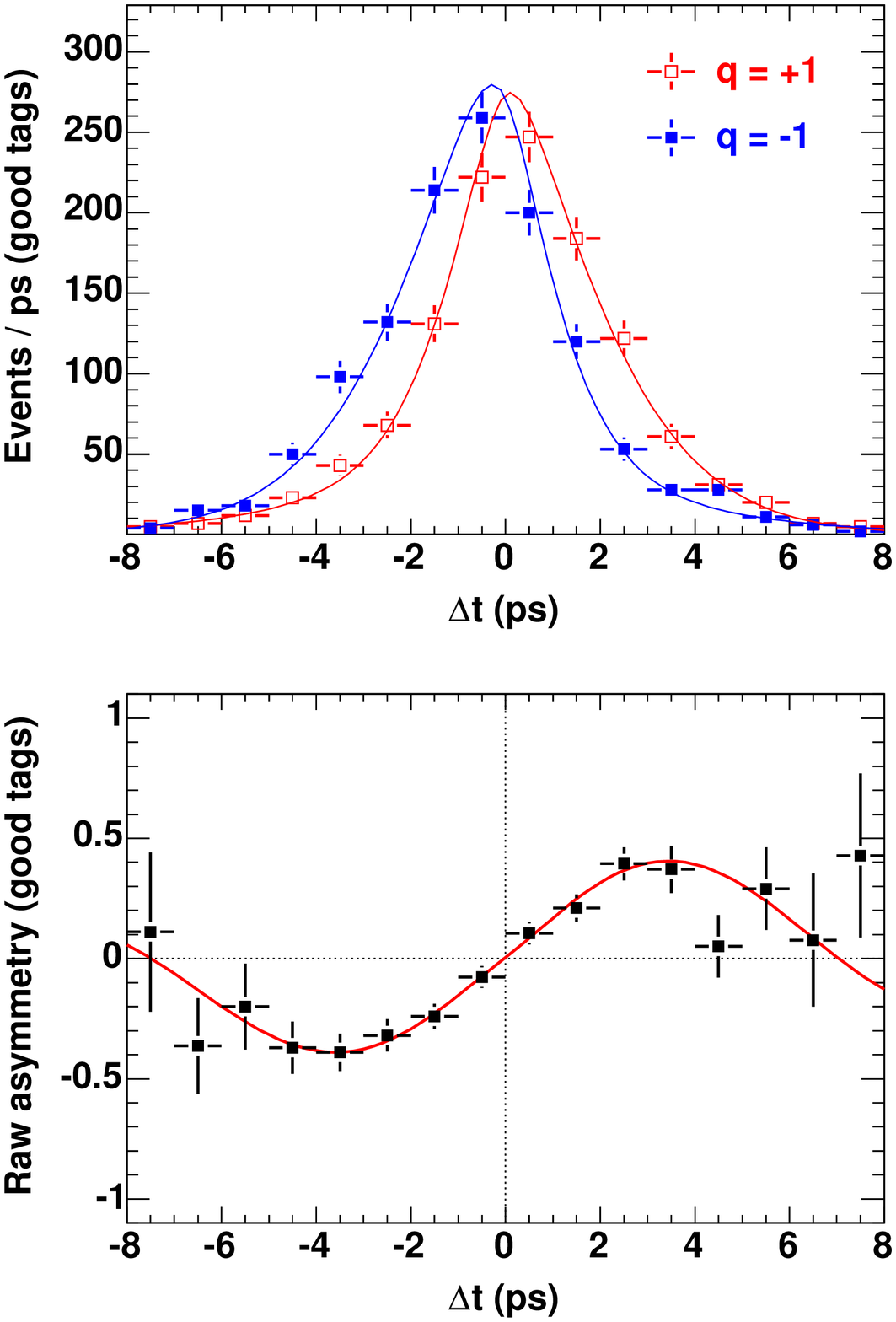} 
\includegraphics[width=0.48\textwidth]{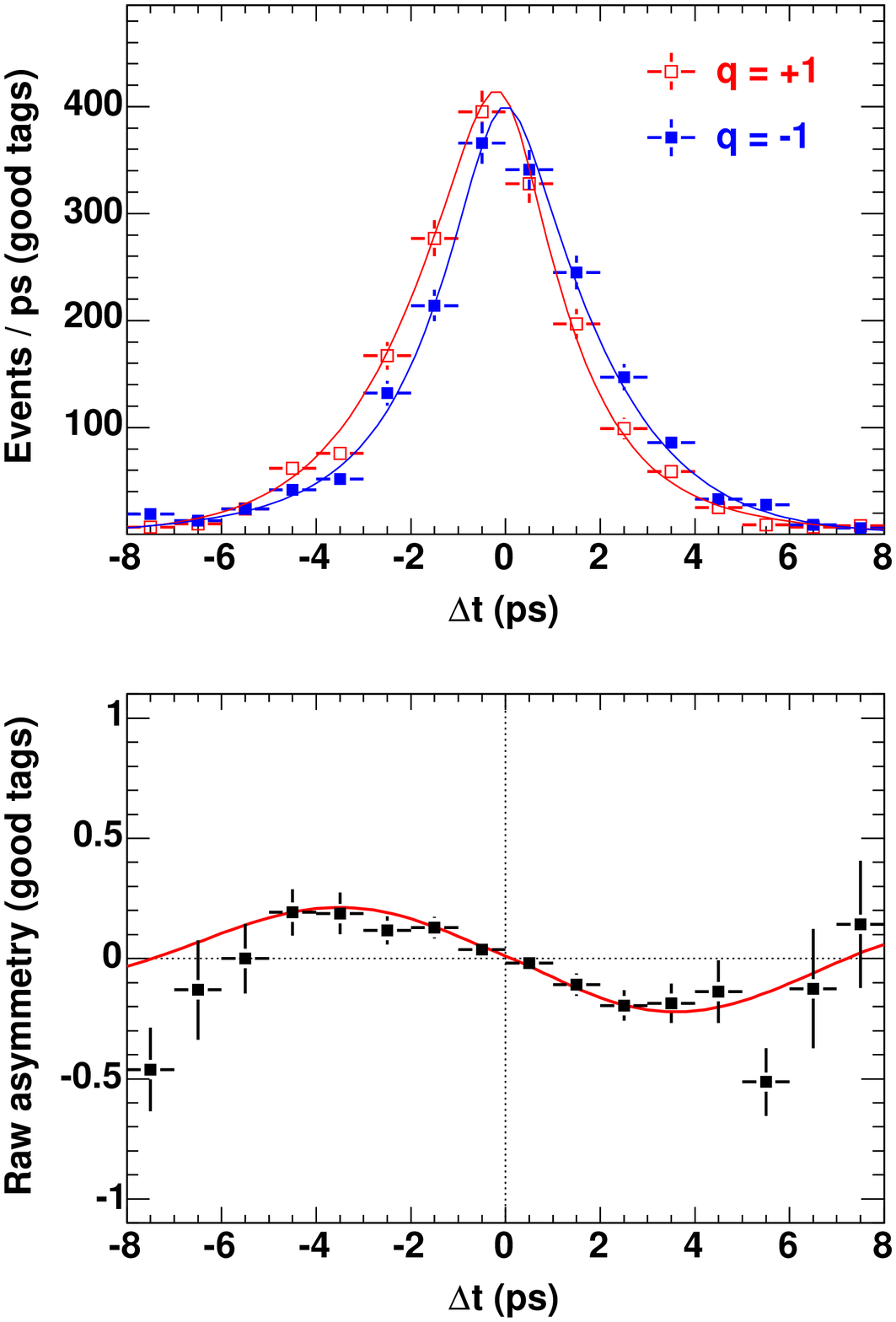} 
\caption{
$\Dt$ distributions in (a) $\bz\to\jpsi\ks$, (b) $\bz\to\jpsi\kl$,
and raw asymmetries in (c) $\bz\to\jpsi\ks$ and (d) $\bz\to\jpsi\kl$
with $0.5 < r \le 1$. 
The curves show the results of the 
unbinned maximum-likelihood fits.}
\label{fig:asym-jpsikz-good}
\rput[l]( -6.5,  12.9)  {\large(a) $\jpsi\ks$}
\rput[l](  1.5,  12.9)  {\large(b) $\jpsi\kl$}
\rput[l]( -6.5,  7.3)  {\large(c) $\jpsi\ks$}
\rput[l](  1.5,  7.3)  {\large(d) $\jpsi\kl$}
\end{figure}

\begin{figure}
\includegraphics[width=0.8\textwidth]{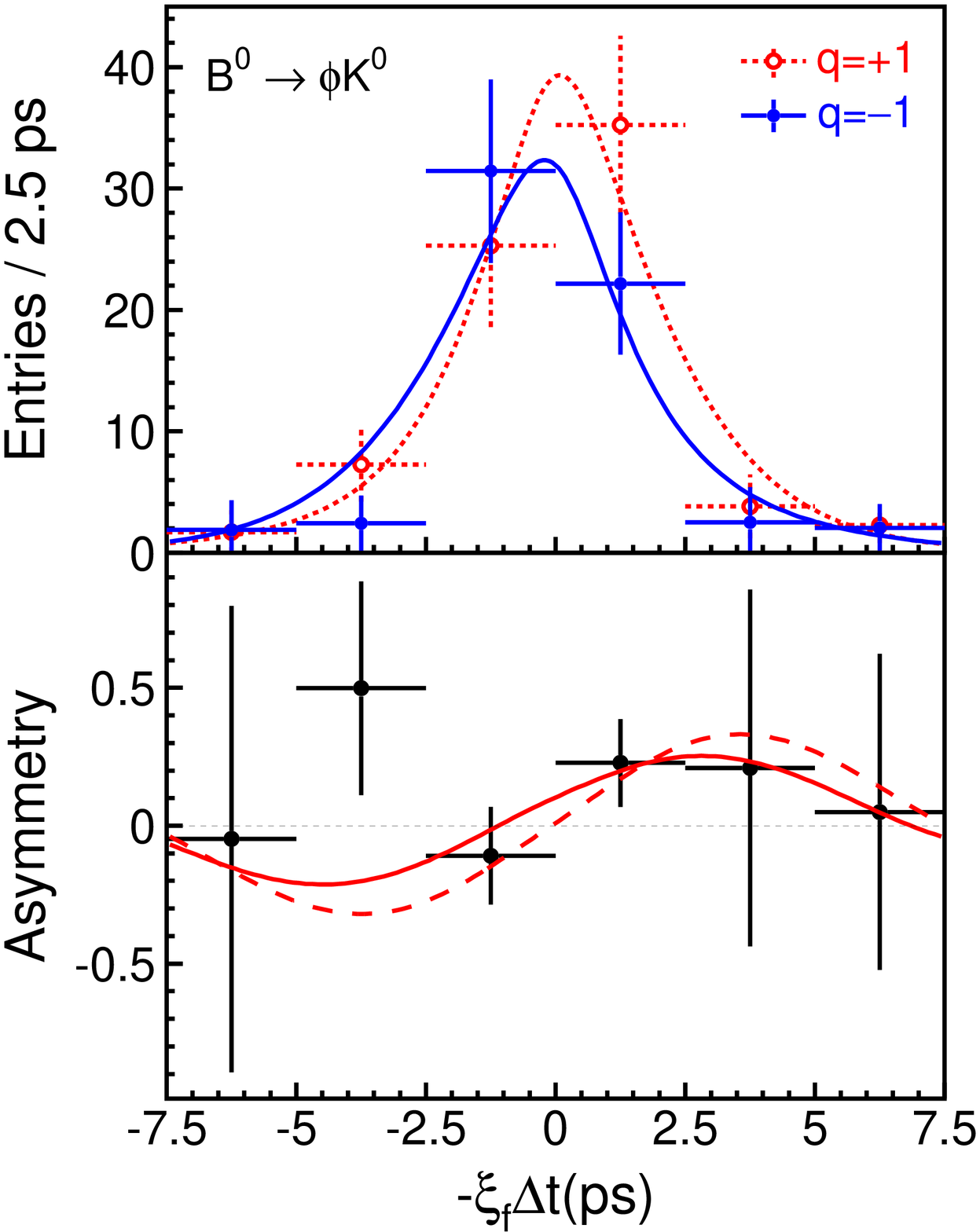} 
\caption{
Background-subtracted $\Dt$ distribution (top) and asymmetry 
in each $\Dt$ bin (bottom) with $0.5 < r \le 1.0$ for $\bz\to\phi\kz$.
The result of the 
unbinned maximum-likelihood fit is also shown.
The dashed curve in the bottom figure shows the
SM expectation with
our measurement of $CP$-violation parameters for the 
$\bz\to\jpsi\kz$ mode
($\sinbb = \SjpsikzVal$ and $\cala = \AjpsikzVal$).}
\label{fig:dt-asym-bgsub-phikz}
\end{figure}

\begin{figure}
\includegraphics[width=0.8\textwidth]{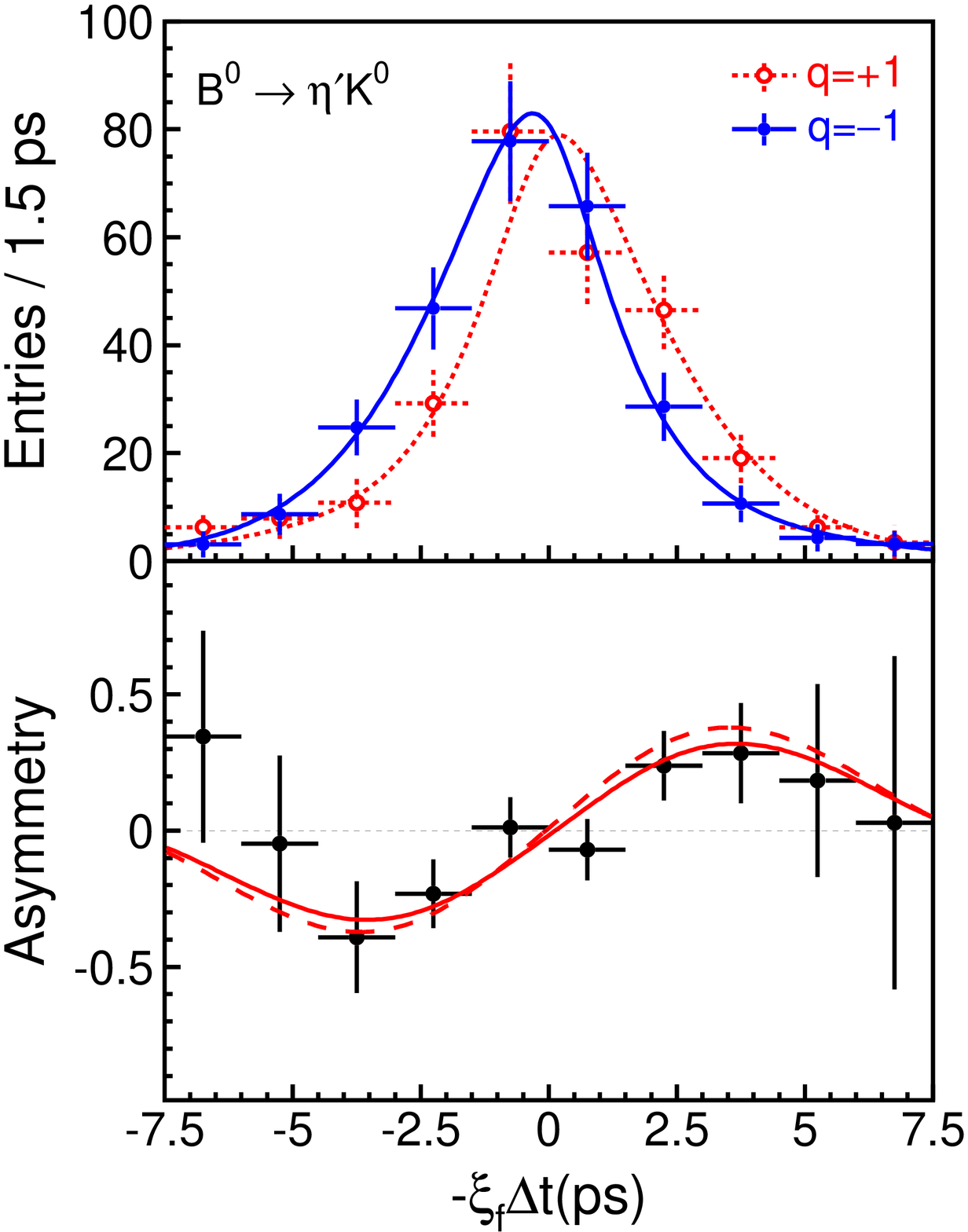} 
\caption{
Background-subtracted $\Dt$ distributions and asymmetry 
in each $\Dt$ bin with $0.5 < r \le 1.0$ for $\bz\to\eta'\kz$.
The result of the unbinned maximum-likelihood fit
is also shown.
The dashed curve in the bottom figure shows the
SM expectation with
our measurement of $CP$-violation parameters for the 
$\bz\to\jpsi\kz$ mode
($\sinbb = \SjpsikzVal$ and $\cala = \AjpsikzVal$).}
\label{fig:dt-asym-bgsub-etapkz}
\end{figure}

\begin{figure}
\includegraphics[width=0.8\textwidth]{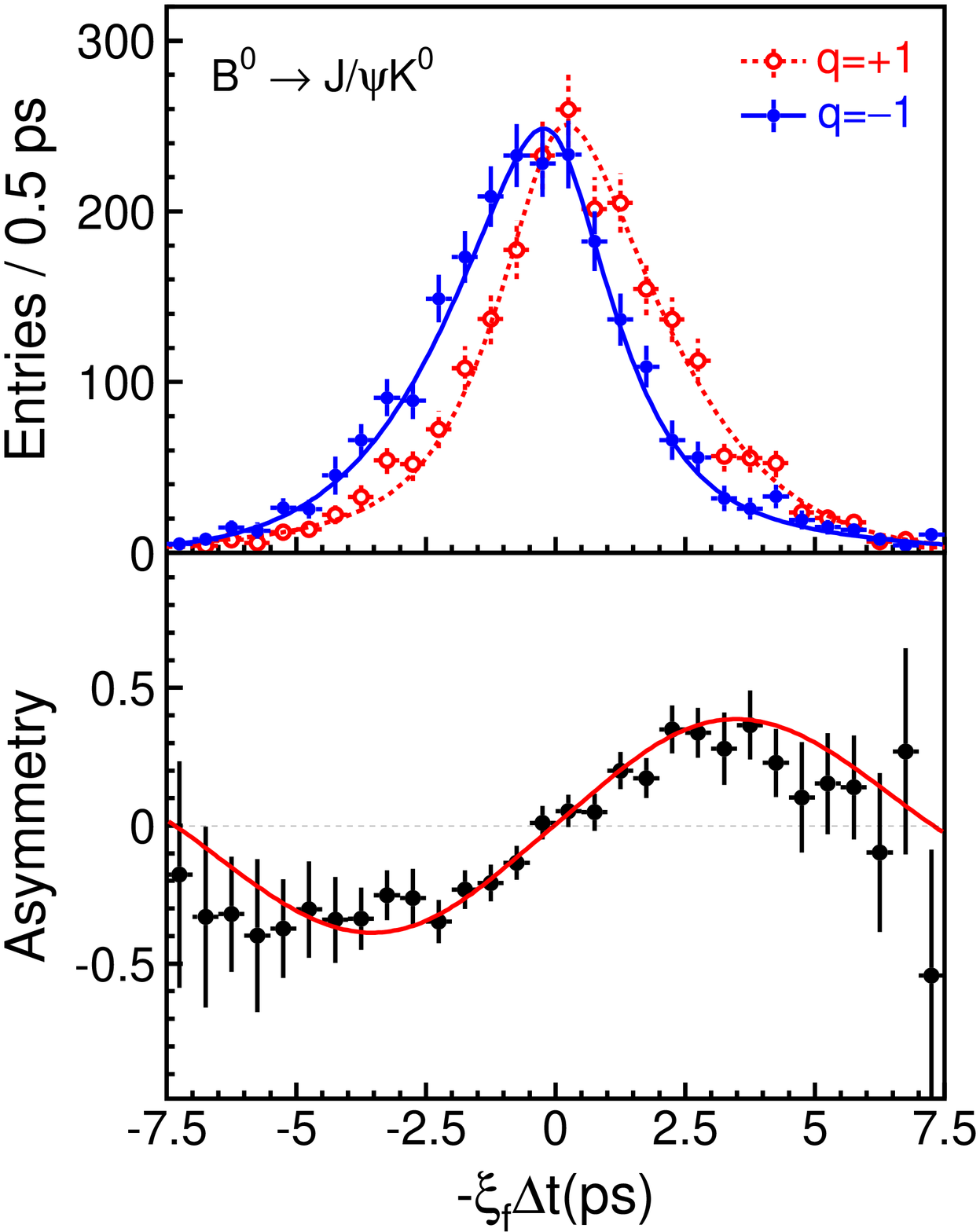} 
\caption{
Background-subtracted $\Dt$ distributions and asymmetry 
in each $\Dt$ bin with $0.5 < r \le 1.0$ for $\bz\to\jpsi\kz$.
The result of the
unbinned maximum-likelihood fit is also shown.}
\label{fig:dt-asym-bgsub-jpsikz}
\end{figure}

The only free parameters in the final fits
are $\cals$ and $\cala$, which are determined by maximizing the
likelihood function
$L = \prod_iP_i(\Dt_i;\cals,\cala)$
where the product is over all events.
Table~\ref{tab:result} summarizes
the fit results of $\cals$ and $\cala$.
We define the raw asymmetry in each $\Dt$ bin by
$(N_{q=+1}-N_{q=-1})/(N_{q=+1}+N_{q=-1})$,
where $N_{q=+1(-1)}$ is the number of 
observed candidates with $q=+1(-1)$.
%
Figures~\ref{fig:asym}-\ref{fig:asym-jpsikz-good}
show the raw asymmetries for each decay mode
in two regions of the flavor-tagging
parameter $r$~\cite{footnote:rawasym}.
%
%
\begin{table}
\caption{Results of the fits to the $\Dt$ distributions.
The first error is statistical and the second
error is systematic.}
\label{tab:result}
\begin{ruledtabular}
\begin{tabular}{lcll}
\multicolumn{1}{c}{Mode} &  
SM expectation for $\cals$ &
\multicolumn{1}{c}{$\cals$} & 
\multicolumn{1}{c}{$\cala$} \\
\hline
$\phi\kz$    & $+\sinbb$   & $\SphikzResult$    & $\AphikzResult$    \\
~~~$\phi\ks$ & $+\sinbb$   & $\SphiksResultStat$& $\AphiksResultStat$\\
~~~$\phi\kl$ & $-\sinbb$   & $\SphiklResultStat$& $\AphiklResultStat$\\
$\eta'\kz$   & $+\sinbb$   & $\SetapkzResult$   & $\AetapkzResult$   \\
~~~$\etap\ks$& $+\sinbb$   & $\SetapksResultStat$&$\AetapksResultStat$\\
~~~$\etap\kl$& $-\sinbb$   & $\SetapklResultStat$&$\AetapklResultStat$\\
$\ks\ks\ks$  & $-\sinbb$   & $\SksksksResult$   & $\AksksksResult$   \\
$\ks\piz$    & $+\sinbb$   & $\SkspizResult$    & $\AkspizResult$    \\
$\fzero\ks$  & $-\sinbb$   & $\SfzeroksResult$  & $\AfzeroksResult$  \\
$\omega\ks$  & $+\sinbb$   & $\SomegaksResult$  & $\AomegaksResult$  \\
$\kp\km\ks$  & $-(2f_{+}-1)\sinbb$& $\SkpkmksResult$ & $\AkpkmksResult$   \\
\hline
$\jpsi\kz$   & $+\sinbb$   & $\SjpsikzResult$    & $\AjpsikzResult$    \\
~~~$\jpsi\ks$& $+\sinbb$   & $\SjpsiksResultStat$& $\AjpsiksResultStat$\\
~~~$\jpsi\kl$& $-\sinbb$   & $\SjpsiklResultStat$& $\AjpsiklResultStat$\\
\end{tabular}
\end{ruledtabular}
\end{table}
%
%
%
%
%
%
%
%
%
%
Figures~\ref{fig:dt-asym-bgsub-phikz}-\ref{fig:dt-asym-bgsub-jpsikz}
also show $\Dt$ distributions and
asymmetries for $\bz\to\phi\kz$, $\etap\kz$ and $\jpsi\kz$ decays
after subtracting background contributions,
where 
the sign of each $\Dt$ measurement for the final
states with $\kl$ is inverted
to combine final states with $\ks$ and $\kl$. 
%
%
%
%
%

%
%
%


Tables~\ref{tab:ssyserr} and \ref{tab:asyserr}
list the systematic errors on $\cals$ and $\cala$, respectively.
The total systematic errors are obtained
by adding each contribution in quadrature,
and are smaller than the statistical errors for all 
$b\to s$ modes.

To determine the systematic error that arises from
uncertainties in the vertex reconstruction,
the track and vertex selection criteria
are varied to search for possible systematic biases.
Small biases in the $\Dz$ measurement 
are observed in $e^+e^-\to\mu^+\mu^-$ and other control
samples. Systematic errors 
are estimated by applying special correction functions
to account for the observed biases, repeating
the fit, and comparing the obtained values with the nominal results.
The systematic error due to the IP constraint 
in the vertex reconstruction is estimated by
varying ($\pm10 \mu$m) the smearing used to account for the
$B$ flight length.
Systematic errors due to imperfect SVD alignment
are determined
from MC samples that have artificial misalignment effects
to reproduce impact-parameter resolutions observed in data.

Systematic errors due to uncertainties in the wrong tag
fractions are studied by varying
the wrong tag fraction individually for each $r$ region.
Systematic errors due to uncertainties in the resolution function
are also estimated by varying each resolution parameter obtained from
data (MC) by $\pm 1\sigma$ ($\pm 2\sigma$), repeating the fit
and adding each variation in quadrature.
Each physics parameter such as $\taubz$ and $\dmd$
is also varied by its error.
A possible fit bias is examined by fitting a large number of MC events.

Systematic errors from uncertainties in the background fractions
and in the background $\Dt$ shape
are estimated by varying each background parameter obtained
from data (MC) by $\pm 1\sigma$ ($\pm 2\sigma$).

The PDF's for $\bz\to\phi\kl$ and $\etap\kl$ assume
no correlation among $\rsigbkg$, $\pbstar$ and $r$.
To estimate systematic errors due to possible correlations between
$\rsigbkg$ and $\pbstar$, we repeat a fit to obtain
$CP$ parameters using signal fractions
determined by the $\rsigbkg$ distribution for 
each $\pbstar$ region separately. The difference from our
nominal result is included in the 
systematic error.
Systematic errors due to other possible correlations are estimated
from events in the $\pbstar$ sideband, events with $r<0.25$ and
off-resonance data.

Additional sources of systematic errors are 
considered for $B$ decay backgrounds
that are neglected in the PDF.
We consider uncertainties both in their fractions
and $CP$ asymmetries.
The effect of backgrounds from $\kp\km\ks$ and 
$\fzero\ks~(\fzero\to\kp\km)$
in the $\bz\to\phi\ks$ sample is considered.
Uncertainties from
$B\to\phi K^*$ and other rare $B$ decay backgrounds
in the $\bz\to\phi\kl$ sample
are also taken into account.
For the $\bz\to\etap\ks$ sample, non-resonant $B$ decay backgrounds
are studied using events in the sideband of the reconstructed $\etap$ mass distribution.
Effects of possible $CP$ asymmetries in $B$ decay backgrounds for $\ks\piz$ 
and $\fzero\ks$ are evaluated.
The peaking background fraction in the $\bz\to\fzero\ks$ sample
depends on the functions used to fit to the 
$\pip\pim$ invariant mass distribution.
The systematic errors due to the uncertainties of the masses and widths of
the resonances used in the fit are also included.
The width of $\fzero$ as well as the mass and the width of $\fx$ are varied
by their errors.
The effect of possible interference between resonant and non-resonant
amplitudes, which is neglected in the nominal analysis, is also evaluated.
We perform a fit to the $\pip\pim$ distribution
of a MC sample generated with
interfering amplitudes and phases for $B\to K\pi\pi$ decays measured
from data~\cite{Garmash:2003er}. The observed difference in the
signal yield from the true value is taken into account in the 
systematic error determination.
We also repeat the fit to the $\Dt$ distribution
ignoring the contribution of the peaking background.
The differences in $\cals$ and $\cala$ from
our nominal results are included in the systematic error.

Finally, we investigate the effects of interference between
CKM-favored and CKM-suppressed $B\to D$ transitions in
the $\ftag$ final state~\cite{Long:2003wq}.
A small correction to the PDF for the signal distribution
arises from the interference.
We estimate the size of the correction using the $\bzdslnu$ 
sample. We then generate MC pseudoexperiments
and make an ensemble test to obtain systematic biases
in $\cals$ and $\cala$. 
In general, we find effects
on $\cals$ are negligibly small, while
there are sizable possible shifts in $\cala$.
%
%
\begin{table*}
\caption{Summary of the systematic errors on $\cals$.}
\begin{ruledtabular}
\begin{tabular}{lrrrrrrr|r}
 & $\phi\kz$ 
 & $\eta'\kz$
 & $\ks\ks\ks$ 
 & $\ks\piz$
 & $\fzero\ks$ 
 & $\omega\ks$
 & $\kp\km\ks$ 
 & $\jpsi\kz$
\\
\hline
Vertex reconstruction 
 & $\SphikzSystVtx$ 
 & $\SetapkzSystVtx$
 & $\SksksksSystVtx$
 & $\SkspizSystVtx$
 & $\SfzeroksSystVtx$
 & $\SomegaksSystVtx$
 & $\SkpkmksSystVtx$
 & $\SjpsikzSystVtx$
\\
Flavor tagging  
 & $\SphikzSystFbtg$ 
 & $\SetapkzSystFbtg$ 
 & $\SksksksSystFbtg$ 
 & $\SkspizSystFbtg$ 
 & $\SfzeroksSystFbtg$ 
 & $\SomegaksSystFbtg$ 
 & $\SkpkmksSystFbtg$ 
 & $\SjpsikzSystFbtg$ 
\\
Resolution function 
 & $\SphikzSystResol$ 
 & $\SetapkzSystResol$ 
 & $\SksksksSystResol$ 
 & $\SkspizSystResol$ 
 & $\SfzeroksSystResol$ 
 & $\SomegaksSystResol$ 
 & $\SkpkmksSystResol$ 
 & $\SjpsikzSystResol$ 
\\
Physics parameters
 & $\SphikzSystPhys$
 & $\SetapkzSystPhys$
 & $\SksksksSystPhys$
 & $\SkspizSystPhys$
 & $\SfzeroksSystPhys$
 & $\SomegaksSystPhys$
 & $\SkpkmksSystPhys$
 & $\SjpsikzSystPhys$
\\
Possible fit bias
 & $\SphikzSystFit$
 & $\SetapkzSystFit$
 & $\SksksksSystFit$
 & $\SkspizSystFit$
 & $\SfzeroksSystFit$
 & $\SomegaksSystFit$
 & $\SkpkmksSystFit$
 & $\SjpsikzSystFit$
\\
Background fraction
 & $\SphikzSystBG$
 & $\SetapkzSystBG$
 & $\SksksksSystBG$
 & $\SkspizSystBG$
 & $\SfzeroksSystBG$
 & $\SomegaksSystBG$
 & $\SkpkmksSystBG$
 & $\SjpsikzSystBG$
\\
Background $\Dt$ shape
 & $\SphikzSystBGdt$
 & $\SetapkzSystBGdt$
 & $\SksksksSystBGdt$
 & $\SkspizSystBGdt$
 & $\SfzeroksSystBGdt$
 & $\SomegaksSystBGdt$
 & $\SkpkmksSystBGdt$
 & $\SjpsikzSystBGdt$
\\
Tag-side interference 
 & $\SphikzSystTagIntrfr$
 & $\SetapkzSystTagIntrfr$
 & $\SksksksSystTagIntrfr$
 & $\SkspizSystTagIntrfr$
 & $\SfzeroksSystTagIntrfr$
 & $\SomegaksSystTagIntrfr$
 & $\SkpkmksSystTagIntrfr$
 & $\SjpsikzSystTagIntrfr$
\\
\hline      
Total    
 & $\SphikzSyst$
 & $\SetapkzSyst$
 & $\SksksksSyst$
 & $\SkspizSyst$
 & $\SfzeroksSyst$
 & $\SomegaksSyst$
 & $\SkpkmksSyst$
 & $\SjpsikzSyst$
\\
\end{tabular}
\end{ruledtabular}
\label{tab:ssyserr} 
\end{table*}
%
%
\begin{table*}
\caption{Summary of the systematic errors on $\cala$.}
\begin{ruledtabular}
\begin{tabular}{lrrrrrrr|r}
 & $\phi\kz$ 
 & $\eta'\kz$
 & $\ks\ks\ks$ 
 & $\ks\piz$
 & $\fzero\ks$ 
 & $\omega\ks$
 & $\kp\km\ks$ 
 & $\jpsi\kz$
\\
\hline
Vertex reconstruction 
 & $\AphikzSystVtx$ 
 & $\AetapkzSystVtx$
 & $\AksksksSystVtx$
 & $\AkspizSystVtx$
 & $\AfzeroksSystVtx$
 & $\AomegaksSystVtx$
 & $\AkpkmksSystVtx$
 & $\AjpsikzSystVtx$
\\
Flavor tagging  
 & $\AphikzSystFbtg$ 
 & $\AetapkzSystFbtg$ 
 & $\AksksksSystFbtg$ 
 & $\AkspizSystFbtg$ 
 & $\AfzeroksSystFbtg$ 
 & $\AomegaksSystFbtg$ 
 & $\AkpkmksSystFbtg$ 
 & $\AjpsikzSystFbtg$ 
\\
Resolution function 
 & $\AphikzSystResol$ 
 & $\AetapkzSystResol$ 
 & $\AksksksSystResol$ 
 & $\AkspizSystResol$ 
 & $\AfzeroksSystResol$ 
 & $\AomegaksSystResol$ 
 & $\AkpkmksSystResol$ 
 & $\AjpsikzSystResol$ 
\\
Physics parameters
 & $\AphikzSystPhys$
 & $\AetapkzSystPhys$
 & $\AksksksSystPhys$
 & $\AkspizSystPhys$
 & $\AfzeroksSystPhys$
 & $\AomegaksSystPhys$
 & $\AkpkmksSystPhys$
 & $\AjpsikzSystPhys$
\\
Possible fit bias
 & $\AphikzSystFit$
 & $\AetapkzSystFit$
 & $\AksksksSystFit$
 & $\AkspizSystFit$
 & $\AfzeroksSystFit$
 & $\AomegaksSystFit$
 & $\AkpkmksSystFit$
 & $\AjpsikzSystFit$
\\
Background fraction
 & $\AphikzSystBG$
 & $\AetapkzSystBG$
 & $\AksksksSystBG$
 & $\AkspizSystBG$
 & $\AfzeroksSystBG$
 & $\AomegaksSystBG$
 & $\AkpkmksSystBG$
 & $\AjpsikzSystBG$
\\
Background $\Dt$ shape
 & $\AphikzSystBGdt$
 & $\AetapkzSystBGdt$
 & $\AksksksSystBGdt$
 & $\AkspizSystBGdt$
 & $\AfzeroksSystBGdt$
 & $\AomegaksSystBGdt$
 & $\AkpkmksSystBGdt$
 & $\AjpsikzSystBGdt$
\\
Tag-side interference 
 & $\AphikzSystTagIntrfr$
 & $\AetapkzSystTagIntrfr$
 & $\AksksksSystTagIntrfr$
 & $\AkspizSystTagIntrfr$
 & $\AfzeroksSystTagIntrfr$
 & $\AomegaksSystTagIntrfr$
 & $\AkpkmksSystTagIntrfr$
 & $\AjpsikzSystTagIntrfr$
\\
\hline      
Total    
 & $\AphikzSyst$
 & $\AetapkzSyst$
 & $\AksksksSyst$
 & $\AkspizSyst$
 & $\AfzeroksSyst$
 & $\AomegaksSyst$
 & $\AkpkmksSyst$
 & $\AjpsikzSyst$
\\
\end{tabular}
\end{ruledtabular}
\label{tab:asyserr} 
\end{table*}

\clearpage

Various crosschecks of the measurements are performed.
We reconstruct charged $B$ meson decays
that are the counterparts of the $\bz\to\fCP$ decays
and apply the same fit procedure.
All results for the $\cals$ term are consistent with no 
$CP$ asymmetry, as expected. 
Lifetime measurements are also performed for 
the $\fCP$ modes and the corresponding charged $B$ decay modes.
The fits yield
$\taubz$ and $\taubp$ values consistent with the world average values.
MC pseudoexperiments are generated for each decay mode to
perform ensemble tests.
We find that the statistical errors obtained
in our measurements are all consistent
with the expectations from the ensemble tests.

The results in this report are consistent with those in our previous
publications~\cite{Chen:2005dr,Sumisawa:2005fz,bib:BELLE-CONF-0436}
within statistical fluctuations and supersede them.
Among our new results,
the largest difference from the previous measurement
is observed in the $\bz\to\ks\ks\ks$ decay.
A fit to the $\lintlastsummer$ fb$^{-1}$ data sample, which 
contains the entire DS-I and a part of DS-II and
were used in the previous publication,
yields $\cals = +1.05\pm 0.64$(stat) and $\cala = +0.51\pm 0.30$(stat),
where a small change from the previous measurement is due to
an improvement in the selection of $\ks$ candidates.
A fit to an additional $\lintthisyear$ fb$^{-1}$ data sample alone
yields $\cals = -2.95\pm 0.53$(stat) and $\cala = +0.64\pm 0.36$(stat).
From MC pseudoexperiments, 
the probability that the significance of difference 
is larger than the observed difference
is estimated to be 1.0\%.
To check if this arises due to a difference
in the SVD1 and SVD2 detectors,
we perform separate fits to DS-I and DS-II.
We obtain $\cals = -0.38\pm 0.82$(stat) and
$\cala = +0.16\pm 0.43$(stat) for DS-I and
$\cals = -0.92\pm 0.49$(stat) and
$\cala = +0.72\pm 0.28$(stat) for DS-II, which are
consistent to each other.
As all the other checks mentioned above also yield
results consistent with expectations,
we conclude that the observed change in the $CP$-violation parameters
for the $\bz\to\ks\ks\ks$ mode
is due to a statistical fluctuation.

Table~\ref{tab:sinbbeff} summarizes
the $\sinbbeff$ determination based on our $\cals$ measurements.
For each mode, the first error shown in the table
is statistical and the second error
is systematic.
For the $\bz\to\kp\km\ks$ decay,
the SM prediction is given by $\cals = -(2f_{+}-1)\sinbbeff$.
The third error is an additional systematic error arising from the
uncertainty of the $CP$-even fraction.
The results for each individual decay mode are consistent with
$\sinbb$ obtained from the $\bz\to\jpsi\kz$ decay
within one standard deviation.

%
%
\begin{table}
\caption{Results of the $\sinbbeff$ measurements. The first errors are statistical, the second
errors are systematic, and the third error for the $\kp\km\ks$ mode arises from
the uncertainty in the $CP$-even fraction.}
\label{tab:sinbbeff}
\begin{ruledtabular}
\begin{tabular}{ll}
Mode & $\sinbbeff$ \\ 
\hline
$\phi\kz$    & $\sinbbphikzResult$    \\
$\eta'\kz$   & $\sinbbetapkzResult$   \\
$\ks\ks\ks$  & $\sinbbksksksResult$   \\
$\ks\piz$    & $\sinbbkspizResult$    \\
$\fzero\ks$  & $\sinbbfzeroksResult$  \\
$\omega\ks$  & $\sinbbomegaksResult$  \\
$\kp\km\ks$  & $\sinbbkpkmksResult$ \\
\hline
$\jpsi\kz$   & $\SjpsikzResult$ \\
\end{tabular}
\end{ruledtabular}
\end{table}
%
%
%
%

\section{Summary}
We have performed improved measurements of 
$CP$-violation parameters $\sinbbeff$ and $\cala$ for 
$\bz \to \phi \kz$, 
$\etap \kz$,
$\ks\ks\ks$,
$\ks\piz$,
$\fzero\ks$,
$\omega\ks$
and
$\kp\km\ks$
decays.
These charmless decays
are dominated by $b\to s$ flavor-changing neutral currents
and are sensitive to possible new $CP$-violating phases.

We have also measured $CP$ asymmetries in 
$\bz\to\jpsi\kz$ decays using the same data sample. 
The same analysis procedure as that used for the $b\to s$ modes
yields $\sinbb = \SjpsikzResultSS$, which serves as a SM reference point,
and $\cala = \AjpsikzResultSS$.

We do not see any significant deviation between the results for each
$b\to s$ mode and those for $\bz\to\jpsi\kz$.  
Since some models of new physics
predict such effects, our results can be used to constrain these
models.  However, many models predict smaller deviations which we cannot
rule out with the current experimental uncertainty.  Therefore, further
measurements with larger data samples are required in order to search for
new, beyond the SM, $CP$-violating phases in the $b\to s$ transition.


%% file: lp05ack.tex
We thank the KEKB group for the excellent operation of the
accelerator, the KEK cryogenics group for the efficient
operation of the solenoid, and the KEK computer group and
the National Institute of Informatics for valuable computing
and Super-SINET network support. We acknowledge support from
the Ministry of Education, Culture, Sports, Science, and
Technology of Japan and the Japan Society for the Promotion
of Science; the Australian Research Council and the
Australian Department of Education, Science and Training;
the National Science Foundation of China under contract
No.~10175071; the Department of Science and Technology of
India; the BK21 program of the Ministry of Education of
Korea and the CHEP SRC program of the Korea Science and
Engineering Foundation; the Polish State Committee for
Scientific Research under contract No.~2P03B 01324; the
Ministry of Science and Technology of the Russian
Federation; the Ministry of Higher Education, Science and Technology of the Republic of Slovenia;  the Swiss National Science Foundation; the National Science Council and
the Ministry of Education of Taiwan; and the U.S.\
Department of Energy.

%% file: sss05.bbl
\begin{thebibliography}{999}
\bibitem{footnote:CC}
Throughout this paper,
the inclusion of the charge conjugate decay mode is implied
unless otherwise stated.
\bibitem{footnote:costhetaH}
The signal yield for the $\bz\to\phi\ks~(\ks\to\piz\piz)$ decay
is determined from
an unbinned two-dimensional maximum-likelihood fit to the
$\dE$-$\mb$ distribution because of the limited statistics. 
We include $\cos\theta_H$ in the $\rsigbkg$ calculation
for this mode to obtain the best sensitivity.
The $\bz\to\phi\ks~(\ks\to\piz\piz)$ candidates
are included in Fig.~\ref{fig:phiks}(a) and (b), but
are not in (c).
\bibitem{footnote:rawasym}
While the numbers of signal events in the two $r$ regions are similar,
the effective tagging efficiency is much larger 
and the background dilution is smaller in the region $0.5 < r \le 1.0$.
Note that these projections onto the $\Delta t$ axis do not take into
account event-by-event information (such as the signal fraction, the
wrong tag fraction and the vertex resolution), which is used in the
unbinned maximum-likelihood fit.
\end{thebibliography}
